\documentclass[paper,notoc]{JHEP3}
\usepackage{epsfig,cite}
\usepackage{amsbsy} 
\usepackage{graphicx}
\usepackage{axodraw4j}

\def\beq{\begin{equation}}   
\def\eeq{\end{equation}}
\def\bea{\begin{eqnarray}}  
\def\eea{\end{eqnarray}}

\def\ihixs{{\tt iHixs}}

\title{\boldmath  Total cross-section  for Higgs boson hadroproduction with anomalous Standard-Model interactions}

\author{Charalampos Anastasiou\\
  Institute for Theoretical Physics, ETH Zurich,
  8093 Zurich, Switzerland\\
  E-mail: \email{babis@phys.ethz.ch}}

\author{Stephan Buehler\\
Institute for Theoretical Physics, ETH Zurich,
  8093 Zurich, Switzerland\\
  E-mail: \email{buehler@itp.phys.ethz.ch}}

\author{Franz Herzog\\
  Institute for Theoretical Physics, ETH Zurich,
  8093 Zurich, Switzerland\\
  E-mail: \email{fherzog@itp.phys.ethz.ch}}

\author{Achilleas Lazopoulos\\
  Institute for Theoretical Physics, ETH Zurich,
  8093 Zurich, Switzerland\\
  E-mail: \email{lazopoli@itp.phys.ethz.ch}}

\abstract{ 
We present a new program ({\ihixs}) which computes the inclusive Higgs
boson cross-section at hadron colliders. It  incorporates  QCD
corrections through NNLO, real and virtual electroweak corrections, 
mixed QCD-electroweak corrections, quark-mass effects through NLO in
QCD,  and finite width effects for the Higgs boson and heavy
quarks. {\ihixs}   can be used to obtain the most precise  cross-section
values in fixed order  perturbation theory in the Standard Model.  In
addition, it allows for a consistent evaluation of the cross-section 
in modified Higgs boson sectors with anomalous Yukawa and electroweak interactions
as required  in extensions of the Standard Model.
{\ihixs}  is interfaced with the LHAPDF library and can be used with
all available NNLO sets of parton distribution functions.  
} 

\begin{document}



\section{Introduction} 
\label{sec:introduction}

A major objective of experimental high energy physics is the discovery of the Higgs boson.  
Production cross-sections at hadron accelerator experiments are
expected  to be small. However, they are sufficiently significant in
order for the Standard Model (SM) Higgs boson to be discovered 
soon in a variety of signatures. 
A remarkable progress in this direction has been made after the direct
searches of LEP~\cite{Barate:2003sz}.  
The TEVATRON~\cite{Aaltonen:2011gs} experiments have already
demonstrated   sensitivity to  Higgs boson cross-sections of magnitudes as
in the Standard Model.    ATLAS and CMS have published limits on the
inclusive cross-section using  first LHC
data\cite{Collaboration:2011qi,Chatrchyan:2011tz} accumulated in 2010. 
These searches are expected to yield tight constraints  in their
forthcoming updates with data from the 2011 runs.

Interactions in the Higgs sector  of the Standard Model are only indirectly constrained experimentally. 
Theory-wise, a modified Higgs sector is a prerequisite for extensions
of  the Standard Model with a more  appealing UV completion.  
When computing Higgs boson production cross-sections, it is necessary
to allow for  the possibility of Higgs boson interactions with modified couplings 
from their Standard-Model values.  

At a  hadron collider, the inclusive Higgs boson cross-section is most
sensitive to the values of the quark  Yukawa couplings. 
The main hadroproduction mechanism is gluon fusion which receives
sizable contributions from top and bottom-quark loops.  In Standard Model 
extensions, the mass of elementary particles may not originate
entirely from the Higgs mechanism and, 
in addition, not yet discovered quarks may be postulated which also contribute to the cross-section.  
Very enhanced or very suppressed Higgs boson cross-sections 
may be obtained,  as it occurs for example  in  a  Standard Model with
a fourth generation~\cite{Marciano:1987gz,Hill:1987wq,Glover:1988sr,Barger:1986jt} and 
composite Higgs boson models\cite{Falkowski:2007hz,Giudice:2007fh,betta} correspondingly.  

Direct production from quark partons in hadrons (mainly the bottom and
charm) may also become sizable in models  with more than one  Higgs  
doublet\cite{Cheng:1987rs} or  dynamically generated Yukawa couplings\cite{Babu:1999me,Giudice:2008uua}.    
In models with a perturbative Higgs sector, gluon fusion is likely 
less sensitive  to modifications  of  the Higgs boson couplings to electroweak gauge bosons; these start
contributing to the production of a single Higgs  boson  only at the  two-loop level. 
The  bulk of the generally small electroweak corrections is due to loops with light Standard Model
quarks which have very well constrained electroweak couplings.

A light Higgs boson is  expected to have a small decay width  in the  Standard Model. The width could be  different  in extensions of the Standard Model. For  example, the  Higgs boson could have  a significant invisible decay width in scenaria with hidden sectors~\cite{vanderBij:2007um,vanderBij:2011wy,Georgi:2007ek}. The width of the Higgs boson is  expected to be  large  also for scenaria where a heavy mass  is  not disallowed. 

The  objective  of this publication
is to provide  precise  theory predictions for  the  production
cross-section of a Higgs boson in the Standard Model 
and extensions which alter the interactions  of the   Higgs boson and
quarks or electroweak gauge bosons.  This publication is accompanied
with a computer program, {\ihixs}, for the precise calculation of inclusive
Higgs boson cross-sections at the Tevatron and the LHC in such a general setup. 

\section{Features of {\ihixs} }
\label{sec:features}

In the last two decades many theoretical studies lead to improved estimates of the Standard Model Higgs boson total 
cross-section at hadron colliders with a level of $ 10-20\%$ precision. Using, extending and  combining the available theoretical 
calculations has  been proven to be a non-trivial task for both theorists and experimentalists. 
{\ihixs}  is an easy to use program which produces  accurate  predictions  for the Higgs  cross-section in the  Standard Model and 
can be adapted readily for a large  class  of extensions of  the
Standard Model.  

For the numerical evaluation of loop and phase-space integrals  
we have used a new Fortran package~\cite{chaplin} of harmonic polylogarithms with complex 
arguments.   For  one-loop box and triangle master integrals with different internal masses  we have used the library {\tt OneLOop} of 
Ref.~\cite{vanHameren:2010cp,vanHameren:2009dr} which allows for complex masses.  We have tested  our implementation of  all other one-loop master  integrals against both {\tt OneLOop} and  the numerical program of Ref.\cite{Ellis:2007qk}.

{\ihixs}  computes the inclusive  Higgs boson cross-section through   next-to-next-to-leading-order (NNLO) 
in perturbative QCD. 
A  large  source  of uncertainty for Higgs  boson cross-sections at hadron colliders  is  the precision in the  determination of the parton densities.  It is  therefore important to compare the effect of  diverse existing determinations of parton densities  on the  Higgs cross-section, as  well as future  sets  which will incorporate refined measurements  and  theory. 
{\ihixs}  allows these studies  effortlessly.
It is interfaced  through the  LHAPDF library~\cite{lhapdf} 
with all available parton distribution functions with a  consistent  evolution at NNLO~\cite{Alekhin:2009ni,Martin:2009iq,JimenezDelgado:2009tv}. 
Other sets of the library can be employed by simple modifications of the code.

{\ihixs}  allows  the study of the Higgs boson invariant mass distribution  
for a finite  width of the Higgs boson  and compute the cross-section sampling over a 
Breit-Wigner distribution.   This assumes  that matrix-elements for the production of a Higgs boson decay 
factorize in matrix-elements for the production of a Higgs  boson times matrix-elements  for the decay. 
A grid of values for the branching rations and the decay width of the Higgs boson in the mass range of interest is a necessary ingredient for {\ihixs}. 
For the Standard Model (or models with very similar Higgs decay rates), {\ihixs}  uses a  grid  of values 
for  the width and  branching ratios that we  produced with the program {\tt HDECAY version 3.532}~\cite{Djouadi:1997yw}. 
To study models where the width and branching ratios differ significantly from the SM, the user has to provide his own grid, in the form of a simple text file.

{\ihixs}  includes perturbative contributions for two production processes: 
\begin{itemize}
\item[-] gluon fusion
\item[-] bottom-quark fusion
\end{itemize}

\subsection{Components of  the gluon fusion cross-section in {\ihixs} }
The  cross-section for the gluon fusion process in {\ihixs}  comprises: 
\begin{enumerate}
\item  leading order and next-to-leading order (NLO) QCD  effects
 with exact quark-mass dependence. The number  of  quarks  and  their
 Yukawa couplings are arbitrary.  

The  required two-loop amplitude has been first computed in Ref.~\cite{Graudenz:1992pv,Spira:1995rr} 
where it was presented in the  form of an integral
 representation~\footnote{First calculations of the  two-loop amplitude in the infinite top-quark
 mass limit were performed in Refs~\cite{Dawson:1990zj,Djouadi:1991tka}}. 
This was later expressed in terms of harmonic
 polylogarithms  with the method of series expansion and resummation
 in Ref~\cite{Harlander:2005rq}.  Independent  analytic evaluations
 were performed in Refs~\cite{Anastasiou:2006hc,Aglietti:2006tp}.   

The real radiation matrix-elements  have  been computed  in
Ref.~\cite{Ellis:1987xu,Baur:1989cm} and recomputed for the purposes of several 
other publications including this one. Numerical implementations of the NLO QCD cross-section 
with full quark-mass effects in the Standard Model were made in Ref.~\cite{Spira:1995rr} and  in 
Refs~\cite{Bonciani:2007ex,Anastasiou:2009kn}.
  
\item  NNLO QCD corrections, using
  heavy quark effective theory (HQET). 

The NNLO Wilson coefficient for an arbitrary number
of heavy quarks and Yukawa couplings has been computed in
Ref.~\cite{betta}.  In the special case  of Standard-Model Yukawa
couplings this is equivalent to the Wilson coefficient of
Ref.~\cite{Anastasiou:2010bt}, 
while  in the case of  integrating out only a single heavy quark it
yields the Wilson coefficient in Refs\cite{Chetyrkin:1997iv,Kramer:1996iq}.
  
The NNLO phase-space integrated matrix-elements in HQET have  been first  evaluated 
in their threshold limit\cite{Catani:2001ic,Harlander:2001is}. The  complete  NNLO correction has been 
computed in Refs~\cite{Harlander:2002wh,Anastasiou:2002yz,Ravindran:2003um}. 

\item two-loop electroweak corrections at leading order in $\alpha$.
 
These include the full Standard Model contributions to the amplitude  as  computed  
in Ref.\cite{Actis:2008ts,Actis:2008ug}~\footnote{We thank the authors  of Refs~\cite{Actis:2008ts,Actis:2008ug}  for  kindly providing 
a text file with the numerical values of the two-loop amplitude.}. For a light Higgs boson, a  dominant  contribution to the two-loop amplitude is  
due  to loops  with light quarks~\cite{Aglietti:2004nj}. We allow for a common re-scale factor  of  the $HWW$ and $HZZ$ couplings. 
This  should  be a sufficient parameter for  models which have a custodial symmetry protection in order to comply with stringent constraints from 
electroweak precision tests~\footnote{Custodial symmetry protects from
  large corrections to the $\hat{T}$ parameter; see, for example, the
  constraints on custodial symmetry breaking operators of the strongly
  interacting Higgs boson effective Lagrangian in \cite{Giudice:2007fh}.}.  

\item one-loop electroweak corrections for the real radiation
  processes $q \bar{q} \to g h$ and  $q g \to q h$.

This  amplitude has  been first  computed in 
Ref~\cite{Keung:2009bs}.  In this paper, we perform an independent calculation and present analytic formulae in terms of a basis of 
finite master integrals adding also contributions with massive  quarks in the loop.  Our results agree with the limit of zero Higgs boson mass of Ref.~\cite{Keung:2009bs}\footnote{We  were  unable to compare  with  the analytic expressions for the electroweak amplitude of Ref.~\cite{Keung:2009bs} due to the lack of an exact definition of the contributing ``finite parts'' from  the divergent master integrals which were chosen as a basis.}.  Subprocesses with bottom quarks in the initial state have also been studied in~\cite{Brein:2010xj} at a differential level. 

\item mixed QCD and electroweak contributions with light quarks. 

This contribution can  be estimated by means of an effective field theory and the required Wilson coefficient computed in Ref.~\cite{Anastasiou:2008tj}. 

\item arbitrary Wilson coefficient for the  $H {\rm tr} \left( G_{\mu \nu} G^{\mu \nu}\right)$ operator. 

New physics at  scales higher than the electroweak scale may introduce  modifications to the gluon fusion cross-section which  cannot be  accounted for  
by modified Yukawa interactions and  rescaling electroweak corrections.  In such situations, {\ihixs}  allows  to introduce corrections  to the Wilson coefficient  of 
the HQET theory.  Given the non-discovery of  new states with Tevatron and first  LHC  data, it is reasonable to anticipate that the new energy frontier is distant 
enough from the electroweak scale (higher than the top-quark mass) 
in order  for an effective theory approach to be  adequate.  

\end{enumerate}

\subsection{Components of  the bottom-quark fusion cross-section in {\ihixs} }

While the dominant production mode of the Higgs boson in the Standard
Model is gluon fusion  via a top-quark loop, a considerable correction of
about $5\%$ arises 
from bottom-quark loops.
Another production channel based on bottom-quark fusion is $gg
\rightarrow
Hb\bar{b}$~\cite{Campbell:2004pu,Dittmaier:2003ej,Dawson:2003kb}. 
In this channel the bottom-quarks are predominantly produced collinear to
the gluons.  The cross-section at fixed order in perturbation theory
suffers from large logarithmic terms. However, these can be resummed 
into the bottom parton density, leading to another (mostly) single
Higgs-boson production mechanism 
$b\bar{b}\rightarrow H$~\cite{Dicus:1998hs,Boos:2003yi,Maltoni:2003pn}.
We have therefore included this channel into {\ihixs}. 

The  cross-section for bottom-quark fusion process in {\ihixs}  
comprises the NNLO cross-section calculation of Harlander and Kilgore in Ref.~\cite{Harlander:2003ai}. 
This cross-section is  important  when the Yukawa  couplings to bottom
quarks are enhanced, as it may happen in models 
with more  than one Higgs  doublet.  We have  implemented  the  analytic  formulae  of Ref.~\cite{Harlander:2003ai} and computed the scale-dependent  terms  of the cross-section separately. We have  checked  that our numerical code agrees  
with the publicly available  program of  Ref.~\cite{bbh@nnlo}. 

We included the bottom-fusion process in {\ihixs}  for the purposes of facilitating 
the simultaneous study of  enhanced bottom Yukawa couplings in the production of a Higgs  boson from bottom-quarks and in gluon fusion via bottom-quark loops.   
We note that the {\ihixs}  program can be adapted easily in order to compute the
cross-section for Higgs production via the fusion of lighter  quarks,
such as the charm-quark, if necessary~\cite{Giudice:2008uua}.

\section{Higgs  boson interactions}
\label{sec:interactions}
We  consider a Higgs  boson with interactions described by the  Feynman rules:
\begin{center}
 \begin{picture}(320,50) (0,0)
    \SetWidth{1.0}
    \SetColor{Black}
    \Line[arrow,arrowpos=0.5,arrowlength=5,arrowwidth=2,arrowinset=0.2](10,50)(40,25)
    \Line[arrow,arrowpos=0.5,arrowlength=5,arrowwidth=2,arrowinset=0.2](40,25)(10,0)
    \Line[dash,dashsize=2](40,25)(56,25)
    \Text(66,20)[lb]{\small{$=Y_f$}}
    \Line[arrow,arrowpos=0.5,arrowlength=5,arrowwidth=2,arrowinset=0.2](90,50)(120,25)
    \Line[arrow,arrowpos=0.5,arrowlength=5,arrowwidth=2,arrowinset=0.2](120,25)(90,0)
    \Line[dash,dashsize=2](120,25)(136,25)
    \SetWidth{0.2}
    \Line[](140,50)(140,0)
    \SetWidth{1.0}
    \Text(142,-2)[lb]{\small{SM}}
    \Text(146,20)[lb]{\small{,}}
    \Photon(175,50)(205,25){1.5}{5}
    \Photon(205,25)(175,0){-1.5}{5}
    \Line[dash,dashsize=2](205,25)(221,25)
    \Text(231,20)[lb]{\small{$=\lambda_{ewk}$}}
    \Photon(255,50)(285,25){1.5}{5}
    \Photon(285,25)(255,0){-1.5}{5}
    \Line[dash,dashsize=2](285,25)(301,25)
    \SetWidth{0.2}
    \Line[](305,50)(305,0)
    \SetWidth{1.0}
    \Text(307,-2)[lb]{\small{SM}}
    \Text(311,20)[lb]{\small{.}} 
 \end{picture}
\end{center}
The triple-vertex Higgs-quark-quark  is the product of 
an arbitrary,  flavor dependent   factor $Y_f$  and the analogous 
Feynman rule in the Standard Model.  {\ihixs}  permits an arbitrary
number of quark flavors $N_f$ in order to accommodate extensions of
the Standard Model with novel quarks.
  
The Standard Model Feynman rules  for the $H-W-W$ and
$H-Z-Z$ vertices  are  rescaled  by a global factor $\lambda_{ewk}$.  
We did not find it necessary to introduce a  separate re-scaling
factor for the W  and Z boson vertices.  The ratio of the coefficients
of  the corresponding operators is fixed by the custodial symmetry and 
very tightly constrained by electroweak precision tests~\cite{Giudice:2007fh}.  

\section{
		Hadronic and partonic cross sections
		}
\label{section:hadronic-and-partonic-cross-sections}		
We consider the production of a Higgs boson, ${\rm H}(p_H)$,  at a hadron collider, which decays to some combination of final state particles that will be collectively denoted by $\{H_{\rm final}\}$:
\begin{equation}
{\rm hadron}_1(P_1) + {\rm hadron}_2(P_2) \to \left\{ 
{\rm H}(p_H) + {\rm X}, {{\rm (other\; processes)}} 
\right\}
\to \{H_{\rm final}\} + X
\end{equation}
The hadronic  cross-section is given by the factorization theorem as,   
\begin{equation}
\sigma^{full}_{\{H_{\rm final}\}+X} = \sum_{i,j \in {\rm partons}} \int dx_1 dx_2
f_i(x_1, \mu_f)f_j(x_2, \mu_f)  \hat{\sigma}^{full}_{ij \to \{H_{\rm
    final}\}+X}(\hat{s}, \mu_f)   
\end{equation}
where
\begin{equation}
\hat{s} = x_1 x_2 s, \quad s \equiv  (P_1+P_2)^2. 
\end{equation}
The indices $i,j$ run over the flavours of initial state partons. 
The functions $f_i(x, \mu_f)$ are parton distribution functions in the
$\overline{{\rm MS}}-$factorization scheme and $\mu_f$ is the
factorization scale.  

Singling out typically dominant contributions from resonant diagrams as
$p_H^2 \to m_H^2$, we cast the partonic cross-section in the form 
\begin{equation}
\label{plainpropagator}
\hat{\sigma}^{full}_{ij \to \{H_{\rm final}\}+X} 
=\hat{\sigma}_{ij \to \{H_{\rm final}\}+X} 
+ \hat{\sigma}^{signal-bkg}_{ij \to \{H_{\rm final}\}+X} 
+ \hat{\sigma}^{bkg}_{ij \to \{H_{\rm final}\}+X}.
\end{equation}  
The first term on the right hand  side corresponds to the square of the resonant
Feynman-diagrams, the second term corresponds to the interference of resonant and
non-resonant diagrams, and the last term is the square of diagrams
without a resonant Higgs propagator. We shall refer to the first term
as the ``signal cross-section''.  It can be written as:
\begin{equation}
\label{eq:BW0}
\hat{\sigma}_{ij \to \{H_{\rm final}\}+X}(\hat{s}, \mu_f)    = 
\int_{Q_{a}^2}^{Q_{b}^2} 
dQ^2 \frac{Q \Gamma_H(Q)}{ \pi }
\frac{
 \hat{\sigma}_{ij \to H}(\hat{s}, Q^2, \mu_f) {\rm Br}_{H \to \{H_{\rm final} \}}(Q)
}
{ (Q^2 -m_H^2)^2}.
\end{equation} 
$Q_a,Q_b$ define the experimentally accessible range for the invariant
mass of the particle system originating from the decay of the
intermediate Higgs boson.  $m_H$ is identified  with the Higgs-boson mass. 
Away from the resonance region, $Q^2 \sim m_H^2$ the above equation
is  adequate for the signal cross-section. 
 {\ihixs}  is dedicated to the evaluation of the ``signal
 cross-section''. 
 A  full description of
the cross-section for a Higgs final state  can be  obtained  by  adding to the ``signal
cross-section''  the remaining two contributions of Eq.~\ref{plainpropagator}:
a theoretical or  experimental estimate  of the
background cross-section and a theoretical estimate  of the
signal-background interference.  

In the resonant limit,  $Q^2 \to m_H^2$, the signal cross-section is
dominant and it  becomes infinite  at any fixed order in perturbation
theory,  for $Q^2 =m_H^2$ exactly.  A  resummation of  resonant
contributions at all orders is necessary in order to  render  the
propagator finite in this limit.      
We  remark that a resummation of partial perturbative  corrections  from
all perturbative orders into the propagator of an unstable particle
is a delicate theoretical issue~\cite{Beneke:2004km,Denner:2005fg}. 
Historically, it has been treated with various prescriptions in the
literature with varied success (see, for example, 
references in~\cite{Zanderighi:2004qu}). To a  first approximation, the signal cross-section becomes:
\begin{equation}
\label{eq:BWG}
\hat{\sigma}_{ij \to \{H_{\rm final}\}+X}(\hat{s}, \mu_f)    = 
\int_{Q_{a}^2}^{Q_{b}^2} 
dQ^2 \frac{Q \Gamma_H(Q)}{ \pi }
\frac{
 \hat{\sigma}_{ij \to H}(\hat{s}, Q^2, \mu_f) {\rm Br}_{H \to \{H_{\rm final} \}}(Q)
}
{ (Q^2 -m_H^2)^2 + m_H^2 \Gamma_H^2(m_H)}.
\end{equation} 
$\Gamma_H(Q)$ is the decay-width of a Higgs-boson at  rest  with mass
$Q$. 
In the zero Higgs boson width limit it reduces  to the
product of the partonic production cross-section $\hat{\sigma}_{ij \to
  H}$ for  an on-shell Higgs boson times the branching ratio  for
its decay ${\rm Br}_{H \to \{H_{\rm final} \}}$.   
Performing the transformation
\begin{equation}
\label{eq:Qtransform}
\frac{Q^2}{m_H^2} = 1 + \delta \tan(\pi y), \quad
\delta \equiv \frac{\Gamma_H\left(m_H\right)}{m_H}  
\end{equation} 
we obtain  an equivalent integral with a better numerical convergence, 
\begin{equation}
\label{eq:Breit-Wigner}
\hat{\sigma}_{ij \to \{H_{\rm final}\}+X}(\hat{s}, \mu_f)    = 
\int_{y_{a}}^{y_{b}} 
dy \frac{Q \Gamma_H(Q)}{ m_H \Gamma(m_H) }
 \hat{\sigma}_{ij \to H}(\hat{s}, Q^2, \mu_f) {\rm Br}_{H \to \{H_{\rm
     final} \}}(Q). 
\end{equation}
The integration boundaries $y_{a,b}$ are  computed from
Eq.~\ref{eq:Qtransform}. 

A light Higgs boson, as  predicted  in the Standard Model, has a
rather small $\delta$ and it  is often sufficient  to  take the
$\delta =0$ limit of the zero width approximation (ZWA).  
Existing experimental studies at  hadron
colliders~\cite{Aaltonen:2011gs,Chatrchyan:2011tz,Collaboration:2011qi}
have always  reported limits on the Higgs boson cross-section
comparing with expectations in this approximation. 
 However, recent years have witnessed the alarming trend of using this approximation in situations where it may be severely
insufficient\footnote{for example, recent limits on the cross-section
for a heavy Higgs boson~\cite{Collaboration:2011qi}}. 
 We therefore
find it  useful to  dedicate a part of  our  numerical studies  to Higgs bosons with a
non-negligible width. 

The width grows for heavier Higgs boson masses due to
decays into electroweak gauge bosons. In extensions of the Standard
Model, this feature  may be more  pronounced  as  new decay modes may
be  available.  For such situations, the zero width approximation is 
poor. An integration over the Breit-Wigner distribution of 
Eq.~\ref{eq:Breit-Wigner} is a more accurate estimate of the
signal cross-section.  
We  also note  that  Eq.~\ref{eq:Breit-Wigner}  convolutes with the 
branching ratio. For Higgs  masses  close to thresholds  branching
ratios  are  steeply changing; a naive estimate of the signal cross-section in the zero
width approximation could also be unsafe.  

{\ihixs}  allows  the possibility for the calculation of the Higgs signal
cross-section taking into account  finite Higgs width effects by
performing the integral of  Eq.~\ref{eq:Breit-Wigner}. 
Notice that the resummed expression of
Eqs~\ref{eq:BWG},\ref{eq:Breit-Wigner}  is in good agreement with
Eq.~\ref{eq:BW0}  away from the resonant region only when 
the partial width, $\Gamma_H(Q) {\rm Br}_{H \to \{H_{\rm final}
  \}}(Q)$, is computed at a variable Higgs-boson  virtuality $Q$.    
Higgs decay rates are rather sensitive  to the Higgs boson virtuality due  to the many 
decay mechanisms which become  available  at diverse mass values.
In order to use {\ihixs}  with an arbitrary BSM model, the user needs  to provide  a  data file  with the width
and branching ratios  of the Higgs  boson  as  a function  of the
virtuality of the Higgs boson.

For the Standard Model, or  models with similar enough width and
branching ratios, we have  generated  a  grid  of values using the 
program {\tt HDECAY} of  Ref~\cite{Djouadi:1997yw}~\footnote{For studies with {\ihixs}  where
  the HDECAY tabulated  Higgs boson width and branching ratios are
  used we request that Ref~\cite{Djouadi:1997yw} is also cited}. 
We note that  for large  Higgs boson widths  the description of  the
Higgs line-shape   may require further improvements, both in the
resonance region where a more sophisticated resummation framework
could be employed \cite{Beneke:2004km}  and, especially, for
virtualities far from the Higgs boson mass where a dedicated
estimation  of  the signal-background interference cross-section is needed. 
We  believe that {\ihixs}  can provide a flexible enough platform for
such modifications if need arises (e.g. with experimental evidence of a
Higgs boson and the associated heavy particles which render the theory
consistent with electroweak precision tests and unitarity bounds). 

In the Standard-Model, it has been observed that significant
cancelations due  to interference of   resonant and  
non-resonant  diagrams  take place at high invariant  masses 
(Refs~\cite{Glover:1988rg,Glover:1988fe,Baur:1990mr,Valencia:1992ix}). 
The magnitude  of the  ``signal-background'' cross-section is very
important and cannot be neglected.    
{\ihixs}  takes into  account only diagrams with an s-channel Higgs  boson propagator.  
The  line-shape  away from the resonance is  therefore poorly described.  
To improve upon this, we  have implemented a prescription based on the
resummation of  
$VV \to VV$ scattering amplitudes  with  the  dominant contributions
from both resonant and  non-resonant 
Feynman diagrams~\cite{Seymour:1995qg} at the  high energy regime. 
The  contributions which unitarize the scattering  amplitude for
vector-boson scattering  can be  approximated  with the amplitude for
Goldstone boson scattering at  high energies.  Conveniently, the
amplitude in this  regime can be  described  in terms of  an effective
Higgs propagator. Ref~\cite{Seymour:1995qg} performs  a Dyson re-summation of 
the tree-level Goldstone boson scattering amplitude  leading to an 
``improved  s-channel approximation''. In this  framework, the Higgs
propagator is modified according to the prescription:
\begin{equation}
\label{eq:Seymprop}
\frac{i}{\hat{s}-m_H^2}  \to 
\frac{
i \frac{m_H^2}{\hat{s}}
}
{\hat{s}-m_H^2 +i \Gamma_H(m_H^2) \frac{\hat{s}}{m_H}}.
\end{equation} 

This prescription   interpolates  smoothly between two limits which
are  well described either by resummation or by fixed-order
perturbation theory: the resonant region $Q \sim m_H$ and the high
energy limit $Q \gg m_H$.  We do not envisage Eq.~\ref{eq:Seymprop}
as the final step towards  a  precise  description of  the line-shape for
the Higgs boson. However, it  is a very useful diagnostic  tool in
order  to assess how  important the signal-background interference
could  be for a heavy Higgs  boson.   
According to this prescription, the  hadronic cross-section is  computed as, 
\begin{eqnarray}
\label{eq:Seymour}
 \hat{\sigma}_{ij \to \{H_{\rm final}\}+X}(\hat{s}, \mu_f)   & =& 
\int_{y_{a}}^{y_{b}} 
dy \frac{Q \Gamma_H(Q)}{ m_H \Gamma(m_H) }
 \hat{\sigma}_{ij \to H}(\hat{s}, Q^2, \mu_f) {\rm Br}_{H \to \{H_{\rm
     final} \}}(Q) \nonumber \\ 
&&  \times f_{{\rm seym}}\left( Q, m_H\right), 
\end{eqnarray}
with 
\begin{equation}
f_{seym}\left( Q^2, m_H^2\right) \equiv \frac{m_H^4}{Q^4} 
\frac{
\left(1-\frac{Q^2}{m_H^2}\right)^2 + \delta^2
}{
\left(1-\frac{Q^2}{m_H^2}\right)^2 + \delta^2 \frac{Q^4}{m_H^4}
}
\end{equation}

We  emphasize once again  that when Eq.~\ref{eq:Seymour} is  used
({\tt Seymour} option in {\ihixs}) some signal-background interference
effects which are  dominant at  very high invariant masses are taken
into account. In contrast, Eq.~\ref{eq:Breit-Wigner} ({\tt default} option
in {\ihixs}) computes purely the signal cross-section (only the
 square of  resonant Feynman diagrams). 

We  believe that the Higgs boson line-shape will enjoy many future
theoretical studies with improved resummation methods for resonant
diagrams and matching to fixed-order  perturbation theory away from
the resonance region. In the course  of  these  developments, new
prescriptions which are not yet implemented  in {\ihixs}   shall emerge.  
We  have made efforts to be able to
include such improvements readily in {\ihixs}.  For example, a  method 
introduced recently in Ref.~\cite{Passarino:2010qk} requires that all virtual
amplitudes must be computed with a complex Higgs virtuality $Q^2$. 
This requirement is effortless to achieve  in {\ihixs}  where we  evaluate
all QCD one and two-loop amplitudes using CHAPLIN~\cite{chaplin}
for harmonic polylogarithms with complex arguments. We
defer to the future a conceptual appraisal and a numerical
comparison of theoretically appealing approaches beyond the
prescriptions of  Eq.~\ref{eq:Breit-Wigner} and Eq.~\ref{eq:Seymour}. 

We  now define the  dimensionless ratios 
\beq
\tau \equiv \frac{m_H^2}{s}\;\;\;\; 
\hat{\tau} \equiv  \frac{Q^2}{s} \;\;\;\; 
z \equiv \frac{\hat{\tau}}{x_1 x_2}
\eeq
The  hadronic Higgs  signal cross-section can be  cast in the form
\begin{eqnarray}
\sigma_{\{H_{\rm final}\}+X} &=& \sum_{i,j \in {\rm partons}} 
\int_{y_{a}}^{y_{b}} 
dy \frac{Q \Gamma_H(Q)}{ m_H \Gamma(m_H) }
{\rm Br}_{H \to \{H_{\rm
     final} \}}(Q)  f_{{\rm seym}}\left( Q, m_H\right) 
\times  \nonumber \\ 
&& \times 
\sum_{ij} \int \frac{dx}{x} dz {\cal L}_{ij}(x_1, x_2, \mu_f) \left[ 
\frac{ \hat{\sigma}_{ij \to H}(Q^2, z, \mu_f) } 
{z}
\right],
\end{eqnarray}
where 
\begin{equation}
x_1 \equiv x, \quad x_2 \equiv \frac{\hat{\tau}}{x z}, \quad 
{\cal L}_{ij} \equiv \left[ x_1 f_i(x_1)\right] \left[ x_2
  f_j(x_2)\right]. 
\end{equation}
In {\ihixs}  we have  implemented  the partonic cross-sections  for two
Higgs  boson production channels: \\
(i) bottom-quark fusion as in Ref.~\cite{Harlander:2003ai}  \\ 
(ii) gluon fusion. \\
Amplitudes for the two processes do not interfere and the two
cross-sections can be computed independently. 

We have faithfully adopted  the definitions and analytic expressions for the 
partonic quark-fusion cross-sections of Ref.~\cite{Harlander:2003ai},
where they have been computed through NNLO in the strong coupling expansion. 
We shall therefore not discuss this process any  further, except for
the presentation of numerical results with {\ihixs}.    
 
The perturbative evaluation of the gluon fusion cross-section  has
been the topic of numerous  publications. Nevertheless, for the sake
of clarity, we find it important to elaborate on our implementation of
perturbative corrections for this  process.


\section{
		The gluon fusion process through NLO QCD
		}

We cast the partonic  cross-sections in the form
\begin{equation}
\frac{ \hat{\sigma}_{ij \to H}(Q^2, z, \mu_f) } 
{z} =  \frac{G_f \pi}{288\sqrt{2}} \sum_{p=0}^{\infty} 
\left(
\frac{\alpha_s(\mu_r)}{\pi} 
\right)^{2+p}   
n_{ij}^{(p)}\left( Q^2, z, \mu_f, \mu_r\right), 
\end{equation}
where $\mu_r$ is the renormalization scale for the strong coupling. 

The partonic cross sections  $z n_{ij}^{(p)}$ at NLO and higher orders
consist of virtual and real emission parts that are separately
infrared divergent. We expose the infrared singularities of the real
radiation matrix-elements by the method of plus-distribution
subtractions. We then add the
virtual part, the collinear counter term which has been generated
from parton distribution factorization in the $\overline{{\rm MS}}-$factorization scheme and the renormalization counter terms
in the $\overline{{\rm MS}}-$renormalization scheme. 
The resulting cross-section is finite and can be written as the sum of
three distinct terms: a term proportional to $\delta(1-z)$ that corresponds to all contributions from leading order kinematics (virtual part plus $\delta$-proportional terms from the integrated soft-collinear pieces), a `regular' term that corresponds to higher order kinematics, and plus-distribution pieces, proportional to 
$\left[  \frac{f(z)}{1-z}  \right]_+$ for various $f(z)$. 

We therefore write
\beq
n_{ij}^{(p)}\left(Q^2, z, \mu_f, \mu_r\right) 
=
 \sum_{k\in \{\delta,+,R\}} n_{ij;k}^{(p)}\left(Q^2, z, \mu_f, \mu_r\right) 
\eeq

	\subsection{
		LO: $gg\to h$ 
		}
		
At leading order only the gluon gluon subprocess contributes. One obtains
\beq
n_{gg;\delta}^{(0)} = \left| \sum_q Y_q\tau_q\frac{3}{2}A(\tau_q)\right|
\eeq
with 
\begin{equation}
\tau_q \equiv \frac{4 m_q (m_q - i \Gamma_q) }{Q^2}
\end{equation}
and $m_q$ being the mass of the heavy quark in the gluon fusion loop.  
The quantity $\tau_q\frac{3}{2}A(\tau_q)$ has the simple limits 
\beq
\lim_{\tau_q\to\infty } \tau_q\frac{3}{2}A(\tau_q) = 1 \;\;\; \lim_{\tau_q\to0} \tau_q\frac{3}{2}A(\tau_q) = 0
\eeq
The full analytic expression is given by, 
\beq
A(\tau_q)=1-\frac{1}{2}\frac{(1+x_q)^2}{(1-x_q)^2}H(0,0;x_q)
\eeq
with
\beq
x_q=\frac{-\tau_q}{(\sqrt{1-\tau_q}+1)^2}
\eeq
Note that the sum runs over all quarks in the model, and $Y_q$ is as
defined in the Feynman rules of Section~\ref{sec:interactions}.


	\subsection{
		NLO: $gg\to h+g$ 
		}
The $\delta$-part of the NLO correction to the gluon gluon subprocess can be written as
\beq
n_{gg;\delta}^{(1)}=|B|^2\left[ 2\beta_0 \log\left(\frac{\mu_R^2}{\mu_F^2}\right)+\pi^2\right] +  \Re\left[ B \sum_q V_q(\tau_q)^*  \right]
\eeq
where 
\beq
B\equiv \sum_q Y_q \tau_q \frac{3}{2}A(\tau_q)
\eeq
is the LO coefficient, and 
\beq
V_q(\tau_q)=Y_q\frac{3}{8} M_{fin}^{(1)}(\tau_q)=Y_q\frac{{\cal G}_i^{2l}}{-2/3}
\eeq
where $M_{fin}^{(1)}(\tau_q)$ can be found in eq.7.4 of ref.~\cite{Anastasiou:2006hc}  and ${\cal G}_i^{2l}$ can be found in  eq.26-30 of ref.~\cite{Aglietti:2006tp}.
The plus distribution part is 
\beq
n_{gg;+}^{(1)}=|B|^2\left[ -6\log\left(\frac{\mu_F^2}{Q^2}\right) \left[\frac{1}{1-z}\right]_+  +12 \left[\frac{\log(1-z)}{1-z}\right]_+ \right]
\eeq
Finally, the regular part of the NLO gluon gluon correction is 
\bea
n_{gg;R}^{(1)}&=&\frac{3}{z(1-z)\lambda(1-\lambda)}    
\Bigg\{ 
\frac{1}{2} 
z^4 
\sum_{j=1}^4\left| \sum_q A^{jq}_{gggH}(\tau_q)\right|^2-
 (1-z+z^2)^2 |B|^2
\Bigg\} \nonumber\\
& &
+|B|^2
\Bigg\{ 6
\left[ p_{gg}(z)\log\left(\frac{(1-z)^2}{z}\right) - \log\left(\frac{z}{1-z}\right) 
\right] 
-6\log\left(\frac{\mu_F^2}{Q^2}\right)p_{gg}(z)
\Bigg\}
\eea
with $p_{gg}(z)$ the gluon splitting kernel
\beq
p_{gg}(z)=\frac{1}{z}+z(1-z)-2
\eeq
The form factors $A^{jq}_{gggH}(\tau_q)$ can be found in 
the Appendix~\ref{formulas:AgggH}.

\subsection{NLO: $q\bar{q}\to h+X$}

The $q\bar{q}$ initial state starts contributing to the total cross
section at order $a_s^3$ in QCD. There are also non-negligible mixed
QCD-electroweak corrections, at order $a_s^2 a_{ewk}$ which we shall
discuss later. The pure QCD corrections lead to the following coefficients: 
\beq
n_{q\bar{q};\delta}^{(1)}=n_{q\bar{q};+}^{(1)}=0
\eeq 
\beq
n_{q\bar{q};R}^{(1)}=\frac{32}{27}\frac{(1-z)^3}{z}\left| \sum_q Y_q \tau_q A_{q\bar{q}gH}(z)\right|^2
\eeq 
with $A_{q\bar{q}gH}(z)$ given in eq.~\ref{Aqqgh}.

\subsection{NLO: $qg\to h+X$}	 
The gluon-quark initial state also contributes at order $a_s^3$, and it receives similar mixed QCD-electroweak corrections of order $a_s^2a_{ewk}$. The pure QCD coefficient is:
\beq
n_{qg;\delta}^{(1)}=n_{qg;+}^{(1)}=0,
\eeq 
and
\begin{eqnarray}
n_{qg;R}^{(1)}&= &
\Bigg\{  
\left| B  \right|^2 
\left[ 
\frac{C_F}{2} z 
-p_{gq}(z) \log\left( \frac{z }{  (1-z)^2}\right)
-p_{gq}(z) \log\left( \frac{ \mu_F^2}{ Q^2 }\right)
\right] \nonumber \\ 
&&
+\int_0^1 d\lambda \frac{1}{(1-\lambda)_+}
\left[
\left|
\sum_q  
Y_q \tau_q  A_{q\bar{q}gH}\left( y_{\lambda}\right)
\right|^2 \frac{1+(1-z)^2\lambda}{z} 
\right]
 \Bigg\}
\end{eqnarray}
with 
\beq
y_{\lambda}=\frac{-z}{(1-z)(1-\lambda)}
\eeq
\beq
p_{gq}(z)=\frac{C_F}{2}\frac{1+(1-z)^2}{z}
\eeq
and $A_{q\bar{q}gH}(z)$ given in eq.~\ref{Aqqgh}.
\subsection{NLO: mixed QCD-EW corrections to $q\bar{q}\to H+g$ and $qg\to h+g$}

The mixed QCD-electroweak contributions\footnote{Note that these corrections are of order $a_s^2$, so we denote them by $n_{q\bar{q}}^{(0)ewk}$.} have the following structure:
\begin{eqnarray}
n_{q\bar{q};R}^{(0)ewk}&= & \lambda_{ewk} 
\frac{8}{3}\frac{1-z}{z}\sum_q \sum_{X\in \{ W_i,Z_i,H \}} \Re 
\Bigg\{   \tau_q A_{qqgH}(z)c_{X,q}\cdot  \nonumber\\ 
& &\left[ F_{1,X,q}^*(s_{13},s_{23},s_{12}) 
\frac{s_{13}^2}{s_{12}}  + F_{2,X,q}^*(s_{13},s_{23},s_{12})\frac{s_{23}^2}{s_{12}} \right] \Bigg\}
\end{eqnarray}
where $A_{q\bar{q}gH}(z)$ given in eq.~\ref{Aqqgh}. 
The sum over $X$ runs over all $W$ and $Z$-like bosons in the model, 
as well as over the Higgs boson, while the sum over $q$ runs over all
heavy quarks, $b,t,\ldots$, as before.  The coupling $c_{X,q}$
contains Kronecker delta symbols which select specific initial state
quarks,  depending on $X$, as explained below.

\begin{itemize}


\item \underline{Z loops:} 
For the initial state quarks $q \in \{u,d,c,s,b\}$ we have the following couplings
\bea
 \lambda_Z &=&  2 \nonumber\\
 c_{Z,q}&=&(\delta_{qu}+\delta_{qc}) (v_{Z,u}^2+a_{Z_u}^2) + (\delta_{qd}+\delta_{qs}+\delta_{qb}) (v_{Z,d}^2+a_{Z_d}^2)
\nonumber\\
v_{Z,u} &=&\frac{g_w}{\cos\theta_w} (\frac{1}{2}-\frac{4}{3}\sin^2\theta_w )  
\quad , \quad
 a_{Z,u} =\frac{g_w}{2\cos\theta_w}
\nonumber\\
v_{Z,d} &=&\frac{g_w}{\cos\theta_w} (-\frac{1}{2}+\frac{2}{3}\sin^2\theta_w )  
\quad , \quad
 a_{Z,d} =\frac{g_w}{2\cos\theta_w} 
\eea
The form factor is identical for all initial state quarks and yields
\begin{eqnarray}
{\cal F}_1^{Z_q} &=& -m_z^2 A_{ewk}\left(s_{31},s_{23},s_{12},Q,m_z \right) \nonumber \\
{\cal F}_2^{Z_q} &=& -m_z^2 A_{ewk}\left(s_{23},s_{31},s_{12},Q,m_z \right)
\end{eqnarray}
with $A_{ewk}(s,t,u,m_h,m_z)$ given in the Appendix in eq.~\ref{Aewk} 

\item \underline{W loops:} 
The couplings are given by 

\begin{equation}
\lambda_W =2   ,\quad
c_{W,q} =(v_{W,q}^2+a_{W,q}^2) 
,  \quad v_{W,q} =\frac{g_w}{\sqrt{2}}, \quad   a_{W,q} =\frac{g_w}{\sqrt{2}}, 
\end{equation}
\begin{eqnarray}
{\cal F}_1^{W} &=& -m_w^2 A_{ewk}\left(s_{31},s_{23},s_{12},Q^2,m_w \right)\sum_L\delta_{qL}  -m_w^2 A_{ewk}^{m_t}\left(s_{31},s_{23},s_{12}\right)\delta_{qb} \nonumber \\
{\cal F}_2^{W} &=&  -m_w^2 A_{ewk}\left(s_{23},s_{31},s_{12},Q^2,m_w \right)\sum_L\delta_{qL}  -m_w^2 A_{ewk}^{m_t}\left(s_{23},s_{31},s_{12}\right)\delta_{qb}
\end{eqnarray}
where $L$ sums over all light quark states $u,d,c,s$. 
Note that here we have summed over the \emph{ internal} light quark flavors. This yields a CKM coefficient of 
$$
\sum_{j=1,2} |V_{ij}|^2\approx 1 \quad \rm {for} \quad i=1,2 
$$
Measurements show this to be true to about 1 in 10000, so  it is a good enough approximation to make.\\
For $q\in\{b\}$, we have to take the internal quark to be a top. 
The couplings are unchanged if we use that $|V_{33}|^2\approx 1$, which is also a good approximation to make. 
The form factor $A_{ewk}^{m_t}\left(s,t,u\right)$ is given in the Appendix in eq.~\ref{Aewk_mt}.


\item \underline{Higgs in the loop:} 
In the standard model this gives a  non negligible contribution for $q=b$,
other quarks may be considered as well if their Yukawas are enhanced.
The couplings are given by 
\begin{equation}
\lambda_H =3  ,\quad  c_{H,q}=\delta_{qb}(v_{H,q}^2+a_{H,q}^2),\quad v_{H,q} =\frac{m_qY_q}{v}, \quad   a_{H,q} =0, 
\end{equation}
and the form factors are
\begin{eqnarray}
{\cal F}_1^{H} &=& -m_h^2 A_{H}\left(s_{31},s_{23},s_{12} \right) \nonumber \\
{\cal F}_2^{H} &=& -m_h^2 A_{H}\left(s_{23},s_{31},s_{12} \right).
\end{eqnarray}

\end{itemize}


\section{Beyond the NLO QCD}

The leading order and next to leading oder cross-section are known
exactly. Beyond that we can only take the limit of heavy electroweak 
gauge bosons and top-quarks. Bottom quark contributions are also
unknown beyond NLO. 

In the effective theory approximation, the Higgs gluon interaction is
described by an operator of the form 
\begin{equation}
{\cal L}_{{\rm eff}}=  -\frac{1}{3\pi} C_w \cdot H G_{\mu\nu} G^{\mu \nu} ,
\end{equation}
where the Wilson coefficient has a perturbative expansion
\begin{equation}
C_w = C_0 +C_1 \frac{\alpha_s(\mu)}{\pi} + C_2 \left(
    \frac{\alpha_s(\mu)}{\pi} \right)^2  + \ldots
\end{equation}
The cross-section calculated in the effective theory is, 
\begin{equation}
\sigma_{{\rm eff}} = \left| C_0 +C_1 \frac{\alpha_s(\mu)}{\pi} + C_2 \left(
    \frac{\alpha_s(\mu)}{\pi} \right)^2  + \ldots \right|^2  \left[
    \eta_0 + \frac{\alpha_s}{\pi} \eta_1 + \left(
      \frac{\alpha_s}{\pi}\right)^2    \eta_2 + \ldots  \right]
\end{equation}
Expanding in $\alpha_S$ we obtain 
\begin{equation}
\sigma_{{\rm eff}} = \sigma_{\rm eff}^{(0)} + 
\left(\frac{\alpha_s}{\pi}\right) \sigma_{\rm eff}^{(1)} 
+\left(\frac{\alpha_s}{\pi}\right)^2 \sigma_{\rm eff}^{(2)} + \ldots 
\end{equation}
with 
\begin{eqnarray}
\sigma_{\rm eff}^{(0)} &=& \left| C_0 \right|^2 \eta_0, \\
\sigma_{\rm eff}^{(1)} &=& \left| C_0 \right|^2 \eta_1 + 
2 {\rm Re}\left( C_0 C_1 \right) \eta_0,   \\
\sigma_{\rm eff}^{(2)} &=& \left| C_0 \right|^2 \eta_2  
+2 {\rm Re}\left( C_0 C_1 \right) \eta_1  
+ \left(  \left| C_1 \right|^2  +2 {\rm Re}\left( C_0 C_2 \right) \right) \eta_0.
\end{eqnarray}
The integrated  cross-sections  $\eta_i$ have  been computed through NNLO in 
Refs~\cite{Harlander:2002wh,Anastasiou:2002yz,Ravindran:2003um}.

Resorting to an effective theory calculation is necessary only for important QCD  and electroweak 
corrections  which cannot be  evaluated  in the full  theory.  In the previous  sections,  we have listed 
results  in the exact theory for the LO  and NLO QCD perturbative expansion as  well as  one-loop 
electroweak  corrections. In addition, two-loop electroweak corrections are  also known 
exactly~\cite{Actis:2008ug}.  We will keep these  corrections  with their full mass dependence as in the exact theory 
calculations and  use  the effective theory approach for contributions at higher orders in the  strong and  electroweak couplings, 
namely for the NNLO correction in QCD  and mixed  QCD  and  electroweak corrections.  
 
We  match the effective theory and full theory perturbative expansions  as follows. 
Let us assume  that we can compute the contributions to the cross-section
exactly through some perturbative order for only some of  the heavy particles  
which contribute to Higgs production amplitudes: 
\begin{equation} 
\sigma_{\rm partial}= 
\sigma_{\rm partial}^{(0)} + 
\left(\frac{\alpha_s}{\pi}\right) \sigma_{\rm partial}^{(1)} 
+\left(\frac{\alpha_s}{\pi}\right)^2 \sigma_{\rm partial}^{(2)} + \ldots
\end{equation}
In an effective theory approach, these  contributions  would factorize as in:
\begin{equation}
\sigma_{{\rm partial, eff}} = 
\left| C_0^{\rm partial } +C_1^{\rm partial } \frac{\alpha_s(\mu)}{\pi} + C_2^{\rm partial } \left(
    \frac{\alpha_s(\mu)}{\pi} \right)^2  + \ldots \right|^2  \left[
    \eta_0 + \frac{\alpha_s}{\pi} \eta_1 + \left(
      \frac{\alpha_s}{\pi}\right)^2    \eta_2 + \ldots  \right]
\end{equation}
We then write the  cross-section as 
\begin{equation}
\sigma  = \sum_{n=0}^\infty 
\left(\frac{\alpha_s}{\pi}\right)^n 
\left[
\sigma_{\rm partial}^{(n)}  
\Theta(n \leq {\rm Norder} ) 
+\delta \sigma_{\rm _eff}
\right]
\end{equation}
where ${\rm Norder}$ is the last perturbative order that the partial
contributions are known in the full theory and 
\begin{eqnarray}
\delta \sigma_{\rm eff}^{(0)} &=& \left\{ \left| C_0 \right|^2
- \Theta(0 \leq {\rm Norder}) \left| C_0^{\rm partial} \right|^2 \right\}  \eta_0, \\
\delta  \sigma_{\rm eff}^{(1)} &=& 
\left\{ \left| C_0 \right|^2
- \Theta(1 \leq {\rm Norder}) \left| C_0^{\rm partial} \right|^2 \right\}
\eta_1 
\nonumber \\ && \quad 
+ 
\left\{ 2 {\rm Re}\left( C_0 C_1 \right) - \Theta(1 \leq {\rm Norder})
    2 {\rm Re}\left( C_0^{\rm partial} C_1^{\rm partial} \right)  
\right\} \eta_0,   \\
\delta  \sigma_{\rm eff}^{(2)} &=& 
\left\{ \left| C_0 \right|^2
- \Theta(2 \leq {\rm Norder}) \left| C_0^{\rm partial} \right|^2 \right\}
\eta_2 
\nonumber \\ && \quad 
+ 
\left\{ 2 {\rm Re}\left( C_0 C_1 \right) - \Theta(2 \leq {\rm Norder})
    2 {\rm Re}\left( C_0^{\rm partial} C_1^{\rm partial} \right)  
\right\} \eta_1
\nonumber \\ && \quad 
+ \left(  
\left\{ \left| C_1 \right|^2
- \Theta(2 \leq {\rm Norder}) \left| C_1^{\rm partial} \right|^2
\right\}
\right. 
\nonumber \\ 
&& + \left. 
\left\{ 2 {\rm Re}\left( C_0 C_2 \right) - \Theta(2 \leq {\rm Norder})
    2 {\rm Re}\left( C_0^{\rm partial} C_2^{\rm partial} \right)  
\right\}
\right)
\eta_0.
\end{eqnarray}

\subsection{The Standard Model Wilson coefficient with anomalous Yukawa and electroweak couplings}

In our theory,  we can integrate out the top-quark, the new  heavy quarks,  and the electroweak gauge  bosons W and Z. 
This yields a Wilson coefficient which is 
\begin{eqnarray}
C_0  &=& \lambda_{\rm QCD} \cdot 1  + \lambda_{\rm EWK} \cdot 1 \\
C_1  &=& \lambda_{\rm QCD} \cdot  \frac{11}{4} + \lambda_{\rm EWK} \cdot \frac{7}{6} \\ 
C_2  &=& \lambda_{\rm QCD} \cdot  C_{2q} + \lambda_{\rm EWK} \cdot  C_{2w} \, .
\end{eqnarray} 
The  factor  for the electroweak component is given by
\begin{eqnarray}
\lambda_{\rm EWK} &=& \lambda_{ewk} \frac{3 \alpha}{16 \pi s_w^2} \left\{ 
4+\frac{2}{c_w^2}
\left[ 
\frac{5}{4}-\frac{7}{3} s_w^2 + \frac{22}{9} s_w^4
\right]
\right\}, 
\end{eqnarray}
with $s_{w}, c_{w}$ the sine and the cosine of the Weinberg mixing angle.
The  factor  for the QCD component  is given by the sum of the  anomalous Yukawa coupling 
re-scaling factors  for all heavy quarks  
\begin{equation}
\lambda_{QCD} = \sum_{q \in {\rm heavy}} Y_q \, .
\end{equation}
The  three-loop QCD  Wilson coefficient $C_{2q}$ has  been computed recently in
Ref.~\cite{betta} and can be read from Eq.~(3.35) of the same reference.  
The  four-loop electroweak Wilson coefficient is not  yet known. 
\begin{equation}
C_{2w} = \;  {\rm  unknown} 
\end{equation}
One can attempt a  rough estimation since it is conceivable that the
perturbative series for the QCD  component and the electroweak
component  follow a  similar pattern:  $|C_{2w} |\sim |C_{2q}| \sim 10$.  In our studies,  we vary  $C_{2w} \in [-30, 30]$ as an estimate of the higher order 
mixed QCD and electroweak corrections.

In {\ihixs}  we  have  implemented the  exact contributions  up to NLO, i.e.  ${\rm Norder} =1$, for the heavy quarks and the electroweak gauge-bosons. These contributions should not be counted twice in the effective theory calculation. 
\begin{eqnarray}
C_0^{\rm partial}  &=& \lambda_{\rm QCD}  \cdot 1  \\
C_1^{\rm partial}  &=& \lambda_{\rm QCD} \cdot \frac{11}{4} \\ 
C_2^{\rm partial}  &=& \lambda_{\rm QCD}  \cdot C_{2q}.  
\end{eqnarray} 

\subsection{Improving on the effective theory approximation}

It has been observed that the effective theory works  better for the K-factors rather than 
the absolute cross-section.  
At next-to-leading order, all real and virtual amplitudes 
in the soft or collinear limit have the  same dependence on  the masses of the heavy quarks as at leading order. 
It appears that the factorization of the 
cross-section in the infrared limit closely  resembles the factorization of the cross-section in the limit of 
infinitely heavy massive particles.
Finite quark-mass effects are important   for ``hard radiation'' terms, but these are expected to have a typical perturbative expansion where 
an $\alpha_s$  suppression occurs from one order to the other. Top-quark mass effects have been studied with explicit calculations with operators 
of higher dimension in HQET demonstrating the validity of the approach at NNLO\cite{Harlander:2009mq,Pak:2009dg}.   
We can then improve on the effective theory approximation by making the replacement
\begin{equation}
\lambda_{\rm QCD} \to  \sum_{q \, \in \, {\rm heavy}} Y_q  \frac{3}{2} \tau_q A(\tau_q), 
\end{equation}
and 
\begin{equation}
\lambda_{\rm EWK} \to  \lambda_{ewk} {{\cal M}_{ggh,{\rm EWK}}}\left(m_H^2, M_W^2, M_Z^2, \ldots \right)
\end{equation}
the two-loop electroweak amplitude for $gg \to H$, computed fully in \cite{Actis:2008ts,Actis:2008ug}. 
In Fig. 21 of Ref.~\cite{Actis:2008ts} we find the quantity
\begin{equation}
\delta_{\bf EWK}/100 = \frac{\sigma_{EWK +QCD}^{\rm born}}{\sigma_{QCD}^{\rm born}}\Bigg|_{\rm top \, only} -1
=  \left|1+ 
\frac{
{\cal M}_{ggh,{\rm EWK}}
}{
\frac{3}{2} \tau_q A\left(\tau_{top}\right)
}
\right|^2-1.
\end{equation}
We  then substitute, 
\begin{equation}
\lambda_{\rm EWK} \to \lambda_{ewk} 
\frac{3}{2} \tau_q A\left(\tau_{top}\right) \times \lambda_0,  
\end{equation}
with 
\begin{equation}
\lambda_0 = \sqrt{1 + \frac{\delta_{EWK}}{100}} -1. 
\end{equation}
The two-loop electroweak corrections  were kindly provided to us in a data file, 
{\tt electroweak.h}, by the authors  of Ref.~\cite{Actis:2008ts}.

\section{Numerical results in gluon fusion}
\label{sec:numericsgluon}		
		
In this section, we present numerical results for the Higgs  boson cross-section via gluon fusion.  
We  will first make  a short discussion of  the stability of the perturbative expansion and the scale variation uncertainty. 
Then we shall compare predictions from all available NNLO sets of parton distribution functions.  We will proceed with a study of 
finite width effects for the heavy quarks in the gluon fusion loops.  Finally we shall discuss the finite width effects on the Higgs boson 
total cross-section. To the best of our knowledge, there have been no published result for NNLO K-factors for the signal cross-section 
beyond the zero width approximation for the Higgs boson.  
		
\subsection{Perturbative convergence and scale uncertainty}
\label{sec:numericsgluon_scale}

\begin{figure}[h]
\begin{center}
\includegraphics[width=0.7\linewidth]{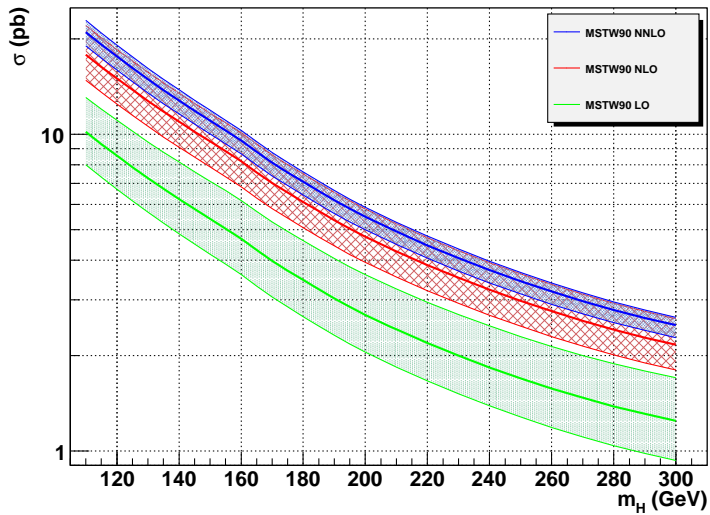}
\end{center}
\caption{
\label{fig:LHC7mstwallorders}
Inclusive Higgs cross section at LO, NLO and NNLO, with scale uncertainty bands, calculated in the range $\mu\in \left[\mu_0/2,2\mu_0\right]$ for MSTW PDFS.}
\end{figure}

The gluon fusion cross-section exhibits a rather  slow  convergence
of the perturbative series  in the strong  coupling  constant.  {\ihixs} 
computes  the cross-section through NNLO in perturbative QCD.  
The perturbative behavior of the total higgs cross section with its
scale uncertainty, when using MSTW08 parton densities, is shown in Fig.~\ref{fig:LHC7mstwallorders}. 
for LHC at $7{\rm TeV}$ collision energy.  One notices that radiative
corrections are sizable, where neither the NLO nor the NNLO
corrections can be neglected.   
\begin{figure}[h]
\begin{minipage}[b]{0.49\linewidth}
\includegraphics[width=\linewidth]{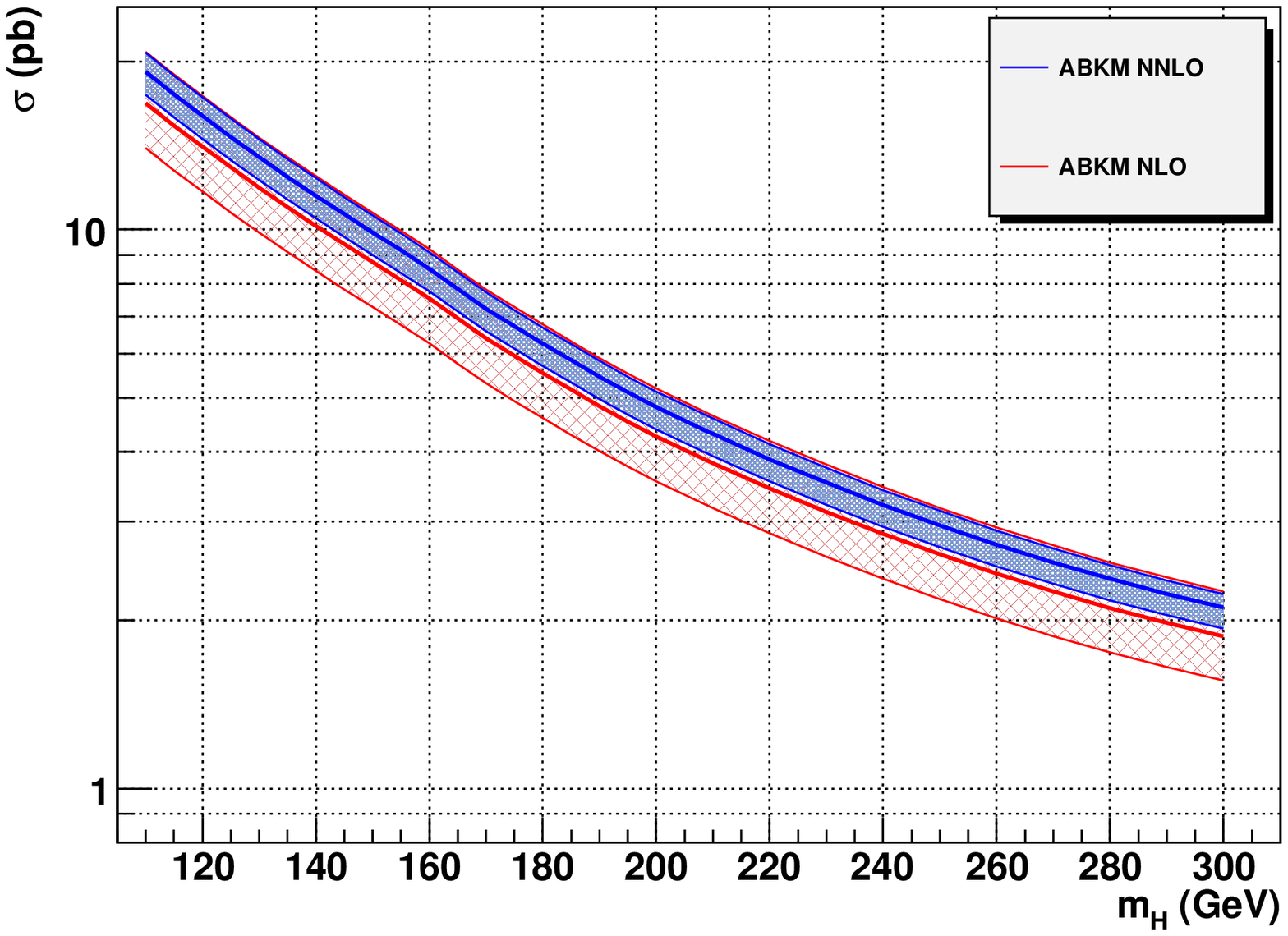}
\end{minipage}
\begin{minipage}[b]{0.49\linewidth}
\includegraphics[width=\linewidth]{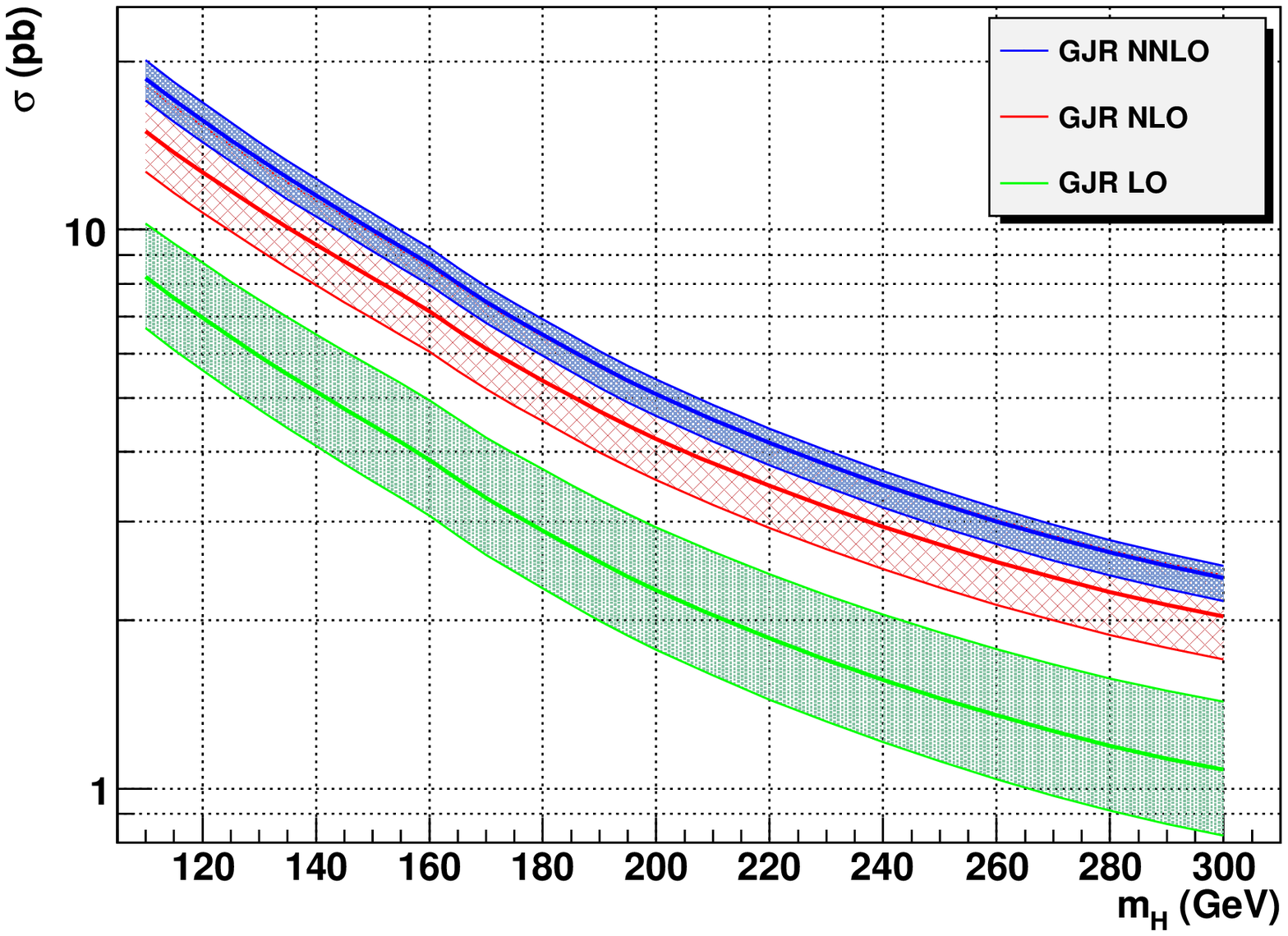}
\end{minipage}
\caption{
\label{fig:LHC7scaleabkmgjrallorders}
Inclusive Higgs cross section at LO, NLO and NNLO, with scale uncertainty bands, calculated in the range $\mu\in \left[\mu_0/2,2\mu_0\right]$ for ABKM and GJR PDF sets.}
\end{figure}

A major source of theoretical uncertainty in Higgs production via
gluon fusion is due to the choice of the factorization and 
renormalization scales. We evaluate the scale uncertainty with a
central scale of $\mu_F=\mu_R=\mu_0=m_H/2$, and a variation in the
range  $\mu\in \left[\mu_0/2,2\mu_0\right]$. 
We note that the NLO scale uncertainty band engulfs the NNLO band in 
almost the entire mass region depicted. We  also notice that the
magnitude of the perturbative corrections (K-factor) is  larger  for lower  values of the Higgs boson mass. 

Similar behavior is observed when using the ABKM set, but not with the
GJR set, Fig.~\ref{fig:LHC7scaleabkmgjrallorders}, where the NLO and NNLO overlap is only partial.

\subsection{PDF comparison}
\begin{figure}[h]
\begin{minipage}[b]{0.5\linewidth}
\includegraphics[width=80mm]{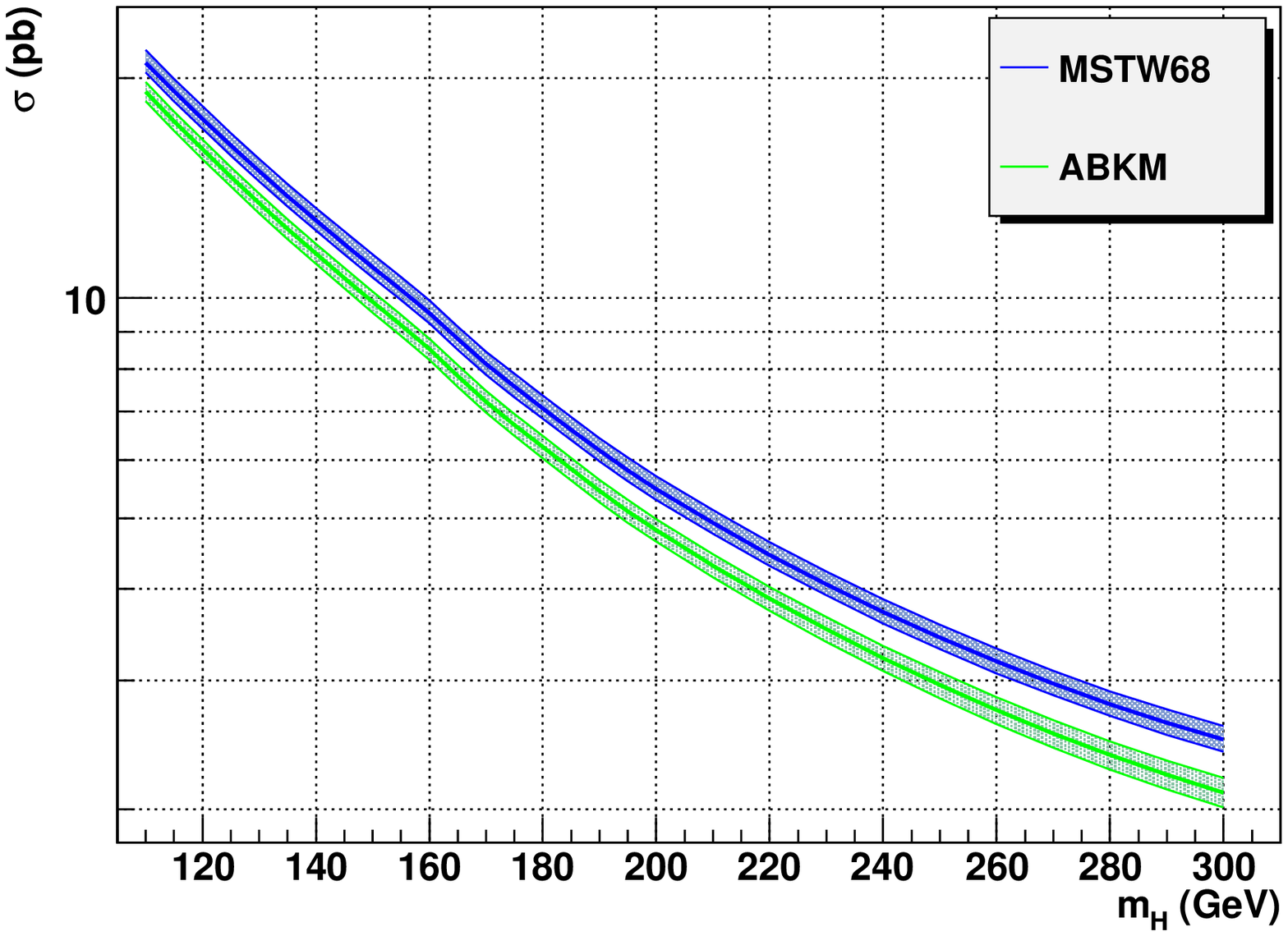}
\end{minipage}
\begin{minipage}[b]{0.5\linewidth}
\includegraphics[width=80mm]{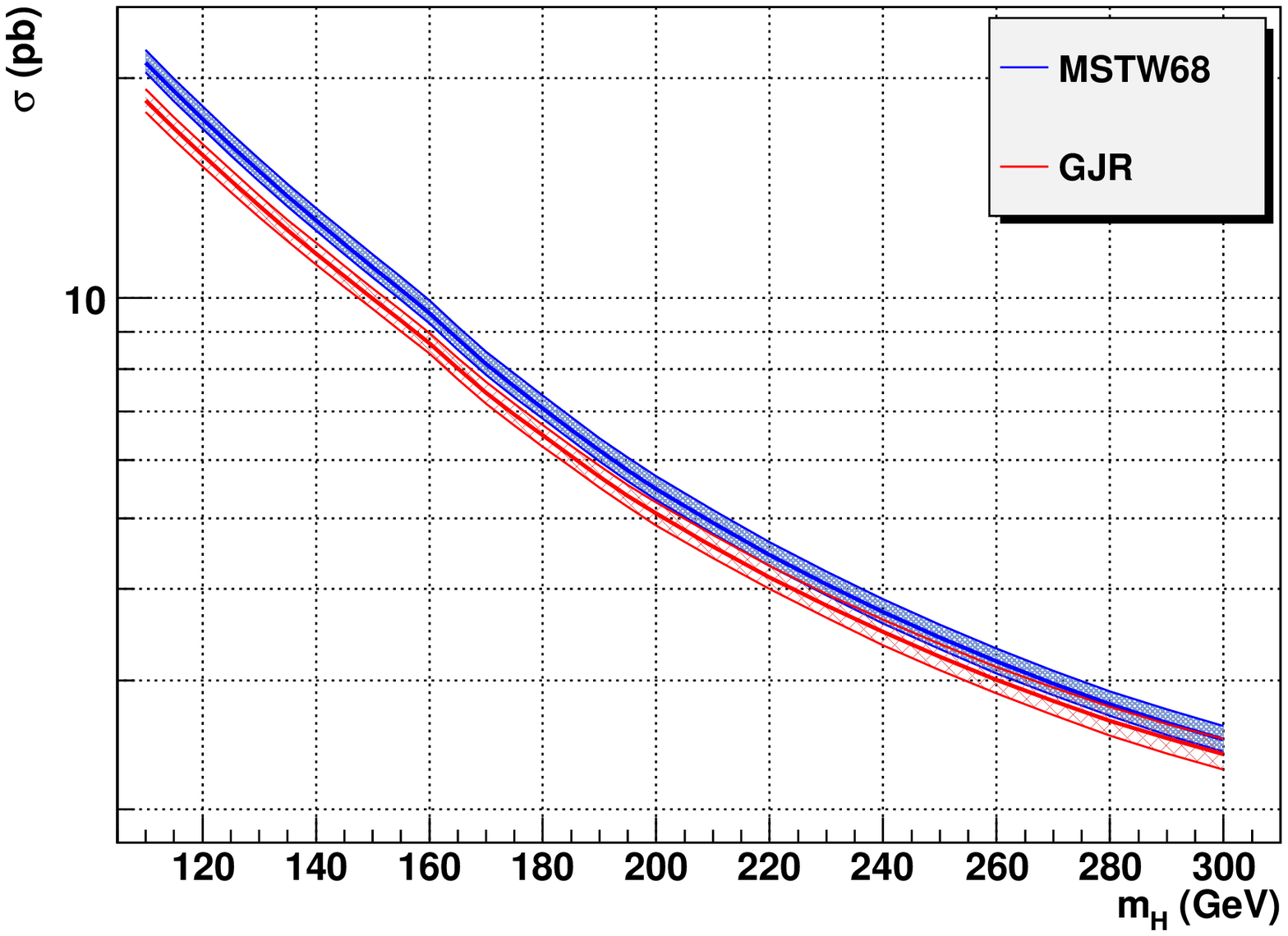}
\end{minipage}
%
\begin{minipage}[b]{0.5\linewidth}
\includegraphics[width=\linewidth]{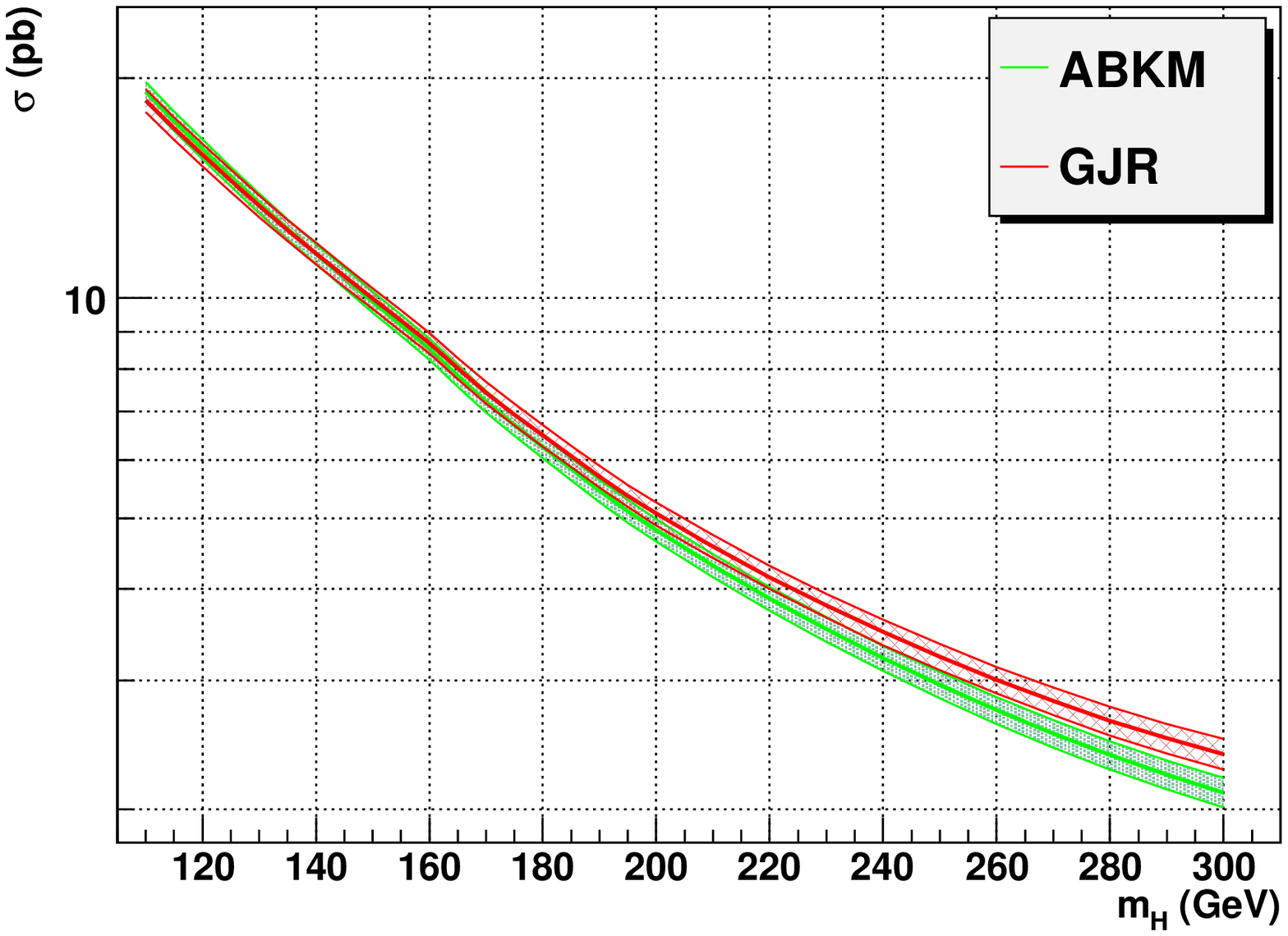}
\end{minipage}
\begin{minipage}[b]{0.5\linewidth}
\includegraphics[width=\linewidth]{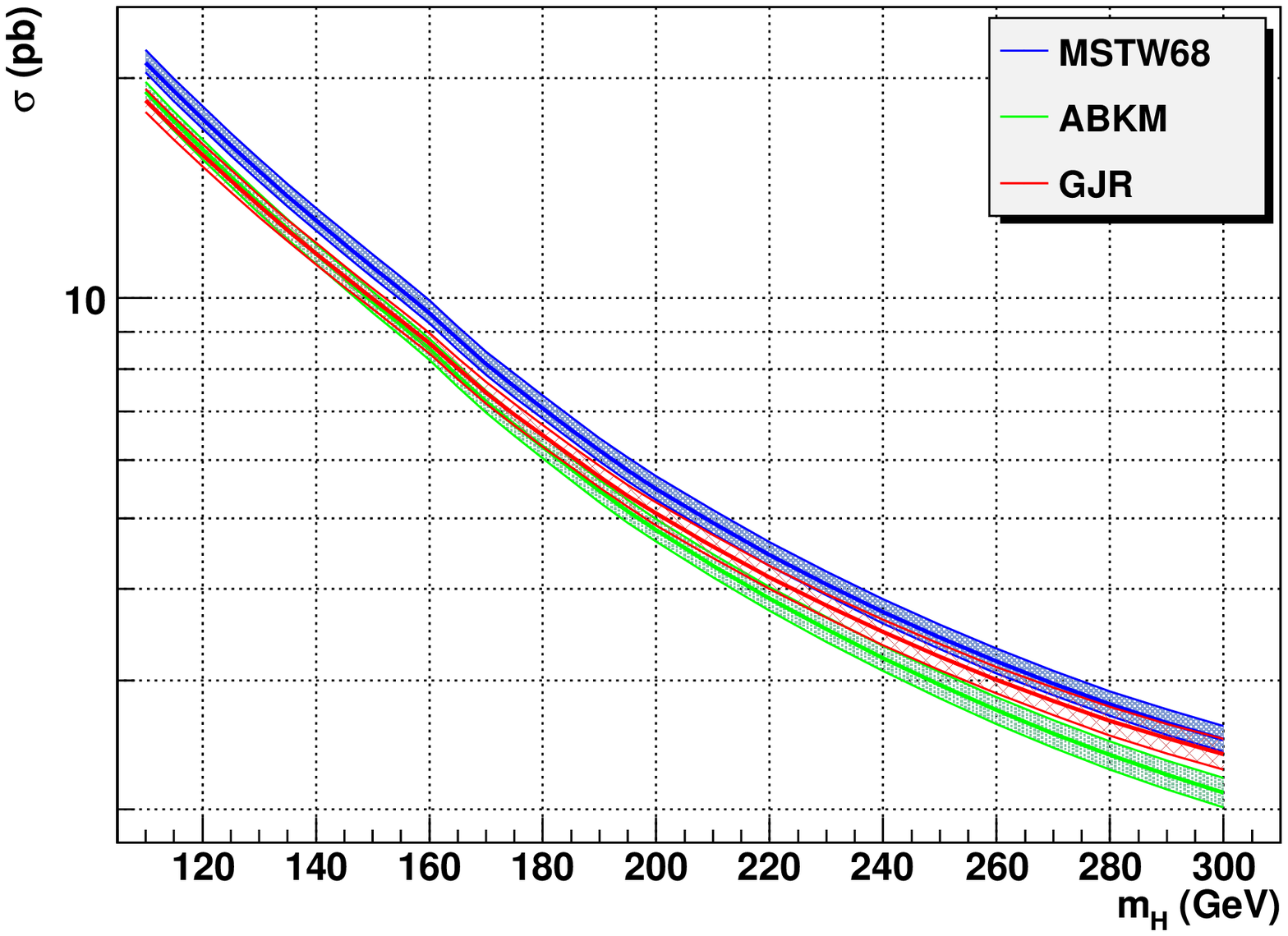}
\end{minipage}
\caption{
\label{LHC7pdfcomparison}
The Higgs production cross section at NNLO with three different PDF sets and their $68\%$CL PDF$+a_s$ uncertainty bands. }
\end{figure}
An enormous progress has  been made  in the  last  decade  towards improving and estimating  reliably 
the  precision of  parton densities. 
While uncertainties are  generally small, the gluon fusion process requires  the less constrained  
gluon density.  The  extraction of this  quantity is an active  field
of  research. It  is very important to compare  the effect of  various
gluon density determinations on the Higgs  boson 
cross-section. We have  enabled  the possibility for such studies  in {\ihixs}.

In tables~\ref{mstw},\ref{abkm},\ref{gjr} we present the inclusive
cross section for $pp\to H+X$ for Higgs masses ranging from $110$ to
$300$ GeV, using three different PDF sets available at NNLO, namely,
MSTW2008\cite{Martin:2009iq}, ABKM09\cite{Alekhin:2009ni} and
GJR09\cite{JimenezDelgado:2009tv} . We  have  restricted  our  choice
of pdf sets to the ones with  NNLO DGLAP evolution, consistently with 
 the {\ihixs}  computation of the partonic cross-sections through the
 same order.  We believe that these {\ihixs}   results constitute the most precise
predictions  in fixed order perturbation theory for the Higgs boson cross-section.  

\begin{figure}[h]
\begin{minipage}[b]{0.5\linewidth}
\includegraphics[width=80mm]{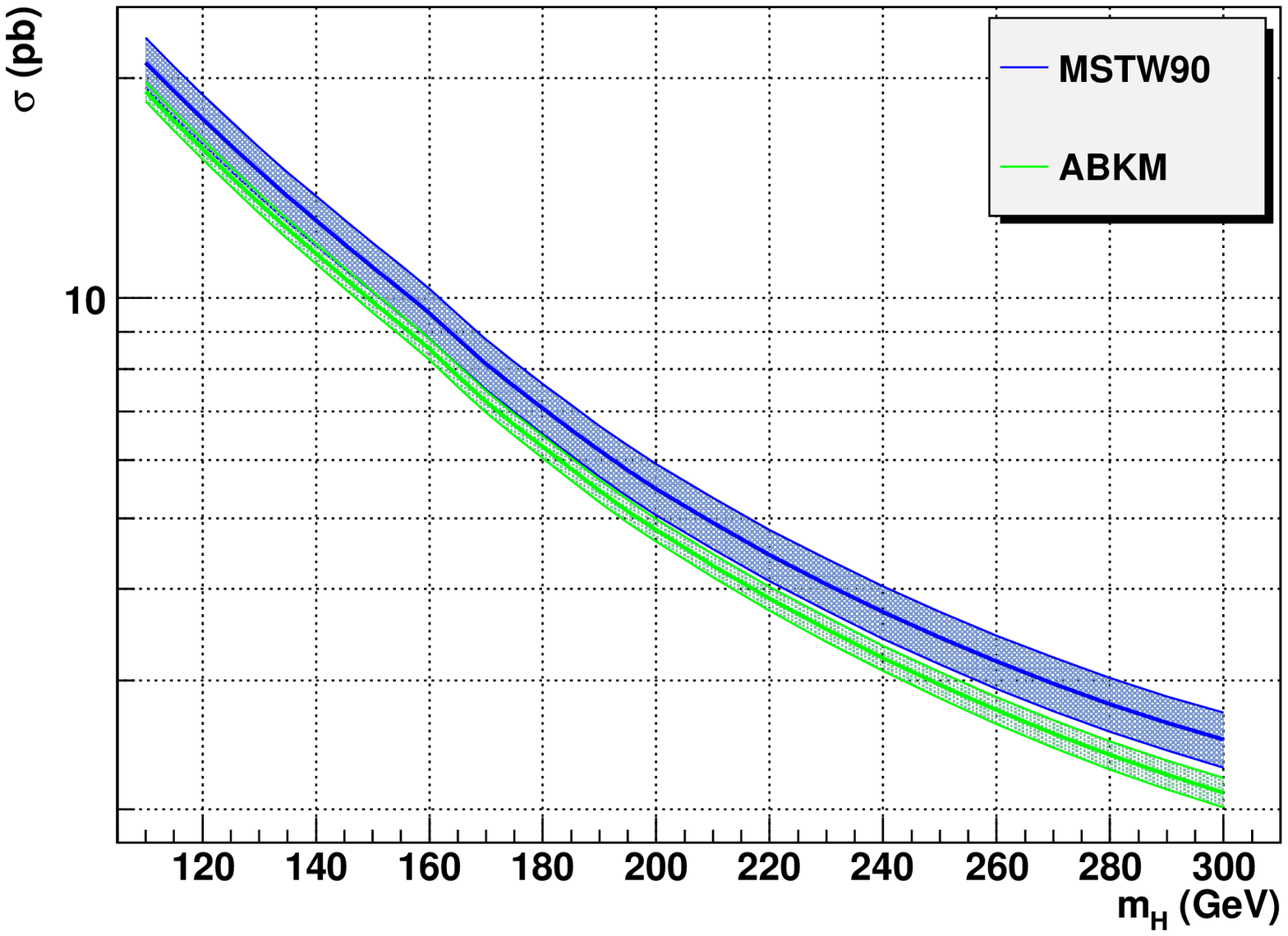}
\end{minipage}
\begin{minipage}[b]{0.5\linewidth}
\includegraphics[width=80mm]{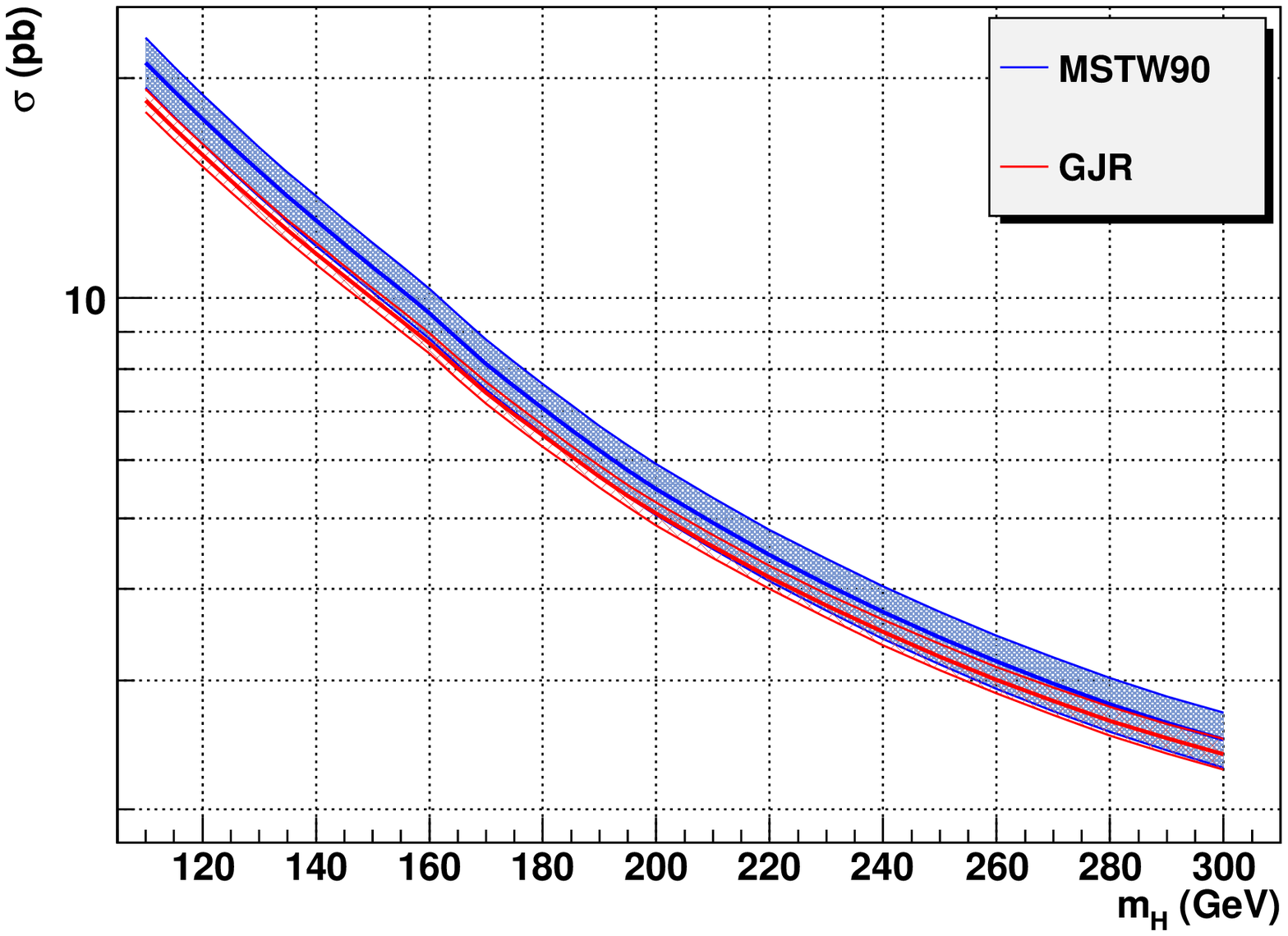}
\end{minipage}
%
\begin{minipage}[b]{0.5\linewidth}
\includegraphics[width=\linewidth]{./ihixs_analysis/root_plots/LHC7_pdf_abkm_vs_gjr.eps}
\end{minipage}
\begin{minipage}[b]{0.5\linewidth}
\includegraphics[width=\linewidth]{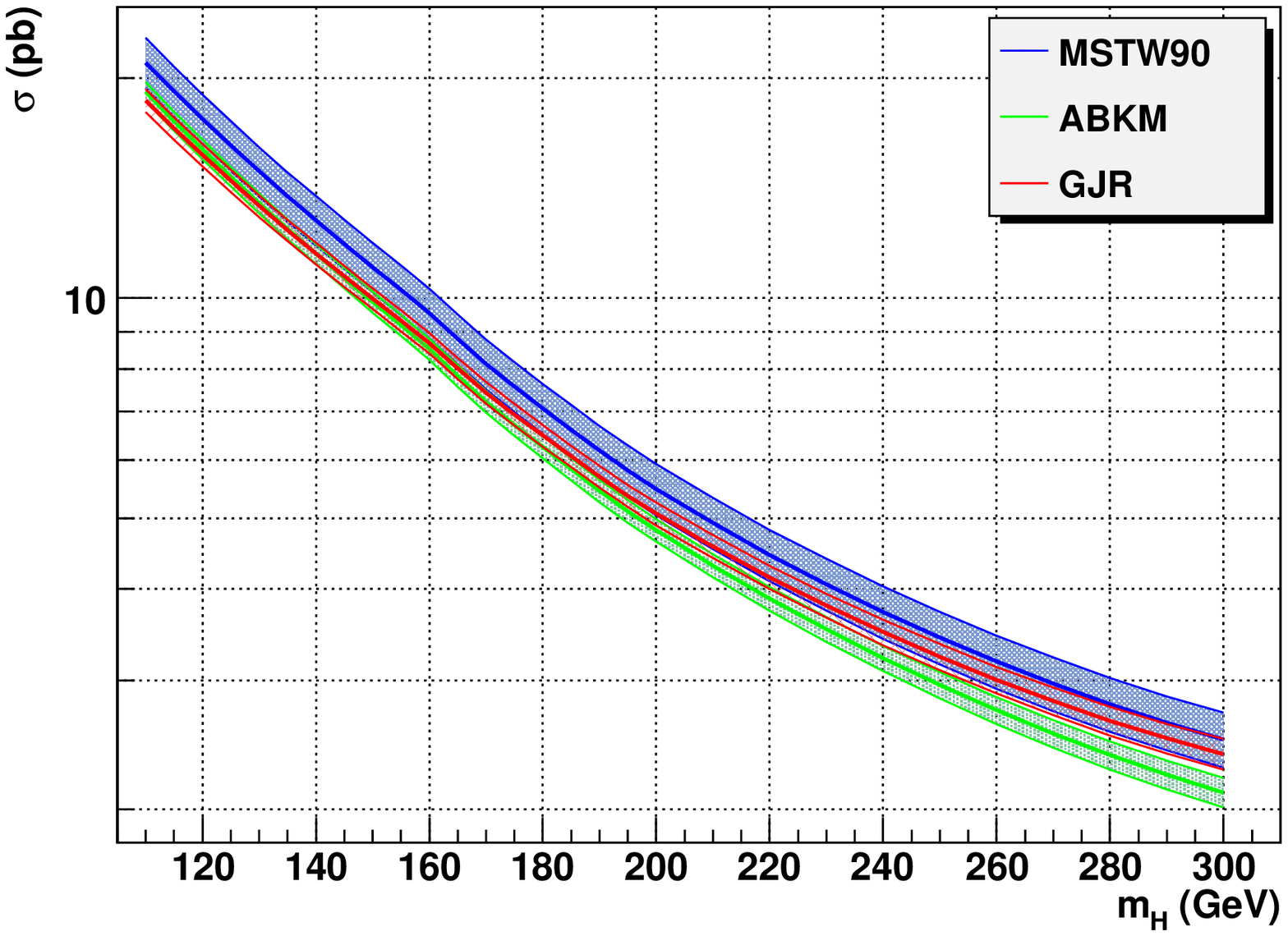}
\end{minipage}
\caption{
\label{LHC7pdfcomparison_mstw90}
The Higgs production cross section at NNLO with three different PDF sets and their PDF$+\alpha_s$ uncertainty bands, using the MSTW $90\%$ CL grids. }
\end{figure}
 A comparison between the three different NNLO PDF sets and the corresponding PDF$+\alpha_s$ uncertainty bands are shown in Fig.~\ref{LHC7pdfcomparison}. The uncertainties quoted include variations of $\alpha_s$ around the preferred value for every set, and are estimated according to the prescriptions of the PDF providers.
A combination of  cross-section values  for  a large set of pdf
parameterizations is necessary,  and these are efficiently computed  in  {\ihixs}   simultaneously.

The reason for the remarkably different predictions between the
different PDF sets (that range from $10\%$ in the low mass region to
$30\%$ in moderately high masses of around $300$GeV) is hard to trace.
The bulk of it may be attributed to differences in the adopted values
of the strong coupling constant $a_s(m_Z)$. The situation is only
partially remedied if one chooses to consider the $90\%$ CL (as
opposed to the one-sigma, $68\%$CL) uncertainty bands provided by the MSTW collaboration.

The comparison, in Fig.~\ref{LHC7pdfcomparison_mstw90}, shows the ABKM
and MSTW uncertainty bands to marginally overlap. 
We finally note that preliminary results~\cite{Alekhin:2010dd} with
the updated ABM10 PDF fit, which includes hadron collider data, show that larger values for the total cross section are obtained, in comparison with ABKM09.

\subsection{Top quark width}
\begin{figure}[h]
\includegraphics[width=\linewidth]{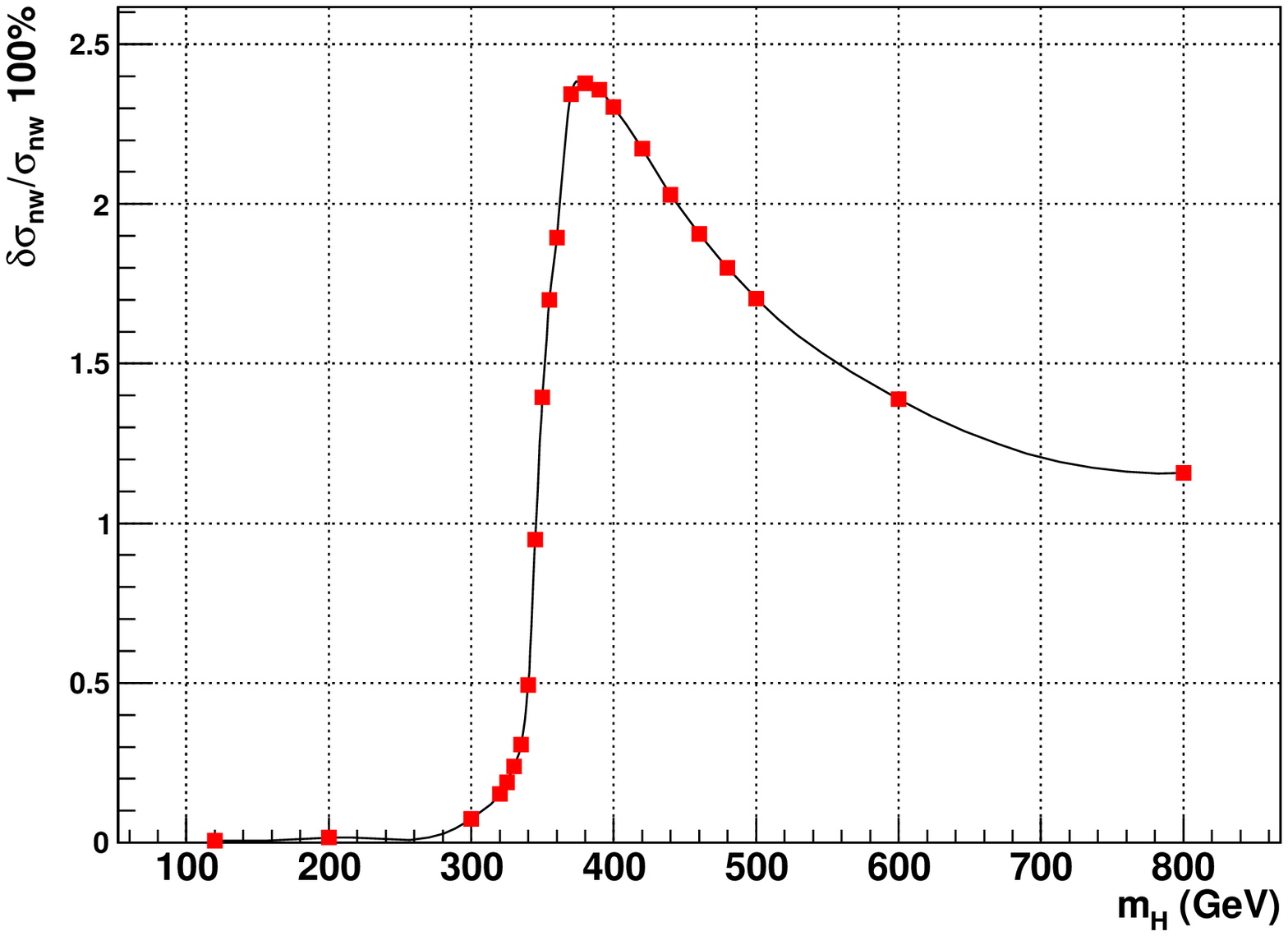}
\caption{
\label{FGwidth}
Relative difference $\delta\sigma_{nw}/\sigma_{nw}=\frac{\sigma
  -\sigma^*}{\sigma^*}\cdot100\%$ of the cross section for the top
quark with a real mass, $\sigma^*$, and in the complex mass scheme
with $\Gamma_{top}=2$ GeV.}
\end{figure}
{\ihixs}   evaluates  two and one-loop amplitudes in all kinematic
regions, permiting  a definition of mass 
and  kinematic  invariants in the full complex plane.   
This is a particularly useful feature when a resummation of  finite
width effects in threshold  regions is  necessary.  
For example, in a  resummation framework using the complex mass 
scheme\cite{Denner:2005fg,Passarino:2010qk}  the  masses  of heavy
quarks  need to be evaluated  according to the prescription, 
\begin{equation}
m_q^2 \to m_q \left( m_q -i \Gamma_q\right), 
\end{equation}
where $\Gamma_q$ is the total decay width of the quark and $m_q$ its mass.  

Using {\ihixs}, we have studied the finite width effects for quarks in fermion loops.  
We  find that   the top width is insignificant (at the level of less
than one per mille) for a Higgs boson mass  
below the $t\bar{t}$ threshold.  Around and above that threshold,  its
effect grows  to the percent order as shown in Fig.~\ref{FGwidth}.

\subsection{Finite Higgs boson width effects}		
	
In Section~\ref{section:hadronic-and-partonic-cross-sections}	we
discussed that there exist various approaches on how to treat the
Higgs propagator when departing from the zero width approximation
(ZWA).  In this section we will present numerical results for the two different prescriptions described in  Section~\ref{section:hadronic-and-partonic-cross-sections}: the default scheme (DEF) of Eq.~\ref{eq:Breit-Wigner} and the Seymour scheme (S) of Eq.~\ref{eq:Seymour}. 

\begin{figure}	
\includegraphics[width=\linewidth]{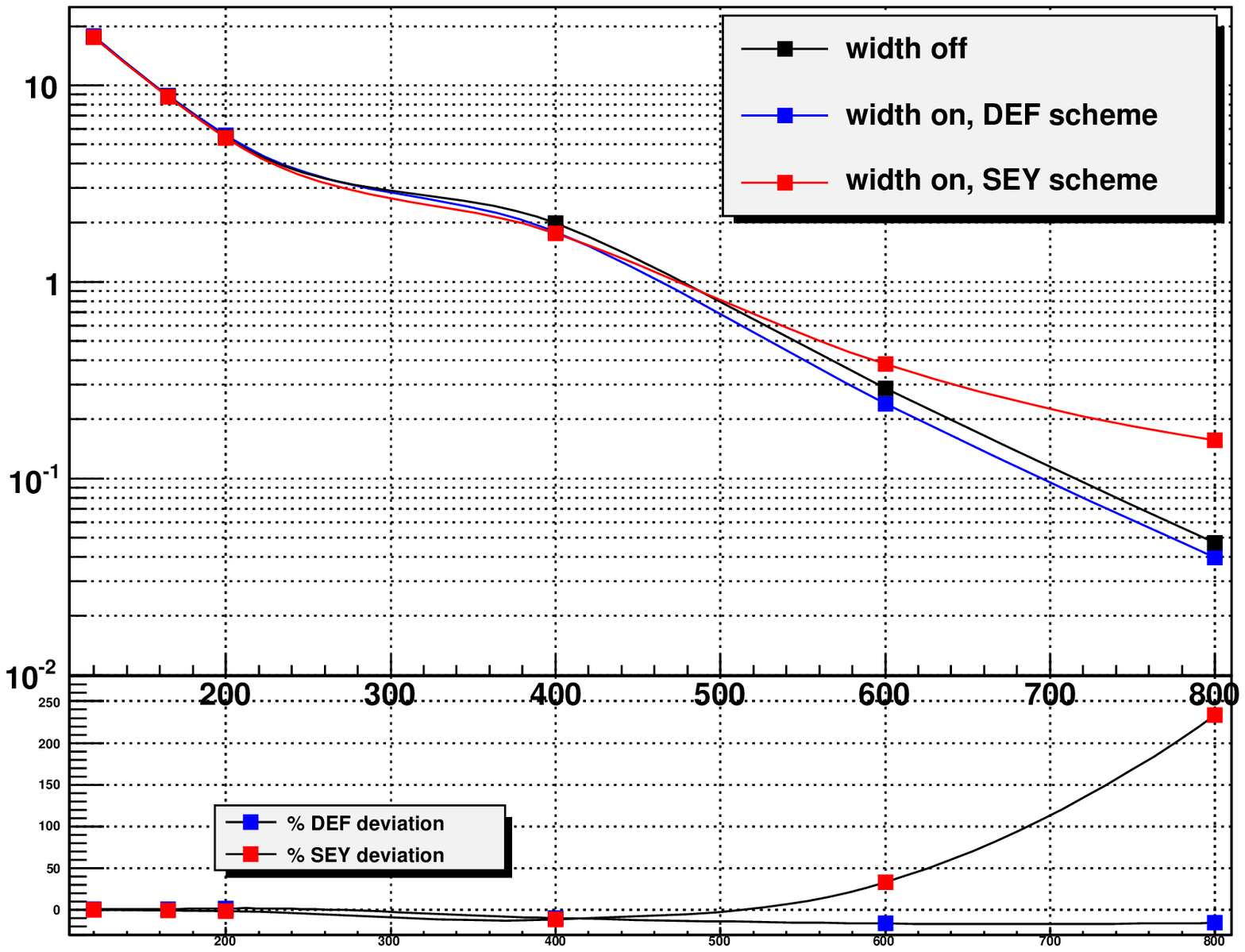}
\caption{
\label{LHC7width}
Comparison of the total cross section in the zero width approximation, $\sigma^{ZWA}$, with a finite width in the default scheme, $\sigma^{DEF}$ and in the Seymour scheme, $\sigma^{SEY}$. In the lower panel we show the relative error one makes when adopting the ZWA, defined as $\frac{\sigma-\sigma^{ZWA}}{\sigma^{ZWA}}\cdot100\%$}
\end{figure}
		
In Fig.~\ref{LHC7width} the inclusive Higgs cross section calculated
within the ZWA and the two finite width schemes is shown, as a
function of the Higgs mass. The width is calculated by interpolating
over a detailed grid\footnote{The precision of the interpolation is
  always better than $2\cdot 10^{-5}$.} constructed with {\tt
  HDECAY}~\cite{Djouadi:1997yw}. The cross sections\footnote{We use
  here the MSTW PDF set. Similar behavior is observed when using the
  other two NNLO PDF sets.}  are shown in table \ref{width_table}. We
note that the three calculations deviate widely for Higgs masses
larger than $300$GeV.  The deviation between the ZWA and the finite
width schemes  is expected since for large Higgs masses the width of
the Higgs boson is comparable to its mass. It is also evident that the
finite width schemes deviate from each other in the high mass region,
indicating a possibly large  contribution due to signal and background  interference 
which the Seymour scheme attempts to simulate.
\begin{table}
 \begin{center}
  \begin{tabular}{| c ||   c || c | c | c | c |}
\hline
$m_H$ & $\Gamma_H$ &$\sigma^{ZWA}$ & $\sigma^{DEF}$  &$\sigma^{SEY}$  \\
\hline\hline
120&	0.0038&		17.57&			17.66 &		 	17.57		\\ \hline
165&	0.2432&		8.78&			8.874 &			8.735	\\\hline
200&	1.43&		5.45&			5.566&			5.390	\\ \hline
400&	29.5&		1.988&			1.799&			1.766	\\ \hline
600&	122&		0.287&			0.2409&			0.3819	\\ \hline
800&	301&		0.04708&	         		 0.03982&			0.15683	\\
\hline
 \end{tabular}
\end{center}
 \caption{
\label{width_table}
Total cross section for LHC at $\sqrt{s}=7$TeV with MSTW PDFs with a finite width in the two schemes, $\sigma^{DEF,SEY}$, and in the zero width approximation denoted by $\sigma^{ZWA}$.}
\end{table}

\begin{figure}
\begin{minipage}[b]{0.5\linewidth}
\includegraphics[width=80mm]{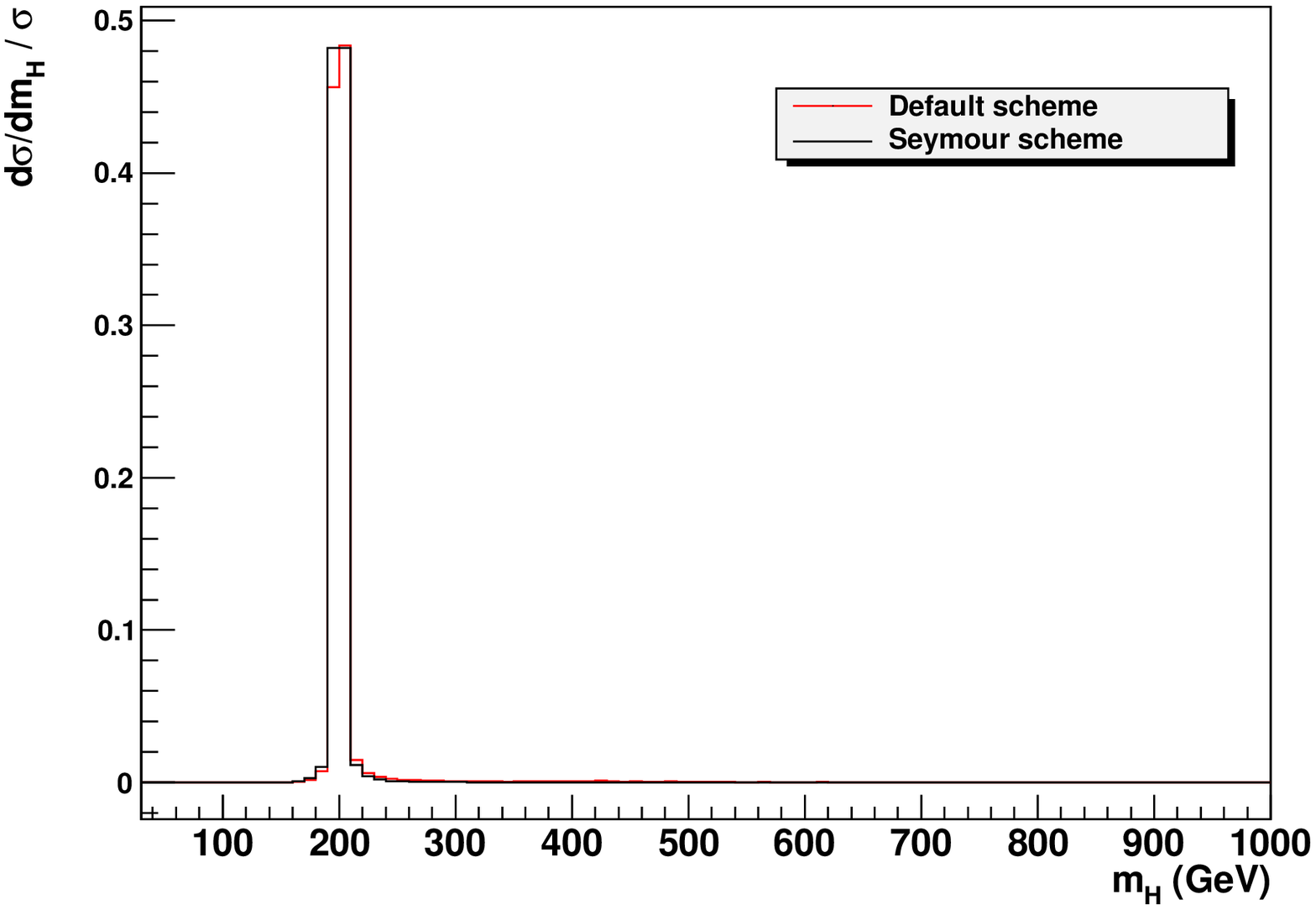}
\end{minipage}
\begin{minipage}[b]{0.5\linewidth}
\includegraphics[width=80mm]{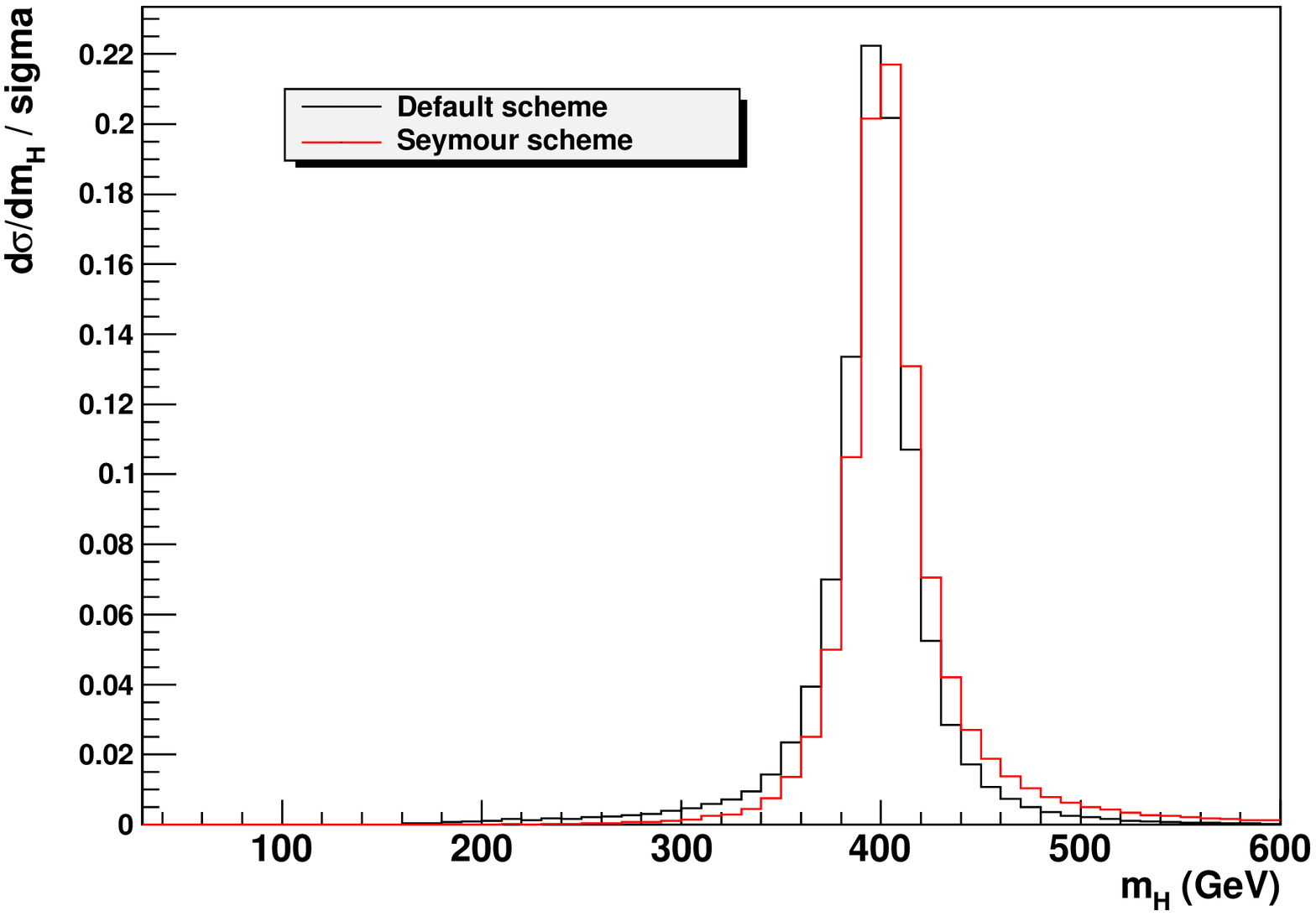}
\end{minipage}
\begin{minipage}[b]{0.5\linewidth}
\includegraphics[width=80mm]{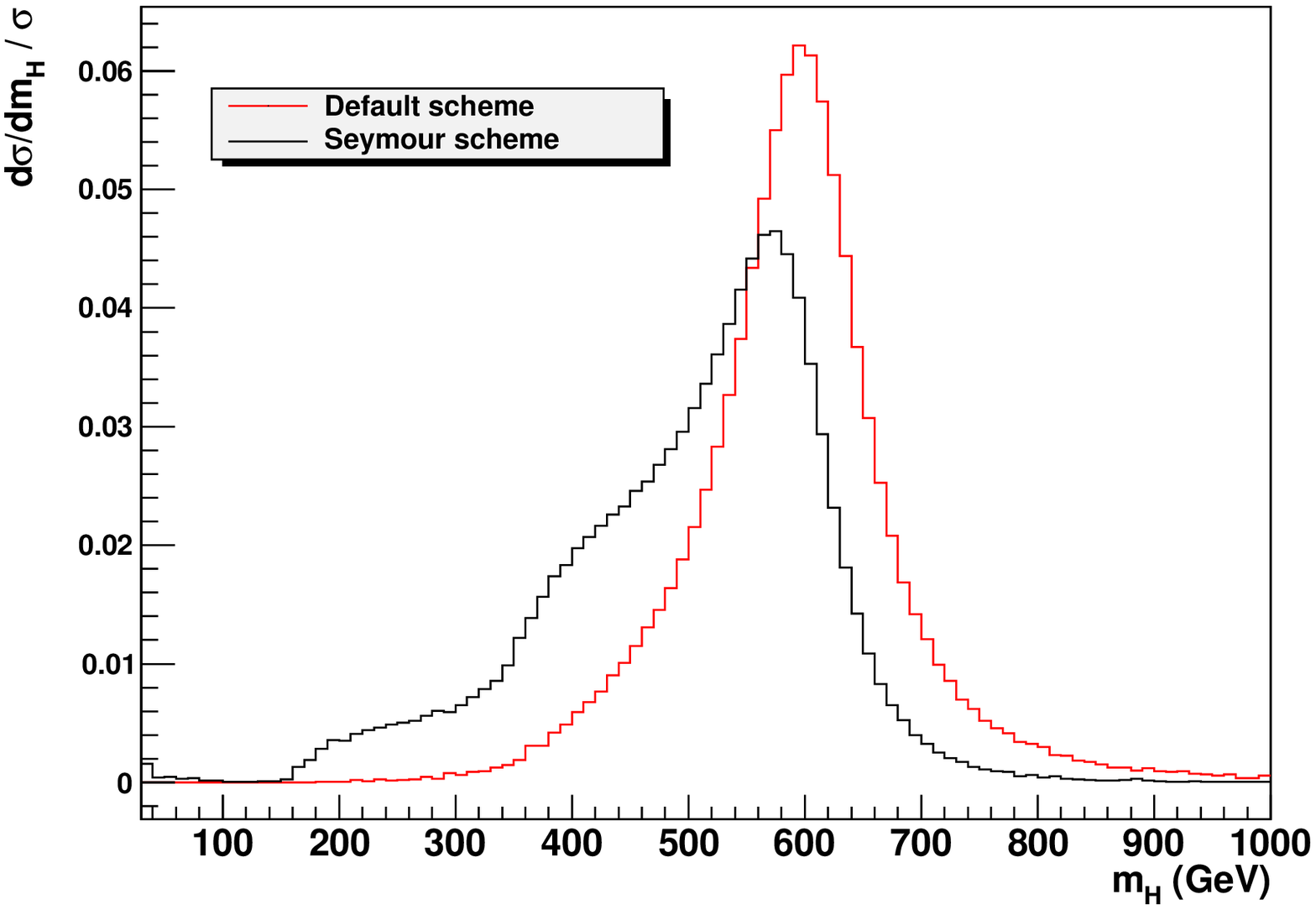}
\end{minipage}
\begin{minipage}[b]{0.5\linewidth}
\includegraphics[width=80mm]{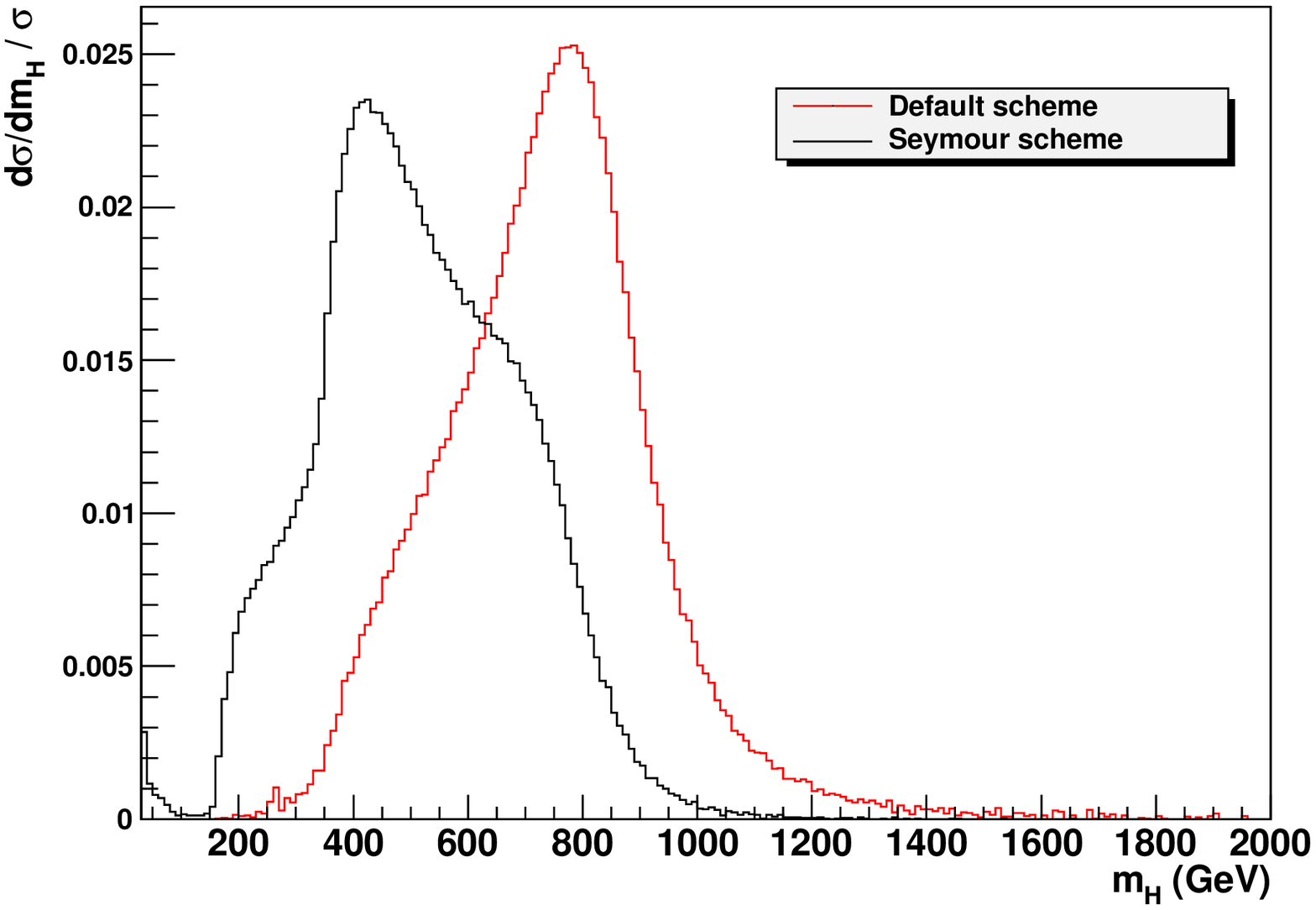}
\end{minipage}
\caption{
\label{invmassdistro}
The invariant mass distribution of the Higgs boson with $m_H=200$,$400$,$600$,$800$ GeV, in the default and the Seymour scheme.}
\end{figure}

Within this context, it is interesting to notice that the invariant mass distribution of the Higgs boson, shown in Fig.~\ref{invmassdistro}, gets significantly distorted in the high mass region, where the Higgs width is large. The distortion is spectacularly stronger in the case of the Seymour scheme, as a consequence of the fact that the scheme tries to simulate the effects of signal-background interference off the resonant peak. These effects become increasingly important for high Higgs masses.

In experimental searches for the Higgs boson where its invariant mass
can be reconstructed from the momenta of the final state partons, as
is the case for $H\to\gamma\gamma$, or $H\to ZZ$, it is beneficial 
for the analysis to impose a kinematical cut on the total invariant
mass of the Higgs decay products. 
The invariant mass of the Higgs boson is then  constrained in a window
around the nominal Higgs boson mass, the size of which depends on the
experimental resolution, see e.g.~\cite{ATLAS_TDR}. As a consequence,
part of the signal is also cut.

The  signal cross section that survives such kinematical cuts on the
Higgs boson invariant mass can be calculated with {\ihixs}  and is
shown in table~\ref{table:windows}, for the window sizes employed
in~\cite{ATLAS_TDR}.  We observe that the reduction in the expected
 signal rate can reach $20\%-40\%$ for window choices smaller than the Higgs
width.  A non-negligible reduction of a few per cent persists even
when the invariant mass window is larger than the nominal Higgs width,
due to contributions from the tail of the Breit-Wigner distribution. 

\begin{table}[h]
 \begin{center}
  \begin{tabular}{| c ||   c || c | c | c | c | c |}
\hline
$m_H$ & $\Gamma_H$ & $\delta Q$&$\sigma^{DEF}$ &  $\sigma^{DEF;w}$ & $\sigma^{SEY}$ & $\sigma^{SEY;w}$ \\
\hline\hline
120&	0.0038&			5	&	17.66	&	17.56&	 	17.57&		17.56	\\ \hline
165&	0.2432&			5	&	8.874  	&  	8.62&	 	8.735&		8.62	\\\hline
200&	1.43&			8	&	5.566	&	5.14&	 	5.390&		5.14\\ \hline
400&	29.5&			34	&	1.799	&	1.448&	 	1.766&		1.447\\ \hline
600&	122&			110	&	0.2409	&	0.1928&		0.3819&		0.2305\\ \hline
800&	301&			300	&	0.03982	&	0.03451&		0.15683&		0.07510\\
\hline
 \end{tabular}
\end{center}
 \caption{
\label{table:windows}
Total cross section, $\sigma^{DEF,SEY}$ compared with the cross section in the invariant mass region $m_H\pm\delta Q$, denoted by $\sigma^{DEF;w}$ or $\sigma^{SEY;w}$, for LHC at $\sqrt{s}=7$ TeV with MSTW PDFs .}
\end{table}

This effect can be estimated by parton shower Monte Carlo simulations
which are the main simulation tools in experimental collaborations.  
It is important that  a realistic line-shape for a heavy Higgs boson is
implemented in these simulations.  
Common practice in experimental studies is to evaluate distributions 
with a LO or NLO Monte Carlo program interfaced with parton showers 
and then rescale the distributions by inclusive K factors. Those
K-factors, however, depend on the scheme adopted for the Higgs width, as well as on the size of the experimental window, if one exists in the analysis, as shown in table~\ref{table:Kfactors}.
\begin{table}[h]
 \begin{center}
  \begin{tabular}{| c ||   c || c | c | c | c | c|c|}
\hline
$m_H$ & $\Gamma_H$ & $\delta Q$&$K_{NNLO}^{DEF}$ &  $K_{NNLO}^{DEF;w}$ & $K_{NNLO}^{SEY}$ & $K_{NNLO}^{SEY;w}$  &$K^{ZWA}_{NNLO}$\\
\hline\hline
120&	0.0038&	5	&	2.05	&2.05&	2.05	&2.05	& 2.05	\\ \hline
165&	0.2432&	5	&	2.02  &2.03&	2.016&2.04	& 2.033	\\\hline
200&	1.43&	8	&	2.00	&2.03&	2.023&2.03	& 2.027 	\\ \hline
400&	29.5&	34	&	1.90	&1.94&	1.94	&1.95	& 1.95	\\ \hline
600&	122&	110	&	1.66	&1.66&	1.87	&1.72	& 1.64	\\ \hline
800&	301&	300	&	1.63	&1.59&	2.07	&1.77	& 1.54	\\
\hline
 \end{tabular}
\end{center}
 \caption{
\label{table:Kfactors}
NNLO K-factors  with the width according to the Default scheme, $K_{NNLO}^{DEF}$,  in the presence of kinematical windows, $K^{DEF;w}_{NNLO}$, within the Seymour scheme, $K_{NNLO}^{SEY}$, and within the Seymour scheme in the presence of kinematical windows, $K^{SEY;w}_{NNLO}$, compared with the K-factors in the zero width approximation, $K_{NNLO}^{ZWA}$.}
\end{table}

When one departs from the zero width approximation, the branching
ratios into the various final states also depend on the virtuality, as
opposed to the nominal mass, of the Higgs boson. Assuming, at first,
that all invariant masses are reconstructed experimentally,
table~\ref{table:BranchingRatios} shows the difference between
convoluting the branching ratio to a WW final state with the
production cross section and the Breit-Wigner distribution, as in
eq. \ref{eq:BWG}, and just multiplying the total cross section with the
branching ratio evaluated at the nominal Higgs mass value. We see
that, in the low mass region,  the relative deviations are very small,
at the per mille level, thanks to the stability of the branching ratio
in the region sampled by the Breit-Wigner. In the high mass region
they can become large, especially when off-resonant effects are taken 
into account, as is the case for the {\tt Seymour} scheme.

\begin{table}
 \begin{center}
  \begin{tabular}{| c ||   c || c | c | c | c | c | c |}
\hline
$m_H$ & $\Gamma_H$ & $BR_{H\to WW}$&$\sigma_{pp\to H \to WW}$ & $\sigma_1$   & 
  $\delta\sigma_1 \%$& $ \sigma_2$ & $\delta\sigma_2 \%$
\\
\hline\hline
120&	0.0038&	0.1354	&	2.441	&2.396	&	-1.8 & 2.384	& -2.3		\\ \hline
165&	0.2432&	0.958	&	8.446  	&8.493	&	0.6	& 8.43	&-0.2		\\\hline
200&	1.43&	0.742	&	4.123	&4.132	&	0.2	& 4.05	&-1.7			\\ \hline
400&	29.5&	0.576	&	1.045	&1.041	&	-0.4	& 1.157	&10.8		\\ \hline
600&	122&	0.560	&	0.132	&0.131	&	-0.8	& 0.163	&23.8		\\ \hline
800&	301&	0.594	&	0.02269	&0.02299	&	1.3	& 0.0285	&25.6		\\
\hline
 \end{tabular}
\end{center}
 \caption{
\label{table:BranchingRatios}
Cross section convoluted with the branching ratio to WW, compared with the product 
 $\sigma_1=\sigma_{pp\to H}\times BR_{H\to WW}$, the relative deviation 
 $\delta\sigma_1=(\frac{\sigma_1}{\sigma_{pp\to H \to WW}}-1)\cdot 100\%$, and the product with the production cross section in the ZWA,  $\sigma_2=\sigma^{ZWA}_{pp\to H}\times BR_{H\to WW}$, with the relative deviation 
 $\delta\sigma_2=(\frac{\sigma_2}{\sigma_{pp\to H \to WW}}-1)\cdot 100\%$. All numbers are for LHC at $\sqrt{s}=7$TeV with MSTW PDFs .}
\end{table}

\begin{figure}[h]
\begin{minipage}[b]{0.5\linewidth}
\includegraphics[width=80mm]{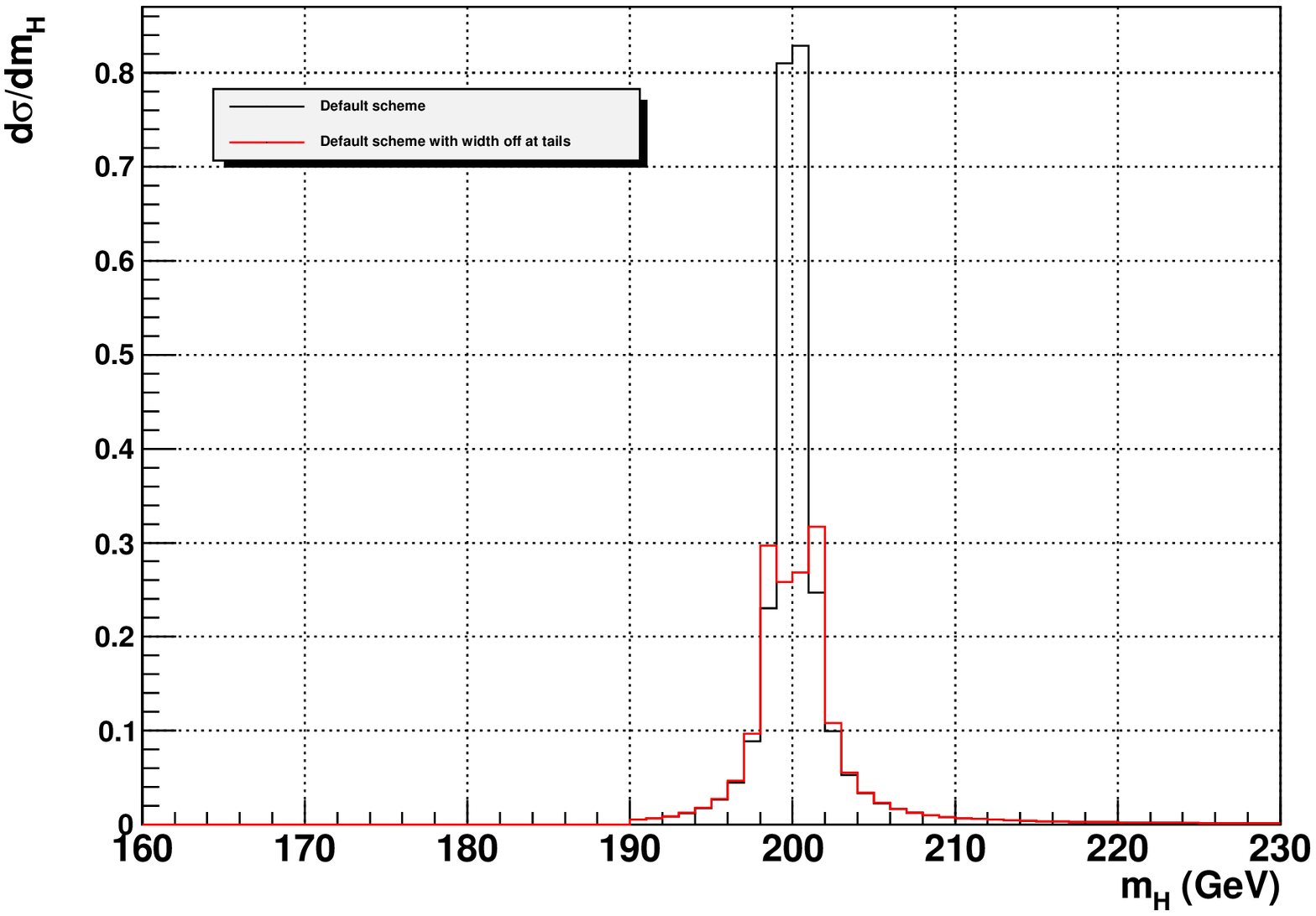}
\end{minipage}
\begin{minipage}[b]{0.5\linewidth}
\includegraphics[width=80mm]{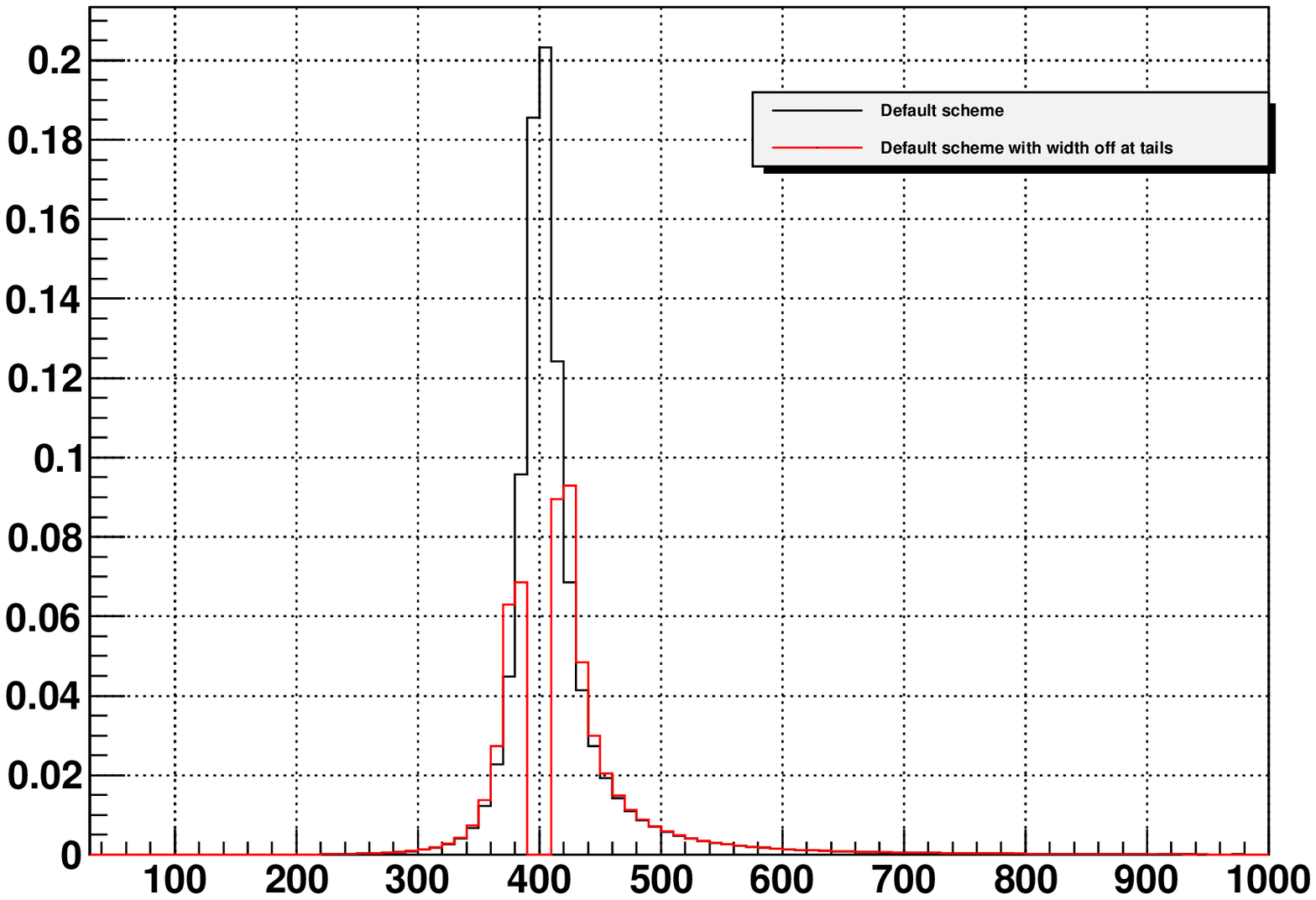}
\end{minipage}
\begin{minipage}[b]{0.5\linewidth}
\includegraphics[width=80mm]{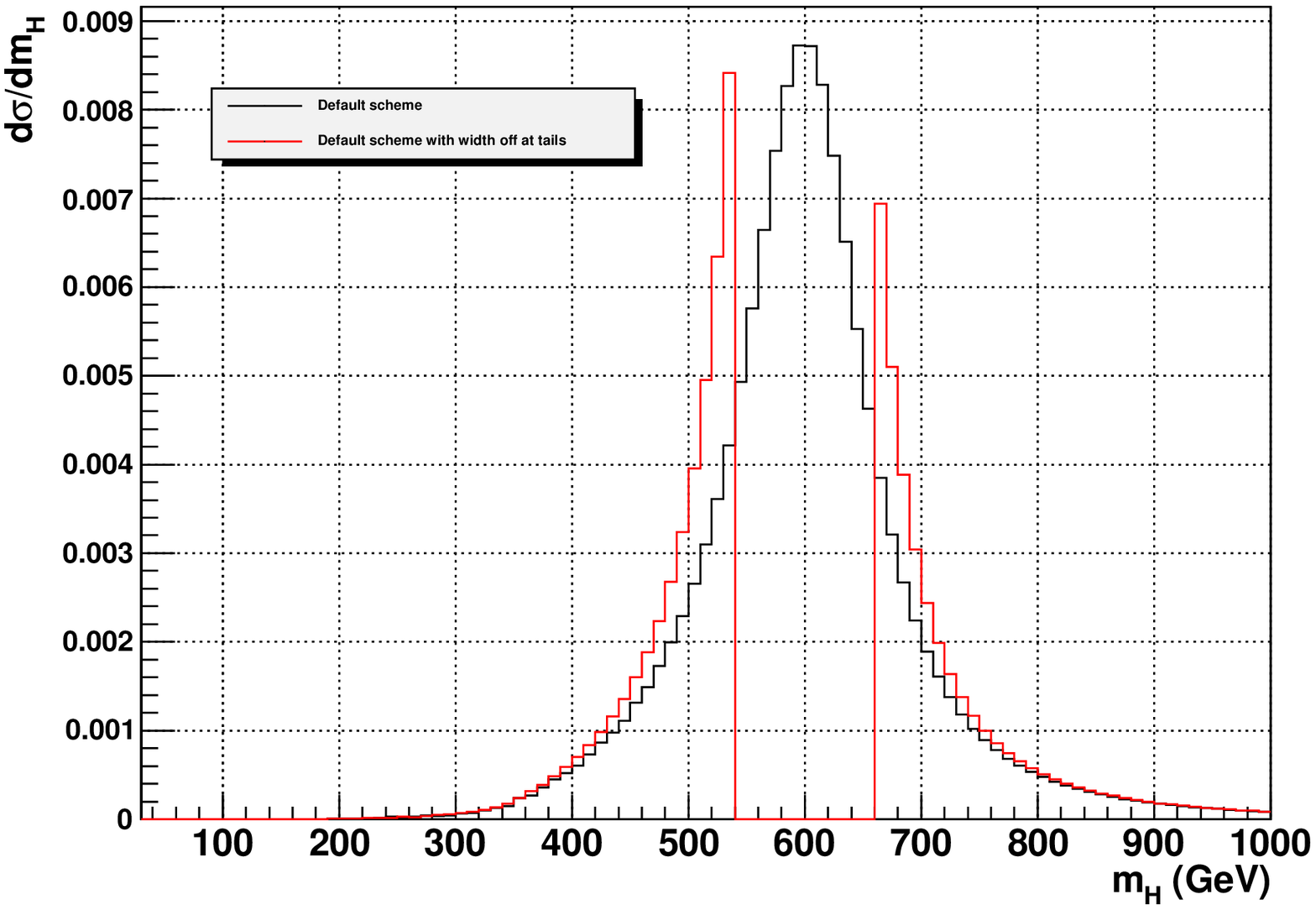}
\end{minipage}
\begin{minipage}[b]{0.5\linewidth}
\includegraphics[width=80mm]{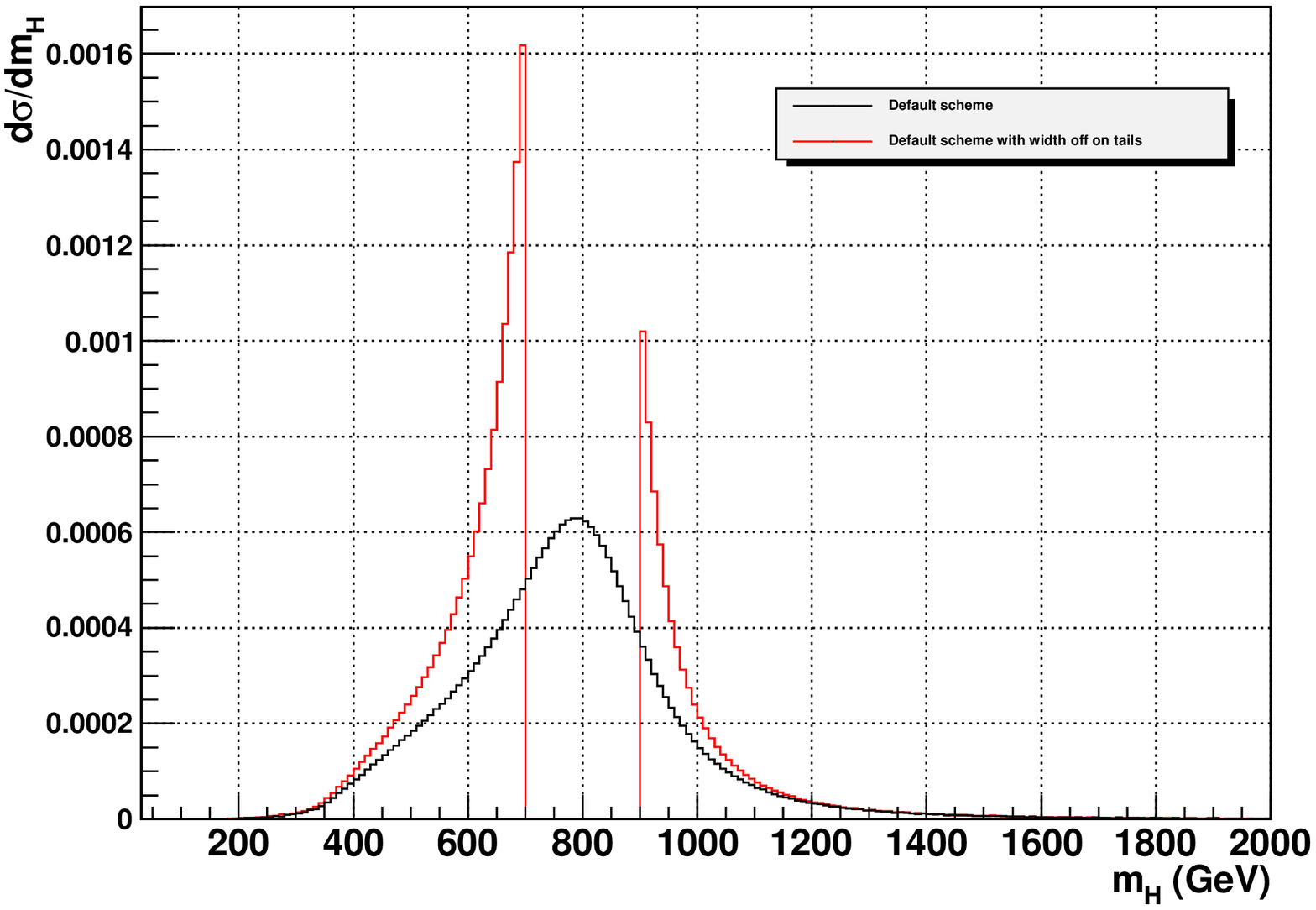}
\end{minipage}
\caption{
\label{figure:tailsDEF}
The invariant mass distribution of the Higgs boson with $m_H=200$,$400$,$600$,$800$ GeV, in the default  scheme, compared to the distribution of the tails computed with the off-resonant propagator.}
\end{figure}

\begin{figure}
\begin{minipage}[b]{0.5\linewidth}
\includegraphics[width=80mm]{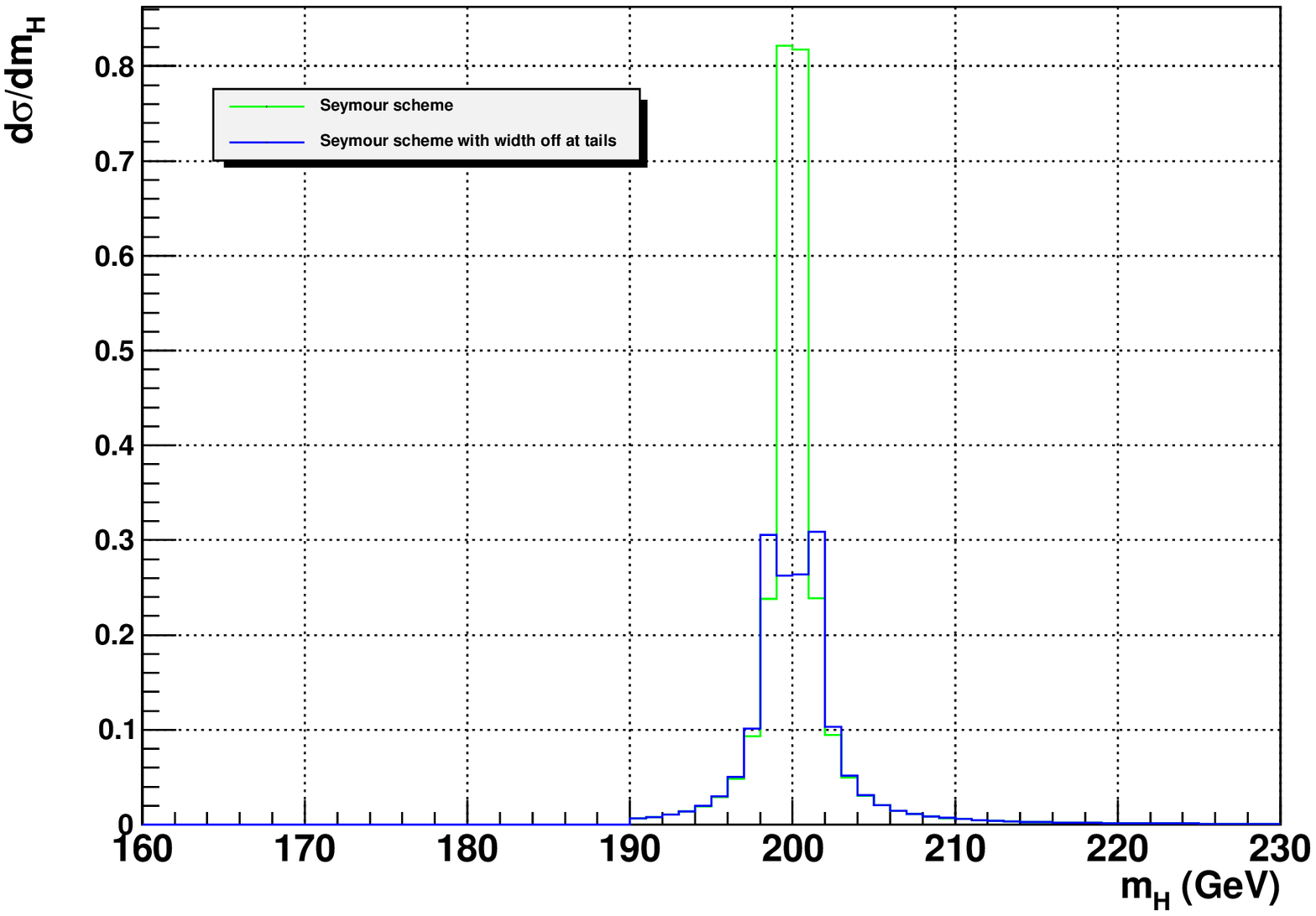}
\end{minipage}
\begin{minipage}[b]{0.5\linewidth}
\includegraphics[width=80mm]{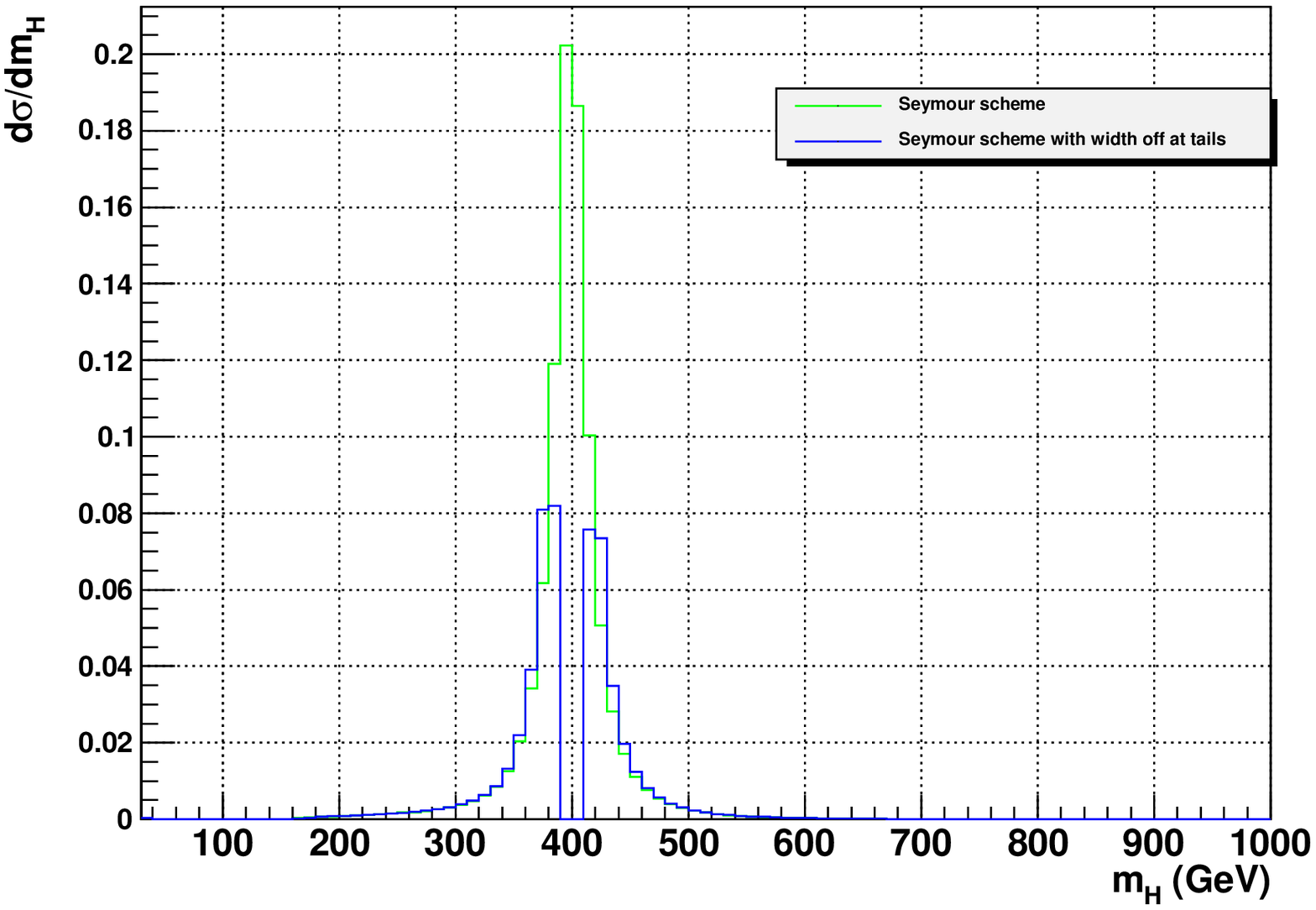}
\end{minipage}
\begin{minipage}[b]{0.5\linewidth}
\includegraphics[width=80mm]{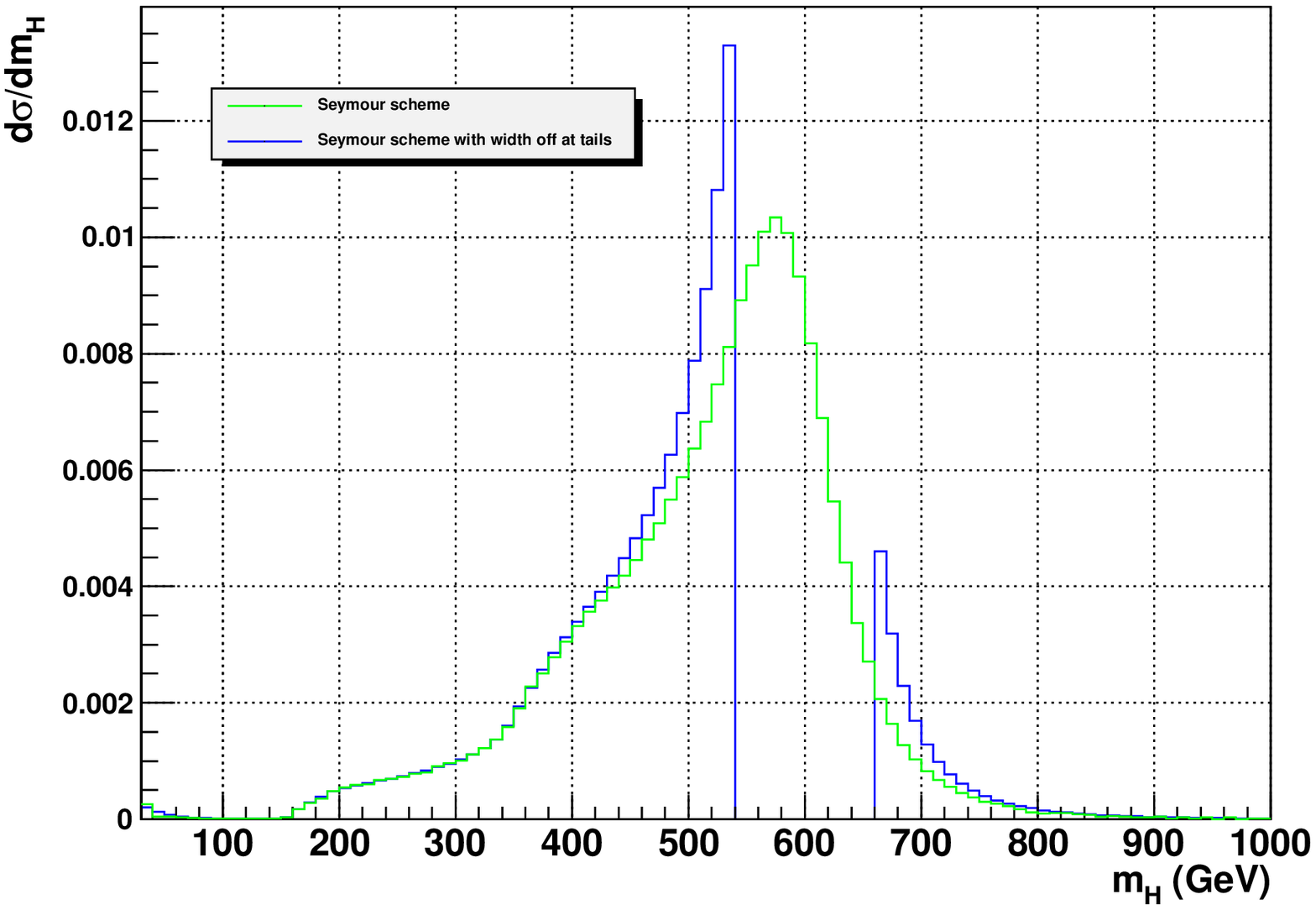}
\end{minipage}
\begin{minipage}[b]{0.5\linewidth}
\includegraphics[width=80mm]{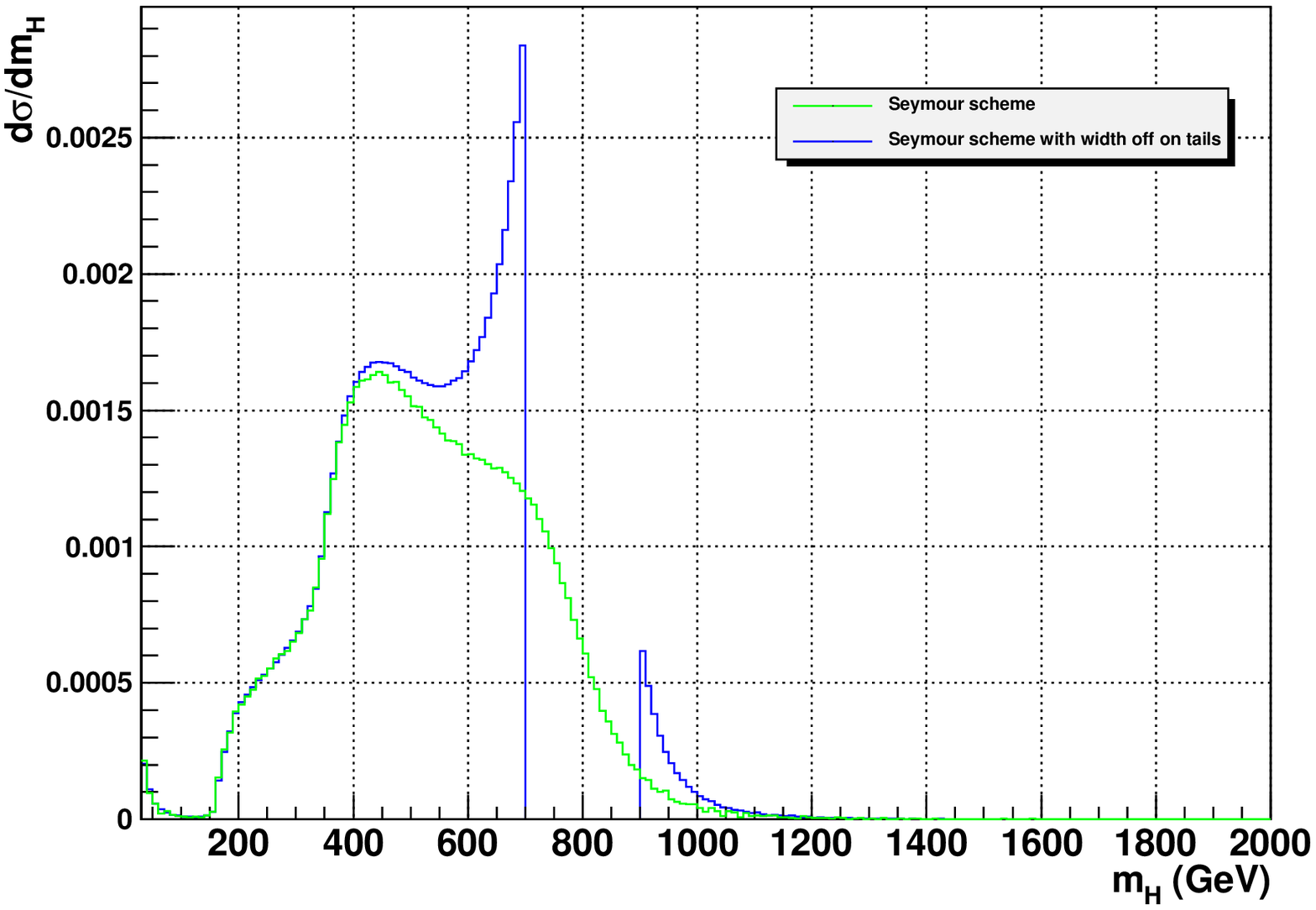}
\end{minipage}
\caption{
\label{figure:tailsSEY}
The invariant mass distribution of the Higgs boson with $m_H=200$,$400$,$600$,$800$ GeV, in the Seymour  scheme, compared to the distribution of the tails computed with the off-resonant propagator.}
\end{figure}

An interesting theoretical question is whether the resummation recipe
for the Higgs propagator employed in the vicinity of the Higgs
resonance affects the tails of the invariant mass distribution, where
we know that the correct propagator is the one appearing in
eq.~\ref{plainpropagator}. To study this,  we compare the invariant
mass distribution with 
the distribution calculated in the region of the tails with the width
of the Higgs boson set to zero in the denominator of the
propagator. This distribution diverges at the peak, as expected, so we
cut off a small region around the peak, of the order of
$\Gamma(m_H)/2$ or smaller, to assist convergence. 
The comparison is shown in fig.~\ref{figure:tailsDEF}, for the default 
scheme and in fig.~\ref{figure:tailsSEY} for the Seymour scheme. In
both cases the tails of the distributions are described well, when $|Q-m_H| > \Gamma(m_H)$ which indicates that the precise prescription for the propagator resummation in the peak region doesn't affect the tails of the invariant mass distribution.

\subsection{Inclusive Higgs boson production in the presence of a fourth generation of quarks}

\begin{figure}
\includegraphics[width=\linewidth]{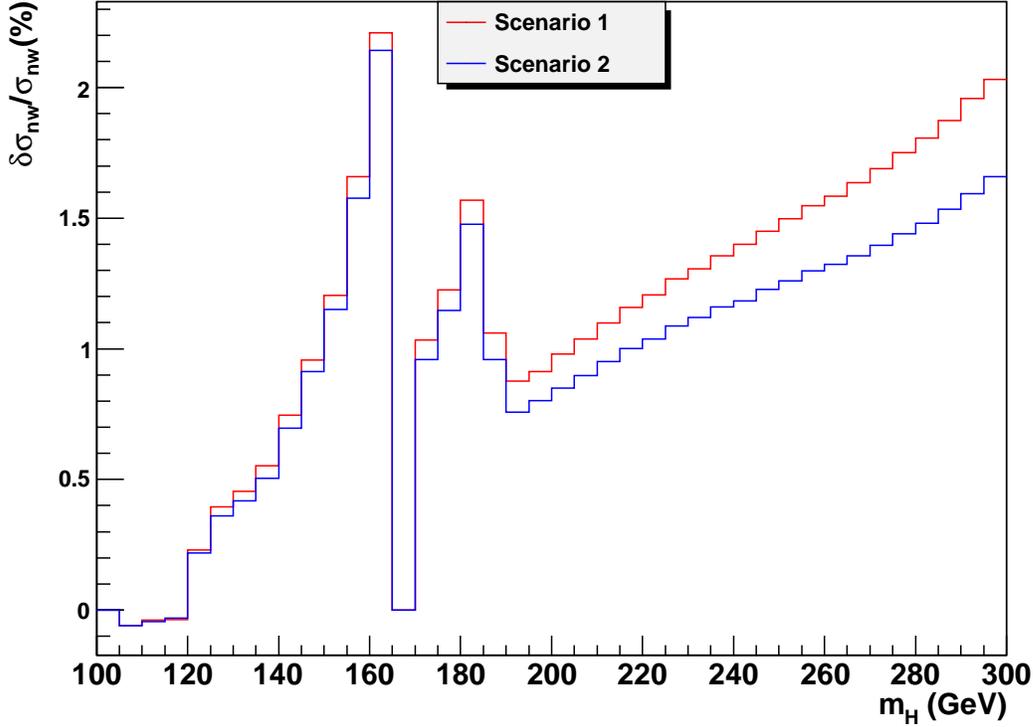}
\caption{
\label{FGwidth4gen}
Relative difference $\frac{\sigma -\sigma^{ZWA}}{\sigma^{ZWA}}\cdot100\%$ of the cross section in SM4, within the two scenaria of ~\cite{Anastasiou:2011qw}, in the zero width approximation, $\sigma^{ZWA}$, and in the approximation where $\Gamma_{SM4}(m_h)=\Gamma_{SM}(m_H)$. Scenario $1$ refers to a fourth generation down quark of mass $m_{d_4}=300$GeV while scenario $2$ assumes $m_{d_4}=300$GeV. In both cases the fourth generation up quark has a mass that is determined by eq.2 in section 3 of~\cite{Anastasiou:2011qw}.}
\end{figure}

An extension of the Standard Model with an additional family of quark
and  leptons yields a large  increase  to the gluon fusion process.  
It is therefore tested more easily at the
TEVATRON and the LHC than the SM scenario .
   
The most accurate computation of the total inclusive cross section for
the production of  the Higgs boson in a model with 
a fourth fermionic generation, for the LHC, has been presented 
recently in~\cite{Anastasiou:2011qw}. This calculation was performed
in the ZWA.  

The value of the physical Higgs width in such a scenario depends on
the details of the model, and is generally bigger than the Standard Model
Higgs width. In this paper, we would like to assess the impact of the
width, making the rough assumption that its  value  is  the  same as
in the SM width. 

With this assumption, the relative difference between the production
cross sections  reported in~\cite{Anastasiou:2011qw} and a computation
with the Higgs width on can reach the level of $2.5\%$ in the high
mass region, but is not significant for masses of $m_H<160$GeV, as
shown in fig.~\ref{FGwidth4gen}. Let us remark, however, that in the
high mass region the width of the Higgs boson  is enhanced  by the
opening of new decay channels to third and  fourth generation
fermions,  in a way that depends on the model. 

Ref.~\cite{Anastasiou:2011qw} refrained  from providing cross-sections
for Higgs boson mass values  higher than $m_h > 300 {\rm GeV}$.
Cross-section predictions in that range  should always take into
account finite width effects.

\section{The Higgs  cross-section for a variable  bottom-quark Yukawa interaction}

In the standard model the bottom-quark Yukawa coupling 
is much smaller than the top-quark counterpartner,
$$
\frac{\lambda_b}{\lambda_t}=\frac{m_b}{m_t}\sim0.02,
$$
and the bottom-quark fusion cross section is  about $2-3\%$ of the gluon fusion cross section. 
This feature may however not be conserved in extensions of the
standard model, which for example contain more than one Higgs field
electroweak doublets. 
In such a scenario more than one physical Higgs boson arise after electroweak symmetry breaking. 
The Yukawa couplings  may then be modified by further mixing angles of
the model and can differ strongly from their Standard Model values.
It is  then possible that the bottom-quark fusion and gluon fusion via
bottom-quarks  become very significant  in comparison to gluon fusion
via top-quark loops.

In this Section, we would like to study the
sum of  the two processes, gluon fusion and bottom-quark fusion,  
which contribute to the inclusive production  of a Higgs boson  as a
function of the bottom Yukawa coupling (we  denote by $Y_b$ its
value, normalized to the Standard Model).

\begin{figure}[htb!]
\centering
\includegraphics{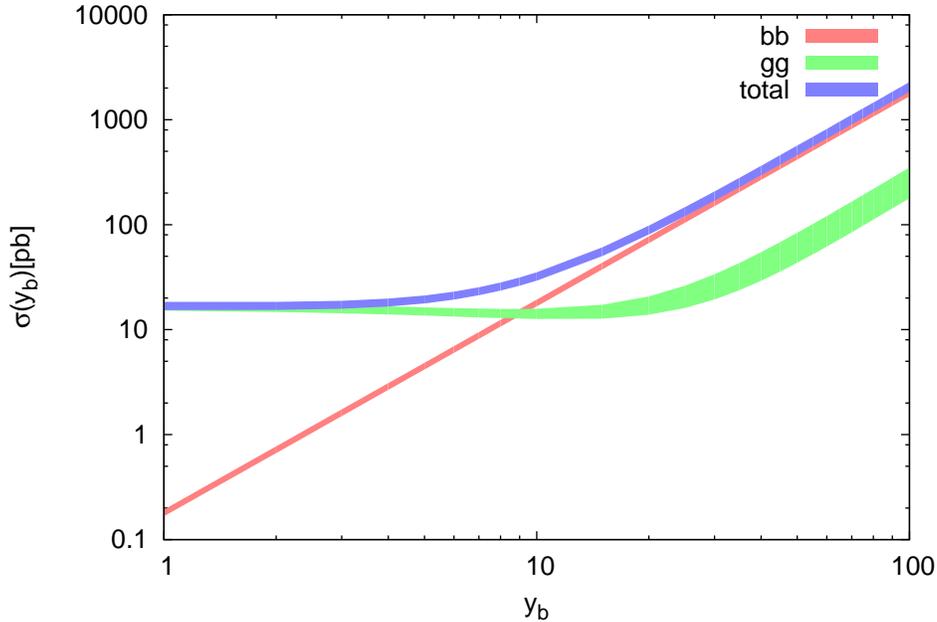}
\caption{
\label{fig:yb1}
Single Higgs production cross sections as a function of the
  re-scaling factor of the bottom-quark  Yukawa coupling $Y_b$. The bands represent the uncertainty due to the choice of  factorization and renormalization scale.}
\end{figure}
In Figure~\ref{fig:yb1} we demonstrate the inclusive  cross-section
for  gluon fusion and bottom-quark fusion as well as  their sum in the
ZWA for a nominal Higgs mass of $120$ GeV at the LHC for $\sqrt{s}=7$ TeV. 
We observe that for  small $Y_b$ the cross-section is  dominated by
gluon fusion. For high $Y_b$, bottom-quark fusion is  dominant but
there is  also  a large  contribution from gluon fusion, however  via
bottom rather  than top-quarks.  Notice  that the gluon fusion
cross-section reduces for  moderate values of $Y_b$ due  to a negative
interference effect  of top and bottom-quark loops.  

It  is often the practice that  uncertainties for beyond the Standard
Model Higgs  bosons are taken over  from studies within the Standard
Model.  This may  not be a very bad option if  new physics only introduces  new
heavy particles and does not alter Higgs couplings to light quarks
significantly, since all such scenaria  can be well described by a
common effective theory operator.  In our scenario however, we  need
to be more attentive.   
Our calculation of the gluon fusion cross section is exact through NLO
for both bottom and top-quark loop contributions. Our NNLO calculation
includes only top-quark loops in the framework of HQET.  For large
$Y_b$ where bottom-quark loops dominate, our evaluation  of the gluon
fusion cross-section is reduced to NLO accuracy ( not NNLO). 

\begin{figure}[htb!]
\centering
\includegraphics{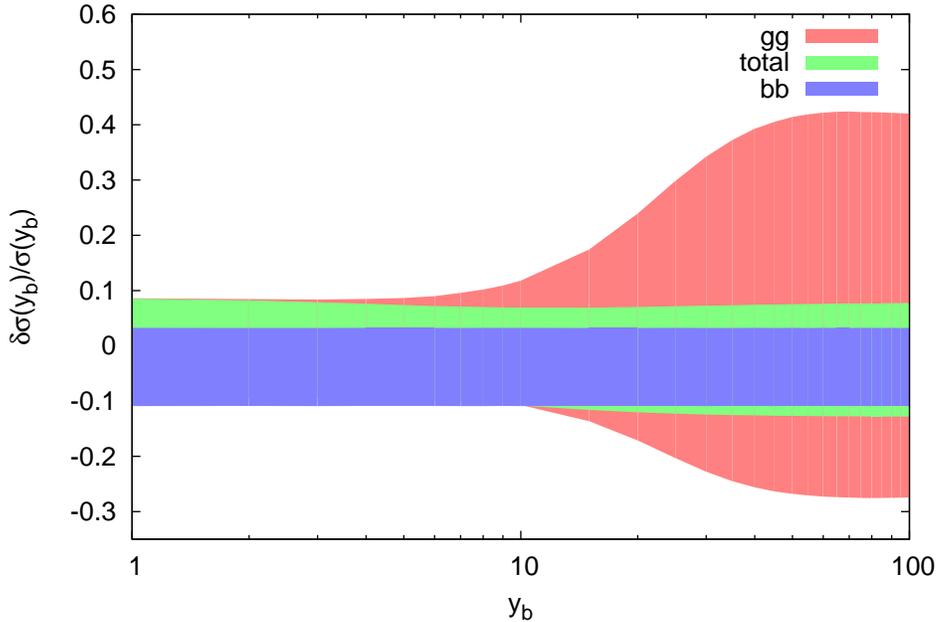}
\caption{
\label{fig:yb2}
Relative scale uncertainties of single Higgs production cross sections as a function of $y_b$.}
\end{figure}

In Fig.~\ref{fig:yb2} we  demonstrate the scale  variation of the
cross-sections for 
gluon fusion , quark-bottom fusion and their sum, where we have combined scale uncertainties linearly.
The scale  uncertainty for gluon fusion increases for  large $Y_b$ as
expected due to the dominance of the NLO only bottom-quark
corrections.  The scale uncertainty of the inclusive cross-section is
dominated by  the largest cross-section contribution.

A large bottom-quark  Yukawa coupling enhances the $H\rightarrow
b\bar{b}$ decay width. The total width of the Higgs  boson can be
derived  from the Standard Model total width and branching ratios as: 
\begin{equation}
\label{eq:Ybwidth}
\Gamma_H(Q) = \Gamma^{{\rm SM}}_H(Q) \times \left[ 
\left(Y_b^2 -1\right) Br^{{\rm SM}}_{H \to b \bar{b}}(Q) + 1
\right].  
\end{equation}

\begin{figure}[htb!]
\centering
\includegraphics{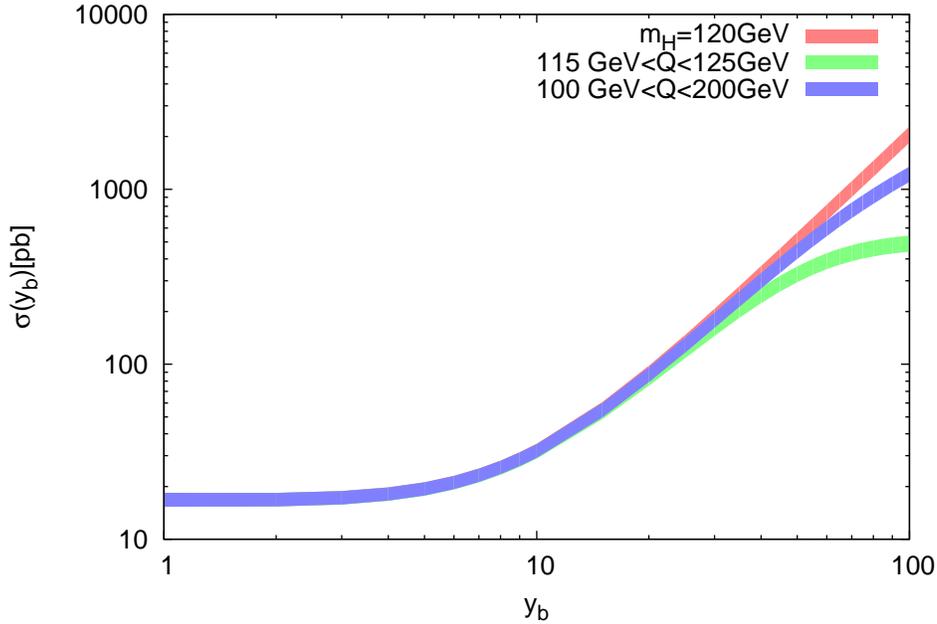}
\caption{Combined gluon and bottom fusion cross section as a function
  of $y_b$ in on-shell and off-shell scheme. The bands represent the scale uncertainty.
  }
\label{fig:yb3}
\end{figure}

In Figure~\ref{fig:yb3}, we have plotted the combined bottom and gluon
fusion cross section for a Higgs boson of $m_H = 120 {\rm GeV}$ 
for two different invariant mass windows.
While the width effects are small up to $Y_b\sim 30$, they significantly change the total cross section for higher $y_b$ values, due to the steep increase of the Higgs width as the mass crosses the vector boson thresholds.


\section{
		The {\ihixs}  program
		}
The source code for {\ihixs}  can be downloaded from its website at 

{\tt http://www.phys.ethz.ch/\textasciitilde pheno/ihixs}

Installation instructions can be found in the website and in the {\tt README} file supplied in the distribution. Here, we briefly describe the main functionality of the code. 

\subsection{Usage}
The various features of {\ihixs}  are controlled by an input {\tt runcard}, a text file that is edited by the user. To run with a given  {\tt runcard} as input type in the installation directory:

{\tt ./ihixs -i runcard\_name -o output\_filename}

When no {\tt runcard} is given, the program runs on the default card (called `runcard') in the installation directory. When no output filename is given, the program writes the output in {\tt runcard\_name.out}. 

The output consists of the total cross sections per perturbative order in QCD, together with the corresponding Monte-Carlo errors achieved and the PDF errors. Those are set to zero if no PDF uncertainty is requested in the {\tt runcard}. The input {\tt runcard} is also appended. 

\subsection{Setting options and variables}
In the {\tt runcard} anything after a hash symbol,`\#', is considered as a comment and is ignored. The following options are available:
\begin{itemize}
\item {\tt pdf\_provider} : sets the PDF grid used. The user can choose between MSTW08, ABKM09 and GJR09. Within the MSTW PDFs there is also the option to switch confidence level from $68\%$ to $90\%$ and to use the MSTW grids with the strong coupling constant varied by one standard deviation from the best fit value.  The exact filenames of the available grids are stated in the default {\tt runcard}.
\item {\tt effective\_theory\_flag } : set to $0$ for the exact LO and NLO QCD effects and HQET approximation for NNLO. Set to $1$ for the improved HQET approximation through LO, NLO, NNLO.
\item {\tt no\_error\_flag}: Set to 0 to calculate with PDF uncertainty, set to $1$ to calculate without PDF uncertainty.
\item {\tt collider}: Set to `LHC' or `TEVATRON' 
\item {\tt Etot}: The total center of mass collider energy. This option is ignored if the collider chosen above is Tevatron.
\item {\tt mhiggs}: The nominal mass of the Higgs boson. 
\item {\tt  higgs\_width\_scheme} : Set to 0 for the default finite width scheme. Set to 1 for the Seymour scheme. for a description of these schemes see section~\ref{section:hadronic-and-partonic-cross-sections}.
\item {\tt higgs\_width\_grid}:  = The path\footnote{Absolute or relative to the run directory.} of the file with the grid for the width of the Higgs, and the branching ratios to $\gamma \gamma$, $WW$, $ZZ$ and $b\bar{b}$ as a function of $m_H$. If no path is set the default grid is used, {\tt HdecayGrid.dat}, constructed with {\tt Hdecay v.3.532} \cite{Djouadi:1997yw}with arguments that can be read in the header of the file. If the user supplies a grid file of his own, operating requirements are that the maximum number of grid points cannot exceed $16200$, that the first three lines of the file are reserved for comments (so they are not read) and that the format of each line is respected, i.e. that the data is given in the order $m_H,\Gamma_H,BR_{\gamma\gamma},BR_{WW},BR_{ZZ},BR_{b\bar{b}}$.  
\item  {\tt min\_mh} : Setting a minimum in the invariant mass of the Higgs boson. This allows the user to study the total cross section in the presence of kinematical cuts. 
\item  {\tt max\_mh} : Setting a maximum in the invariant mass of the Higgs boson.
\item {\tt bin\_flag} : Set to 1 to produce  files with the bin-integrated Higgs invariant mass. Set to 0 not to produce it. The data files produced contain the cross section per bin, with the bin size set to $1$ GeV, from $30$ to $2000$ GeV at LO, NLO and NNLO. The files are named `{\tt masshisto\$$m_H$.\$order}', so e.g. for $m_H=200$GeV the NLO file will be `{\tt masshisto200.1} '.
\item {\tt muf/mhiggs} : The ratio of the factorization scale and the Higgs mass. 
\item {\tt mur/mhiggs} : The ratio of the renormalization scale and the Higgs mass. 
\item {\tt DecayMode}: Set to {\tt no\_width} for the zero width approximation total cross section, to `{\tt total}' for finite width total cross section, or to the decay modes `{\tt gamma gamma}', `{\tt ZZ}', `{\tt WW}', `{\tt b b-bar}'.  
\item {\tt ProductionMode}: Set to `{\tt gg}' for gluon fusion or to `{\tt bb}' for bottom-quark fusion.
\item {\tt K\_ewk}: This is a global rescaling factor for all electroweak corrections. Set to $0.0$ to switch them off. 
\item {\tt K\_ewk\_real}: Set to 0.0 to switch the electroweak corrections to $H+j$ off.
\item {\tt K\_ewk\_real\_b}: Set to 0.0 to switch the electroweak corrections to $H+j$ that include diagrams with massive quarks or Higgs boson in the loop, off.          
\item {\tt m\_top}: the pole mass of the top-quark. 
\item {\tt Gamma\_top}: the width of the top-quark.
\item {\tt Y\_top} : rescaling factor for the SM Yukawa coupling of the top. Note that this can be set to an arbitrarily small positive value, but not to $0.0$ exactly. 
\item {\tt m\_bot}: the $\overline{MS}$ mass of the bottom-quark at $10$GeV. 
\item {\tt Gamma\_bot}: the width of the bottom-quark.
\item {\tt Y\_bot}: rescaling factor for the  SM Yukawa coupling of the bottom-quark. 
\item {\tt heavy quark}: Optional extra quarks in the model. The argument of this option should be formatted as $m_Q:\Gamma_Q:Y_Q$ where $Y_Q$ is the rescaling factor from a SM-like Yukawa coupling $m_Q/v$. For example, adding an extra $300$GeV quark with width $1.2$GeV and a Yukawa coupling that is $5.7\frac{m_Q}{v}$ the user should type: 

`{\tt heavy quark = 300.0 : 1.2 : 5.7}' 

\item {\tt m\_Z}: the mass of the Z boson.
\item {\tt Gamma\_Z}: the width of the Z boson
\item {\tt m\_W}: the mass of the W boson 
\item {\tt Gamma\_W}: the width of the W boson

\item {\tt epsrel}: Sets the relative Monte-Carlo integration error. 
\item {\tt epsabs}: Sets the absolute Monte-Carlo integration error. 
\item {\tt nstart} : Sets the number of points per Vegas iteration.
\item {\tt nincrease}: Set the number of points by which the number of points per iteration increases
\item {\tt mineval}: Set the minimum number of points before ending the Monte-Carlo integration
\item {\tt maxeval}: Set the maximum number of points after which the integration ends.
\item {\tt adapt to central only}: Set to 0 to force Vegas to adapt to all integrand. Set to 1 to adapt to the central integrand only. This is useful when running with PDF errors. Then each member of the PDF grid is treated as a separate integral. Adapting to the central only assumes that the peak structures of all integrals is similar which is a good approximation, and saves some CPU time.
\item {\tt vegas\_verbose}: Set to 0 for silent Vegas output. Set to $2$ to have information about each iteration printed in the standard output (the console).

\end{itemize}

\subsection{Libraries used}
The program uses the following libraries:
\begin{itemize}
\item The {\tt Cuba} library~\cite{Hahn:2004fe}, {\tt v.2.1},  for numerical integration. We use the Vegas algorithm, that employs importance sampling for variance reduction. For details on integration related arguments see~\cite{Hahn:2004fe} or the manual included in the {\tt Cuba-2.1} directory. We distribute {\tt Cuba-2.1} and compile it from source.
\item The LHAPDF library~\cite{Whalley:2005nh}. We assume the library is installed by the user. See 
Ref~\cite{lhapdf}
for details on installation.
\item The package {\tt OneLOop} ~\cite{vanHameren:2010cp,vanHameren:2009dr} for the evaluation of one-loop scalar integrals with complex masses. We use the library for the evaluation of finite box and triangle master integrals with massive propagators, necessary for the electroweak corrections to $H+j$ with massive fermions in the loop. We have checked our implementation of all other integrals against both {\tt OneLOop} and {\tt QCDloop}~\cite{Ellis:2007qk} at the limit of zero width for the massive propagators. We distribute {\tt OneLOop} and compile it from source.
\item The {\tt CHAPLIN} package~\cite{chaplin}, for  evaluating harmonic polylogarithms up to weight four for any complex argument. This package is also distributed. 
\end{itemize}

\section{Conclusions}

In this article, we have presented  a computer program, {\ihixs}, for
the inclusive cross-section of the  Higgs  boson in gluon fusion and
bottom-quark fusion. {\ihixs}  provides the most precise predictions for
the Higgs boson rate at hadron colliders in  fixed order perturbation
theory, including QCD corrections through NNLO and electroweak
corrections for virtual and real radiative partonic processes.  
{\ihixs}  is   interfaced with the LHAPDF library and allows the
assessment of uncertainties due to the various determinations of the
parton densities which are available in the library.  

In a time in which Higgs boson searches are growing in intensity, {\ihixs} 
provides a very flexible tool which can assist in this effort.  {\ihixs} 
provides inclusive cross-section
predictions  in and beyond  the  Standard Model, by allowing
modifications of Yukawa and electroweak couplings as  well as the
introduction of  new quarks  with arbitrary Yukawa couplings  and
masses. In addition, one can readily introduce effective Higgs-gluon
interactions which can account for further beyond  the Standard Model
effects~\footnote{We are grateful to Elisabetta Furlan for extensive
  testings and feedback on implementing complicated extensions of the
  Standard Model in {\ihixs}.}.

The phenomenology of the Higgs boson and its production rates have
been described extensively in the literature (recent  updates can be
found in Refs~\cite{Anastasiou:2008tj,deFlorian:2009hc,Ahrens:2010rs,Baglio:2010ae}). 
We dedicated  a very short analysis to issues
which have been studied earlier at length, 
such as the magnitude  of perturbative  corrections and the
convergence of the perturbative  series beyond NNLO in 
QCD~\cite{Moch:2005ky}. 

We  have  noticed that in recent experimental
studies~\cite{Collaboration:2011qi}  the zero width approximation
which is  used  for an  expected light Higgs boson in
the  Standard  Model is also employed for Higgs bosons or  Higgs
bosons with a sizable decay width. In this publication, we  discuss
finite width effects  on the  cross-section due  to resonant Higgs
boson diagrams. We also employ a  prescription to estimate  the effect
of  the signal-background interference  for high Higgs  boson masses
which can be dramatic.  

We  believe  that a realistic description of the Higgs line shape is
necessary in setting exclusion limits for the Higgs 
boson. We  remark that the description of the line shape in
  parton shower Monte-Carlo generators can be  very different (for a
  comparison see Ref.~\cite{Alioli:2008tz}).  
We have also demonstrated that the  magnitude  of radiative corrections
(K-factor)  differs  from expectations in the zero width
approximation when a large range of  virtualities for  the Higgs boson
is sampled. 

{\ihixs}  allows the  user to perform exhaustive studies of the Higgs
boson cross-section at  hadron colliders. We are looking forward to
comparing {\ihixs}  predictions  with LHC data.

\section*{Acknowledgments}
We are  grateful to  Elisabetta Furlan and  Claude Duhr for numerous
communications concerning their parallel research on issues relevant to the
development  of {\ihixs}.  Special thanks  to Fabian Stoeckli for
detailed  discussions on the line-shape of a heavy Higgs boson and
sharing his insight from parton shower Monte-Carlo simulations, and to Andreas van Hameren for his assistance with {\tt OneLOop} . 
Research supported by the Swiss National
Foundation under contract SNF 200020-126632.

\newpage

\begin{appendix}
\label{sec:appendix}
\section{Tables of Higgs  cross-sections}
\vfill 

\begin{table}[h]
 \begin{center}
  \begin{tabular}{| c ||  c | c | c |c|c|c|}
\hline
$m_H$&$\sigma(pb)$ & $\% \delta_{PDF}^+$ & $\% \delta_{PDF}^-$ & $\% \delta_{\mu_F}^-$ & $\% \delta_{\mu_F}^+$\\ 
\hline\hline
110.0&21.04&4.05&-3.1&8.95&-9.6\\ \hline
115.0&19.22&4.05&-3.11&8.78&-9.55\\ \hline
120.0&17.7&4.05&-3.11&8.63&-9.5\\ \hline
125.0&16.3&4.04&-3.12&8.48&-9.46\\ \hline
130.0&15.04&4.04&-3.12&8.35&-9.42\\ \hline
135.0&13.92&4.03&-3.14&8.23&-9.37\\ \hline
140.0&12.93&4.04&-3.15&8.12&-9.34\\ \hline
145.0&12.03&4.03&-3.16&8.0&-9.32\\ \hline
150.0&11.22&4.04&-3.17&7.89&-9.28\\ \hline
155.0&10.49&4.05&-3.18&7.8&-9.25\\ \hline
160.0&9.77&4.04&-3.2&7.7&-9.22\\ \hline
165.0&8.87&4.05&-3.22&7.65&-9.2\\ \hline
170.0&8.23&4.05&-3.24&7.58&-9.17\\ \hline
175.0&7.69&4.05&-3.26&7.51&-9.15\\ \hline
180.0&7.2&4.06&-3.28&7.43&-9.13\\ \hline
185.0&6.69&4.06&-3.29&7.37&-9.13\\ \hline
190.0&6.26&4.07&-3.31&7.31&-9.12\\ \hline
195.0&5.89&4.07&-3.34&7.24&-9.1\\ \hline
200.0&5.57&4.07&-3.36&7.19&-9.06\\ \hline
210.0&5.01&4.09&-3.39&7.06&-9.02\\ \hline
220.0&4.54&4.1&-3.44&6.92&-8.99\\ \hline
230.0&4.14&4.11&-3.48&6.79&-8.96\\ \hline
240.0&3.8&4.12&-3.53&6.68&-8.91\\ \hline
250.0&3.5&4.14&-3.56&6.57&-8.85\\ \hline
260.0&3.25&4.13&-3.6&6.44&-8.84\\ \hline
270.0&3.04&4.17&-3.65&6.3&-8.79\\ \hline
280.0&2.85&4.18&-3.69&6.18&-8.74\\ \hline
290.0&2.7&4.19&-3.73&6.04&-8.65\\ \hline
300.0&2.57&4.21&-3.78&5.89&-8.58\\ \hline

 \end{tabular}
\end{center}
 \caption{Total cross section for LHC at $\sqrt{s}=7$TeV with MSTW PDF errors (corresponding to $68\%$CL).}
\label{mstw}
\end{table}

\vfill 

\newpage 
		
\vfill 

\begin{table}[h]
 \begin{center}
  \begin{tabular}{| c ||  c | c | c |c|c|c|}
\hline
$m_H$&$\sigma(pb)$ & $\% \delta_{PDF}^+$ & $\% \delta_{PDF}^-$ & $\% \delta_{\mu_F}^-$ & $\% \delta_{\mu_F}^+$\\ 
\hline\hline

110.0&19.2&3.1&-3.1&8.16&-9.19\\ \hline
115.0&17.51&3.1&-3.1&8.02&-9.13\\ \hline
120.0&16.07&3.1&-3.1&7.89&-9.09\\ \hline
125.0&14.76&3.1&-3.1&7.77&-9.06\\ \hline
130.0&13.6&3.1&-3.1&7.65&-9.02\\ \hline
135.0&12.55&3.2&-3.2&7.55&-8.99\\ \hline
140.0&11.63&3.2&-3.2&7.44&-8.95\\ \hline
145.0&10.8&3.2&-3.2&7.33&-8.94\\ \hline
150.0&10.05&3.3&-3.3&7.26&-8.9\\ \hline
155.0&9.37&3.3&-3.3&7.17&-8.88\\ \hline
160.0&8.71&3.3&-3.3&7.1&-8.85\\ \hline
165.0&7.89&3.4&-3.4&7.05&-8.83\\ \hline
170.0&7.3&3.4&-3.4&6.99&-8.82\\ \hline
175.0&6.81&3.4&-3.4&6.93&-8.79\\ \hline
180.0&6.36&3.5&-3.5&6.86&-8.79\\ \hline
185.0&5.9&3.5&-3.5&6.82&-8.79\\ \hline
190.0&5.5&3.5&-3.5&6.77&-8.77\\ \hline
195.0&5.17&3.6&-3.6&6.72&-8.74\\ \hline
200.0&4.88&3.6&-3.6&6.66&-8.71\\ \hline
210.0&4.37&3.7&-3.7&6.54&-8.7\\ \hline
220.0&3.94&3.8&-3.8&6.43&-8.67\\ \hline
230.0&3.58&3.9&-3.9&6.34&-8.62\\ \hline
240.0&3.27&4.0&-4.0&6.22&-8.61\\ \hline
250.0&3.0&4.1&-4.1&6.13&-8.57\\ \hline
260.0&2.77&4.2&-4.2&6.03&-8.52\\ \hline
270.0&2.58&4.3&-4.3&5.94&-8.45\\ \hline
280.0&2.41&4.4&-4.4&5.82&-8.4\\ \hline
290.0&2.27&4.5&-4.5&5.7&-8.35\\ \hline
300.0&2.15&4.6&-4.6&5.55&-8.28\\ \hline

 \end{tabular}
\end{center}
 \caption{Total cross section for LHC at $\sqrt{s}=7$TeV with ABKM PDF errors (corresponding to $68\%$CL).}
\label{abkm}
\end{table}
\vfill 
\newpage
\vfill
		
\begin{table}[h]
 \begin{center}
  \begin{tabular}{| c ||  c | c | c |c|c|c|}
\hline
$m_H$&$\sigma(pb)$ & $\% \delta_{PDF}^+$ & $\% \delta_{PDF}^-$ & $\% \delta_{\mu_F}^-$ & $\% \delta_{\mu_F}^+$\\ 
\hline\hline
110.0&18.66&3.6&-3.6&7.87&-8.63\\ \hline
115.0&17.1&3.5&-3.5&7.73&-8.59\\ \hline
120.0&15.79&3.5&-3.5&7.58&-8.55\\ \hline
125.0&14.58&3.5&-3.5&7.46&-8.5\\ \hline
130.0&13.49&3.4&-3.4&7.35&-8.46\\ \hline
135.0&12.52&3.4&-3.4&7.25&-8.42\\ \hline
140.0&11.66&3.4&-3.4&7.14&-8.39\\ \hline
145.0&10.88&3.4&-3.4&7.04&-8.36\\ \hline
150.0&10.17&3.3&-3.3&6.95&-8.33\\ \hline
155.0&9.53&3.3&-3.3&6.85&-8.31\\ \hline
160.0&8.89&3.3&-3.3&6.77&-8.3\\ \hline
165.0&8.09&3.4&-3.4&6.73&-8.29\\ \hline
170.0&7.53&3.4&-3.4&6.66&-8.31\\ \hline
175.0&7.05&3.4&-3.4&6.58&-8.33\\ \hline
180.0&6.62&3.4&-3.4&6.53&-8.36\\ \hline
185.0&6.17&3.4&-3.4&6.46&-8.42\\ \hline
190.0&5.78&3.5&-3.5&6.4&-8.46\\ \hline
195.0&5.45&3.5&-3.5&6.35&-8.5\\ \hline
200.0&5.16&3.6&-3.6&6.29&-8.54\\ \hline
210.0&4.66&3.6&-3.6&6.18&-8.62\\ \hline
220.0&4.24&3.7&-3.7&6.04&-8.7\\ \hline
230.0&3.88&3.8&-3.8&5.94&-8.75\\ \hline
240.0&3.57&4.0&-4.0&5.83&-8.81\\ \hline
250.0&3.31&4.1&-4.1&5.7&-8.9\\ \hline
260.0&3.08&4.2&-4.2&5.61&-8.92\\ \hline
270.0&2.88&4.4&-4.4&5.49&-8.96\\ \hline
280.0&2.72&4.5&-4.5&5.38&-8.97\\ \hline
290.0&2.58&4.6&-4.6&5.24&-9.0\\ \hline
300.0&2.46&4.8&-4.8&5.09&-9.01\\ \hline

 \end{tabular}
\end{center}
 \caption{Total cross section for LHC at $\sqrt{s}=7$TeV with GJR PDF errors (corresponding to $68\%$CL).}
\label{gjr}
\end{table}	

\vfill 
\newpage

\section{Matrix element coefficients}


\subsection{Master Integral Definitions}
In the following we will use the shorthand notation 
$$
p_{i_1i_2..i_n}=p_{i_1}+p_{i_2}+..+p_{i_n}
$$ 
and define the Mandelstam variables 
$$s=(p_{12})^2, \quad t=(p_{23})^2, \quad u=(p_{13})^2, \quad m_H^2=(p_{123})^2$$ where $p_1^2=p_2^2=p_3^2=0$.
The master integrals are then given by 
\begin{eqnarray}
{\rm Tadp}(m^2) & = & \int \frac{d^Dk}{i \pi^{D/2}} \frac{1}{[k^2-m^2]} \nonumber \\
{\rm Bub}(s,m^2)& = & \int \frac{d^Dk}{i \pi^{D/2}} \frac{1}{[k^2-m^2][(k+p_{12})^2-m^2]} \nonumber \\
{\rm Tria}(s,m^2)& = &s\cdot \int \frac{d^Dk}{i \pi^{D/2}} \frac{1}{[k^2-m^2][(k+p_1)^2-m^2][(k+p_{12})^2-m^2]} \nonumber \\
{\rm Box}(s,t,u,m^2)& = &s\cdot t\cdot \int \frac{d^Dk}{i \pi^{D/2}} \frac{1}{[k^2-m^2][(k+p_1)^2-m^2][(k+p_{12})^2-m^2][(k+p_{123})^2-m^2]} \nonumber \\
{\rm BubE}(s,m_1^2,m_2^2) & = &\int \frac{d^Dk}{i \pi^{D/2}} \frac{1}{[k^2-m_1^2][(k+p_{12})^2-m_2^2]} \nonumber \\
{\rm TriaE}(s,m_H^2,m_1^2,m_2^2) & = &s\cdot \int \frac{d^Dk}{i \pi^{D/2}} \frac{1}{[k^2-m_1^2][(k+p_{12})^2-m_2^2][(k+p_{123})^2-m_1^2]}  \\
{\rm TriaF}(s,m_1^2,m_2^2) & = & s\cdot \int \frac{d^Dk}{i \pi^{D/2}} \frac{1}{[k^2-m_1^2][(k+p_1)^2-m_2^2][(k+p_{12})^2-m_2^2]} \nonumber \\
{\rm BoxE}(s,t,m_H^2,m_1^2,m_2^2) & = &s\cdot t\cdot \int \frac{d^Dk}{i \pi^{D/2}} \frac{1}{[k^2-m_1^2][(k+p_1)^2-m_2^2][(k+p_{12})^2-m_2^2][(k+p_{123})^2-m_1^2]}. \nonumber 
\end{eqnarray}
We also define
$$
{\rm BubD} \left( s,m_q^2,t \right)={\rm Bub} \left( s,m_q^2\right)-{\rm Bub} \left(t,m_q^2\right).
$$

\subsection{$A_{gggH}$}
\label{formulas:AgggH}
\begin{eqnarray}
A^{1q}_{gggH}&&\!\!\!\!(s,t,u,m_q^2) = 2\,{\frac { \left( 2\,st+tu+su \right) {\rm Box} \left( s,t
,u,m_q^2 \right) }{u}}
+2\,{\frac { \left( 2\,tu+st+su \right) {\rm Box}
 \left( t,u,s,m_q^2 \right) }{s}}\nonumber\\
&&
+2\,{\frac { \left( 2\,su+tu+st
 \right) {\rm Box} \left( u,s,t,m_q^2 \right) }{t}}
-8\,{\frac {
 \left( 4\,tu+{t}^{2}+{u}^{2} \right) s{\rm BubD} \left( s,m_q^2,s+t+u \right) }{ \left( u+t \right) ^{2}}}\nonumber\\
&&
-8\,{\frac { \left( {u}^{2}+4
\,su+{s}^{2} \right) t{\rm BubD} \left( t,m_q^2,s+t+u	 \right) }{
 \left( u+s \right) ^{2}}}
-8\,{\frac { \left( 4\,st+{s}^{2}+{t}^{2}
 \right) u{\rm BubD} \left( u,m_q^2,s+t+u \right) }{ \left( t+s
 \right) ^{2}}}\nonumber\\
&&+ \bigg( 4
\frac {5\,{s}^{4}u{t}^{2}+5\,s{u}^{4}{t}^{2}+8\,s{u}^{3}{t}^{3}+10\,{s
}^{2}{u}^{3}{t}^{2}+10\,{s}^{2}{u}^{2}{t}^{3}+10\,{s}^{3}{u}^{2}{t}^{2
}+8\,{s}^{3}u{t}^{3}+5\,{t}^{4}{u}^{2}s}
{t \left( u+t \right) s \left( u+s \right) u \left( t+s
 \right) }\nonumber\\
&&
+\frac {5\,{s}^{2}{t}^{4}u+5\,t{s}^{2}
{u}^{4}+8\,t{s}^{3}{u}^{3}+5\,{s}^{4}{u}^{2}t+2\,{u}^{4}{t}^{3}+2\,{u}
^{3}{t}^{4}+2\,{s}^{3}{u}^{4}+2\,{s}^{4}{u}^{3}+2\,{s}^{3}{t}^{4}+2\,{
s}^{4}{t}^{3}}
{t \left( u+t \right) s \left( u+s \right) u \left( t+s
 \right) }\nonumber\\
&&
-16m_q^2\bigg[\frac{  8\,{t}^{3}{u}^{2}s+4\,t{u
}^{4}s+8\,t{u}^{3}{s}^{2}+8\,{s}^{2}{t}^{3}u+8\,{t}^{2}{u}^{3}s+4\,{t}
^{4}su+4\,{t}^{3}{u}^{3}+3\,{t}^{2}{u}^{4}+3\,{t}^{4}{s}^{2}}
{ \left( u+s
 \right) ^{2} \left( t+s \right) ^{2} \left( u+t \right) ^{2}}\nonumber\\
&&
+\frac{
3\,{s}^{2
}{u}^{4}+3\,{t}^{4}{u}^{2}+4\,{s}^{3}{u}^{3}+3\,{s}^{4}{u}^{2}+3\,{s}^
{4}{t}^{2}+4\,{t}^{3}{s}^{3}+8\,{s}^{3}{u}^{2}t+4\,{s}^{4}ut+6\,{s}^{2
}{t}^{2}{u}^{2}+8\,{s}^{3}{t}^{2}u  }
{ \left( u+s
 \right) ^{2} \left( t+s \right) ^{2} \left( u+t \right) ^{2}}\bigg]\bigg)\nonumber\\
&& \times {\rm Tria} \left( s+t+u,m_q^2 \right)\nonumber\\
&&
 +\left( 16\,{\frac { \left( {u}^{2}+{t}^{2} \right) m_q^2}{
 \left( u+t \right) ^{2}}}-4\,{\frac {2\,{t}^{3}s+3\,s{t}^{2}u+3\,s{u}
^{2}t+2\,s{u}^{3}+{t}^{3}u+{u}^{3}t}{tu \left( u+t \right) }} \right)
{\rm Tria} \left( s,m_q^2 \right)\nonumber\\
&&
 + \left( 16\,{\frac { \left( {s
}^{2}+{u}^{2} \right) m_q^2}{ \left( u+s \right) ^{2}}}-4\,{\frac
{2\,t{s}^{3}+3\,{s}^{2}ut+3\,s{u}^{2}t+2\,{u}^{3}t+s{u}^{3}+{s}^{3}u}{
su \left( u+s \right) }} \right) {\rm Tria} \left( t,m_q^2 \right)\nonumber\\
&&
 + \left( 16\,{\frac { \left( {t}^{2}+{s}^{2} \right) m_q^2
}{ \left( t+s \right) ^{2}}}-4\,{\frac {2\,{s}^{3}u+3\,{s}^{2}ut+3\,s
{t}^{2}u+2\,{t}^{3}u+{t}^{3}s+t{s}^{3}}{st \left( t+s \right) }}
 \right) {\rm Tria} \left( u,m_q^2 \right)\nonumber\\
&&\nonumber
 -16\,{\frac {s{t}^{2}u
+{s}^{2}ut+s{u}^{2}t+{s}^{2}{t}^{2}+{t}^{3}u+{s}^{3}u+{t}^{3}s+{t}^{2}
{u}^{2}+s{u}^{3}+t{s}^{3}+{u}^{3}t+{s}^{2}{u}^{2}}{ \left( u+t
 \right)  \left( u+s \right)  \left( t+s \right) }}
\end{eqnarray}

\begin{eqnarray}
A^{2q}_{gggH}&&\!\!\!\!(s,t,u,m_q^2) = \left( -16\,m_q^2-2\,{\frac {-{u}^{2}-tu+2\,st}{u}} \right) {\rm
Box} \left( s,t,u,m_q^2 \right) \nonumber\\
&&
+ \left( -16\,m_q^2+2\,{
\frac {st+{s}^{2}-2\,tu}{s}} \right) {\rm Box} \left( t,u,s,m_q^2
 \right) \nonumber\\
&&
+ \left( 16\,m_q^2+4\,{\frac {su}{t}} \right) {\rm Box}
 \left( u,s,t,m_q^2 \right)
+8\,{\frac {s \left( -u+t \right) {
\rm BubD} \left( s,m_q^2,s+t+u \right) }{u+t}}\nonumber\\
&&
+8\,{\frac { \left(
{u}^{2}+4\,su+{s}^{2} \right) t{\rm BubD} \left( t,m_q^2,s+t+u
 \right) }{ \left( u+s \right) ^{2}}}
-8\,{\frac { \left( s-t \right) u
{\rm BubD} \left( u,m_q^2,s+t+u \right) }{t+s}}\nonumber\\
&&
+ \bigg( -16\,{
\frac { \left( {u}^{3}t+3\,{t}^{2}{u}^{2}+5\,{s}^{2}ut+3\,{s}^{2}{t}^{
2}-s{u}^{3}+5\,s{u}^{2}t+8\,s{t}^{2}u-{s}^{3}u+t{s}^{3} \right) m_q^2}
{ \left( u+s \right) ^{2} \left( t+s \right)  \left( u+t \right)
}}\nonumber\\
&&
-4\,{\frac {2\,{s}^{3}{t}^{2}-s{t}^{2}{u}^{2}+2\,{t}^{2}{u}^{3}-{s}^
{2}{t}^{2}u-2\,t{u}^{2}{s}^{2}-2\,{s}^{2}{u}^{3}-2\,{s}^{3}{u}^{2}}{
 \left( u+s \right) stu}} \bigg) {\rm Tria} \left( s+t+u,m_q^2
 \right) \nonumber\\
&&
+ \left( 16\,{\frac { \left( -u+t \right) m_q^2}{u+t}}+4
\,{\frac {2\,{t}^{2}s-{t}^{2}u-2\,{u}^{2}s}{tu}} \right) {\rm Tria}
 \left( s,m_q^2 \right) \nonumber\\
&&
+ \left( 16\,{\frac { \left( {u}^{2}+4\,su
+{s}^{2} \right) m_q^2}{ \left( u+s \right) ^{2}}}+4\,{\frac {2\,t
{s}^{3}-2\,{s}^{2}{u}^{2}+{s}^{2}ut+s{u}^{2}t+2\,{u}^{3}t}{su \left( u
+s \right) }} \right) {\rm Tria} \left( t,m_q^2 \right)\nonumber\\
&&
 + \left(
-16\,{\frac { \left( s-t \right) m_q^2}{t+s}}-4\,{\frac {2\,{s}^{2
}u+{t}^{2}s-2\,{t}^{2}u}{st}} \right) {\rm Tria} \left( u,m_q^2
 \right) -16\,{\frac {su+st+tu}{u+s}}\\
A^{3q}_{gggH}&&\!\!\!\!(s,t,u,m_q^2)=A_2(t,u,s,m_q^2) \\
A^{4q}_{gggH}&&\!\!\!\!(s,t,u,m_q^2)=A_2(u,s,t,m_q^2)
\end{eqnarray}
\subsection{$A_{q\bar{q}gH}$}
Defining 
\[
\tau_q \equiv \frac{4 m_q^2}{m_H^2}
\]
we have
\begin{eqnarray}
&&  {\rm A}_{qqgH}\left(\tau_q , y, m_H^2\right) =  
-\frac{3}{4}\Bigg\{   \frac{-2A(\tau_q)
}{1-y}  
\nonumber \\
&&+\frac{2y}{(1-y)^2}
{\rm BubD}(y\cdot m_H^2, m_q^2,m_H^2) 
+
\frac{\tau y }{(1-y)^2}
\left[ 
\frac{{\rm Tria}(y\cdot m_H^2, m_q^2) }{y}
- 
{\rm Tria}(m_H^2, m_q^2) 
\right]
\nonumber \\ 
& & - \frac{1}{1-y}  
(1- \epsilon)  {\rm Tria}(y\cdot m_H^2, m_q^2)
\Bigg\}
\label{Aqqgh}
\end{eqnarray}
Notice that for $y_{12} \to 0$ we have that 
\begin{equation}
 \lim_{y_{12} \to 0} {\rm A}_{qqg}\left(\tau_q , y_{12}, m_H^2\right)
= 
A\left(\tau_q, m_H^2\right),  
\end{equation}
which is the familiar kernel of the born $gg \to h$ amplitude.

\subsection{$A_{ewk}$}

\begin{eqnarray} 
&&A_{ewk}\left(s,t,u,m_H,m_z \right) =
\frac{t\left(m_z^2-s\right)}
{s 
\left[m_z^2\left(s +t \right) -s t \right]
}
{\rm
  BoxE}^{(d=6)}
\left(s, t, m_H^2, m_z^2, 0 \right)  \nonumber \\ 
&&\hspace{2cm} 
+
 \frac{m_z^2\left(m_z^2-s\right)}
{s 
\left[m_z^2\left(s +t \right) -s t \right]
}
{\rm
  TriaE}
\left(s, m_H^2, m_z^2,0 \right)  \nonumber \\ 
&&\hspace{2cm} 
+ 
\frac{1}{s} 
\Bigg[ 
 1 -\frac{m_z^2}{s+u}-\frac{m_z^4}{m_z^2\left(s +t \right) -s t}
\Bigg]
{\rm
  TriaE}
\left(t, m_H^2, m_z^2, 0 \right)  
\nonumber \\ 
&&\hspace{2cm} 
+\frac{{\rm Bub}(m_H^2, m_z^2) - {\rm BubE}(t, m_z^2,0) }{s\left( s+u\right)},
\label{Aewk}
\end{eqnarray}
with $u=m_H^2-s-t$. Where we have expressed the form factor in terms of the 6-dimensional Box 
using the following relation
\begin{eqnarray} 
&& {\rm BoxE^{(d=6)}}\left(s, t, m_H^2, m_z^2,0 \right)=    
-{\frac { \left(  \left( s+t \right) {m_{{z}}}^{2}-2\,st \right) 
 \left(  \left( s+t \right) {m_{{z}}}^{2}-st \right)  \left( -1+\epsilon
 \right) {\rm Tadp} \left( {m_{{z}}}^{2} \right) }{2\epsilon \left( {m_{{z}}}^
{2}-t \right) {m_{{z}}}^{2} \left( {m_{{z}}}^{2}-s \right)  \left( -1+
2\,\epsilon \right) stu}}\nonumber \\ 
&&\hspace{2cm} 
+{\frac { \left(  \left( s+t \right) {m_{{z}}}^{2}-st \right) {\rm BubE} \left( s,{m_{{z}}}^{2},0 \right) }{2tu \left( {
m_{{z}}}^{2}-s \right) \epsilon}}
+{\frac { \left(  \left( s+t \right) {m_{{z}}}^{2}-st \right) {\rm BubE} \left( t,{m_{{z}}}^{2},0 \right) }{2su
 \left( {m_{{z}}}^{2}-t \right) \epsilon}}\nonumber \\ 
&&\hspace{3cm} 
-{\frac { \left(  \left( st+{t}^{2}-su+tu \right) {m_{{z}}}^{2}-st \left( u+t \right)  \right) {\rm
TriaE} \left( s,{m_H}^{2},{m_{{z}}}^{2},0 \right) }{ 2\left( -1+2\,\epsilon
 \right) stu}}\nonumber \\ 
&&\hspace{3cm} 
-{\frac { \left(  \left( su+{s}^{2}-tu+st \right) {
m_{{z}}}^{2}-st \left( s+u \right)  \right) {\rm TriaE} \left( t,{m_{{
h}}}^{2},{m_{{z}}}^{2},0 \right) }{2 \left( -1+2\,\epsilon \right) stu}}\nonumber \\ 
&&\hspace{3cm} 
+{\frac { \left(  \left( s+t \right) ^{2}{m_{{z}}}^{4}-2\,st \left( s+t
 \right) {m_{{z}}}^{2}+{s}^{2}{t}^{2} \right) {\rm BoxE} \left( s,t,{m_H}^{2},{m_{{z}}}^{2},0 \right) }{2 \left( -1+2\,\epsilon \right) stu}}
\end{eqnarray}

\subsection{$A_{ewk}^{m_t}$}
The form factor for the mixed QCD-electroweak corrections to $H+j$ with a top-quark in the loop is:
\begin{eqnarray} 
&&A_{ewk}^{m_t }\left(s,t,u,m_w^2,m_t^2\right)=
  {      \frac {{m_{{t}}}^{2}{\rm Tadp} \left( {m_{{w}}}^{2} \right) -{m_
{{t}}}^{2}{\rm Tadp} \left({m_{{t}}}^{2} \right)}{4{s}^{2}{m_{{w}}}^{4}}} \nonumber\\
&&\hspace{0.5cm} 
{\frac { +{m_{{t}}}^{2}\left( -{m_{{w}}}^{2}+{m_{{t}}}^{2} \right)   {\rm BubE} \left( s,{m_{{
w}}}^{2},{m_{{t}}}^{2} \right) }{4{s}^{2}{m_{{w}}}^{4}}} \nonumber\\
&&\hspace{0.5cm} 
+{\frac { \left( -{m_H}^{2}{m_{{t}}}^{2}-2\,{m_{{t}}}^{2}{m_{{
w}}}^{2}-4\,{m_{{w}}}^{4} \right) {\rm BubE} \left( t,{m_{{w}}}^{2},{m
_{{t}}}^{2} \right) }{4s \left( u+s \right) {m_{{w}}}^{4}}} \nonumber\\
&&\hspace{0.5cm} 
+{\frac {  \left( {m_H}^{2}{m_{{t}}}^{2}+2\,{m_{{t}}}^{
2}{m_{{w}}}^{2}+4\,{m_{{w}}}^{4} \right) {\rm Bub} \left( {m_H}^{2
},{m_{{w}}}^{2} \right) }{4s \left( u+s \right) {m_{{w}}}^{4}}} \nonumber\\
&&\hspace{0.5cm} 
 + \bigg\{ \big[      -2\,t \left( -t+s \right) {m_{{w}}}^{2}{m_{{t}}}^
{6}-t \left( -t+s \right) {m_H}^{2}{m_{{t}}}^{6}-2\,t \left( -t+s
 \right) {m_{{w}}}^{4}{m_{{t}}}^{4}   \nonumber \\
&&\hspace{0.5cm} 
\left. +t \left( 3\,{s}^{2}+us-tu-{t}^{2}\right) {m_{{w}}}^{2}{m_{{t}}}^{4}-{t}^{2}s{m_H}^{2}{m_{{t}}}^{4}
+4\,t \left( -t+s \right) {m_{{w}}}^{6}{m_{{t}}}^{2}\right.\nonumber \\
&&\hspace{0.5cm} 
-2\,t \left( {s}^{
2}-st \right) {m_{{t}}}^{2}{m_{{w}}}^{4}+2\,{s}^{2}{t}^{2}{m_{{w}}}^{2
}{m_{{t}}}^{2} \big] {\rm TriaF} \left( t,{m_{{w}}}^{2},{m_{{t}}}^{2} \right) \nonumber\\
&&\hspace{0.5cm} 
+ \big[ - \left( s+t \right) {m_H}^{2}{m_{{t}}}^{8}+ \left( -2\,t
-2\,s \right) {m_{{w}}}^{2}{m_{{t}}}^{8}-{m_H}^{2} \left( tu+us+{t
}^{2} \right) {m_{{t}}}^{6} \nonumber \\
&&\hspace{0.5cm} 
 + \left( {t}^{2}+3\,{s}^{2}+6\,st+tu+us\right) {m_{{w}}}^{2}{m_{{t}}}^{6}    \nonumber \\
&&\hspace{0.5cm}
+ \left( 2\,s+2\,t \right) {m_{{w}}}^{4}{m_{{t}}}^{6}  +{m_H}^{2} \left( stu+{t}^{2}s \right) {m_{{t}}}^{4}  \nonumber \\
&&\hspace{0.5cm} 
+ \left( t{u}^{2}+{t}^{3}+3\,{s}^{2}u+3\,stu+2\,{t}^{2}s+2\,{t}^{2
}u-{s}^{2}t+s{u}^{2} \right) {m_{{w}}}^{2}{m_{{t}}}^{4}\nonumber \\
&&\hspace{0.5cm}
+ \left( -5\,{t}^{2}-7\,{s}^{2}-5\,us-5\,tu-8\,st \right) {m_{{w}}}^{4}{m_{{t}}}^{4}
+ \left( 6\,t+6\,s \right) {m_{{w}}}^{6}{m_{{t}}}^{4} \nonumber \\
&&\hspace{0.5cm}
-2\,{s}^{2}t\left( u+t \right) {m_{{w}}}^{2}{m_{{t}}}^{2}+ \left( -3\,{s}^{2}t-3
\,stu-2\,{s}^{2}u-3\,{t}^{2}s \right) {m_{{w}}}^{4}{m_{{t}}}^{2}\nonumber \\
&&\hspace{0.5cm} 
+ \left( 12\,st+5\,tu+5\,us+9\,{s}^{2}+5\,{t}^{2} \right) {m_{{w}}}^{6}
{m_{{t}}}^{2}+ \left( -10\,s-10\,t \right) {m_{{w}}}^{8}{m_{{t}}}^{2}+
4\,{m_{{w}}}^{6}{s}^{2}t\nonumber \\
&&\hspace{0.5cm} 
+ \left( -8\,st-4\,{s}^{2} \right) {m_{{w}}}^{8}
+ \left( 4\,t+4\,s \right) {m_{{w}}}^{10} \big] {\rm TriaE}\left( s,{m_H}^{2},{m_{{w}}}^{2},{m_{{t}}}^{2} \right) \nonumber\\
&&\hspace{0.5cm} 
+ \big[ {m_{{t}}}^{8} \left( s+t \right) ^{2} + \left( -{s}^{2}u+3\,{t}^{2}s+2\,{s}^{2}t+5
\,stu-{s}^{3} \right) {m_{{t}}}^{6}\nonumber \\
&&\hspace{0.5cm} 
+ \left( -4\,{s}^{2}-2\,{t}^{2}-2\,
st \right) {m_{{w}}}^{2}{m_{{t}}}^{6}-{s}^{2}t \left( u+s \right) {m_{
{t}}}^{4}\nonumber \\
&&\hspace{0.5cm} + \left( {s}^{2}u+3\,{s}^{3}-stu-2\,{s}^{2}t-3\,{t}^{2}s
 \right) {m_{{w}}}^{2}{m_{{t}}}^{4}+ \left( -{s}^{2}+{t}^{2}+4\,st
 \right) {m_{{w}}}^{4}{m_{{t}}}^{4}\nonumber \\
&&\hspace{0.5cm} 
+2\,{s}^{3}t{m_{{w}}}^{2}{m_{{t}}}^{2}-2\,{s}^{2} \left( -t+s \right) {m_{{w}}}^{4}{m_{{t}}}^{2}+ \left( 
-4\,st+4\,{s}^{2} \right) {m_{{w}}}^{6}{m_{{t}}}^{2} \big] {\rm TriaF} \left( s,{m_{{w}}}^{2},{m_{{t}}}^{2} \right) \nonumber\\
&&\hspace{0.5cm} 
+ \big[ 2\,t \left( s+t \right) {m_{{w}}}^{2}{m_{{t}}}^{6} 
+ t \left( s+t \right) {m_H}^{2}{m_{{t}}}^{6}\nonumber \\
&&\hspace{0.5cm} 
-2\,t \left( st+tu+{t}^{2}+{s}^{2}-us \right) {m_{{t}}}^{4}{m_{{w}}}^{2}
+t{m_H}^{2} \left( st+2\,us \right) {m_{{t}}}^{4}\nonumber \\
&&\hspace{0.5cm} -6\,t \left( s+t \right) {m_{{w}}}^{6}{m_{{t}}}^{2}
+t \left( 5\,{s}^{2}+{t}^{2}+tu+9\,us+8\,st \right) {m_{{t}}}^{2}{m_{{w}}}^{4}\nonumber \\
&&\hspace{0.5cm} 
-t \left( {t}^{2}s+4\,{s}^{2}u+{s}^{2}t+stu \right) {m_{{t}}}^{2}{m_{{w}}}^{2}+4\,t\left( s+t \right) {m_{{w}}}^{8}
-4\,t \left( 2\,st+{s}^{2} \right) {m_{{w}}}^{6}\nonumber \\
&&\hspace{0.5cm} 
+4\,{m_{{w}}}^{4}{s}^{2}{t}^{2} \big] {\rm BoxE}^{(d=6)}
 \left( s,t,{m_H}^{2},{m_{{w}}}^{2},{m_{{t}}}^{2}\right) \bigg\} \nonumber \\
&&\hspace{0.5cm} 
\times \left[ 4s {m_{{w}}}^{4} \left( 4\,{m_{{t}}}^{2}stu+ \left(  \left( s+t \right)  \left( -{m_{{w}}}^{2}+{m_{{t
}}}^{2} \right) +st \right) ^{2} \right) \right]^{-1} \nonumber\\
&&\hspace{0.5cm} 
 +\bigg\{  \left( s+t \right)  \left( 2\,s+u+t \right) {m_{
{h}}}^{2}{m_{{t}}}^{8}+2\, \left( s+t \right)  \left( 2\,s+u+t
 \right) {m_{{w}}}^{2}{m_{{t}}}^{8}\nonumber\\
&&\hspace{0.5cm} 
-2\, \left( s+t \right)  \left( 2\,
s+u+t \right) {m_{{w}}}^{4}{m_{{t}}}^{6}\nonumber\\
&&\hspace{0.5cm} 
+ \left( -3\,stu-9\,{s}^{2}t-4
\,{s}^{3}-s{u}^{2}-6\,{t}^{2}s-3\,{t}^{3}-4\,{t}^{2}u-t{u}^{2}-5\,{s}^
{2}u \right) {m_{{w}}}^{2}{m_{{t}}}^{6}\nonumber\\
&&\hspace{0.5cm} 
+ \left( {t}^{2}u+3\,{t}^{2}s+6
\,stu+t{u}^{2}+3\,{s}^{2}t+s{u}^{2}+{s}^{3}+2\,{s}^{2}u \right) {m_{{h
}}}^{2}{m_{{t}}}^{6}\nonumber\\
&&\hspace{0.5cm}
 -6\, \left( s+t \right)  \left( 2\,s+u+t \right) {
m_{{w}}}^{6}{m_{{t}}}^{4}\nonumber\\
&&\hspace{0.5cm} 
+ \left( 17\,{s}^{2}u+5\,s{u}^{2}+27\,stu+5\,
t{u}^{2}+6\,{t}^{2}u+17\,{s}^{2}t+12\,{t}^{2}s+12\,{s}^{3}+3\,{t}^{3}
 \right) {m_{{w}}}^{4}{m_{{t}}}^{4}\nonumber\\
&&\hspace{0.5cm} 
+ \left( -14\,{t}^{2}su-6\,{u}^{2}s
t-2\,{t}^{3}u-9\,{s}^{2}{t}^{2}-3\,{t}^{2}{u}^{2}-t{u}^{3}-7\,{s}^{3}u
-5\,{s}^{2}{u}^{2}-6\,{t}^{3}s\nonumber \right.\\
&&\hspace{0.5cm} \left.
-s{u}^{3}-8\,{s}^{3}t-3\,{s}^{4}-13\,{s}
^{2}ut \right) {m_{{w}}}^{2}{m_{{t}}}^{4}+ \left( 2\,{s}^{2}{t}^{2}+{t
}^{2}su+{u}^{2}st+2\,{s}^{2}ut+{s}^{3}t \right) {m_H}^{2}{m_{{t}}}
^{4}\nonumber\\
&&\hspace{0.5cm}
+10\, \left( s+t \right)  \left( 2\,s+u+t \right) {m_{{w}}}^{8}{m_
{{t}}}^{2}\nonumber\\
&&\hspace{0.5cm}
+ \left( -19\,{s}^{2}u-5\,s{u}^{2}-41\,stu-5\,t{u}^{2}-8\,{t
}^{2}u-33\,{s}^{2}t-22\,{t}^{2}s-14\,{s}^{3}-{t}^{3} \right) {m_{{w}}}
^{6}{m_{{t}}}^{2}\nonumber\\
&&\hspace{0.5cm}
+ \left( {t}^{3}u+{t}^{2}{u}^{2}+12\,{t}^{2}su+15\,{s
}^{2}{t}^{2}+12\,{u}^{2}st+2\,{s}^{2}{u}^{2}+3\,{t}^{3}s+4\,{s}^{3}u \nonumber\right.\\ 
&&\hspace{0.5cm}\left.
+10\,{s}^{3}t+2\,{s}^{4}+22\,{s}^{2}ut \right) {m_{{w}}}^{4}{m_{{t}}}^{2}\nonumber\\
&&\hspace{0.5cm}
-st \left( 2\,s{u}^{2}+2\,{s}^{3}+{t}^{2}u+t{u}^{2}+4\,{s}^{2}u+3\,s
tu+2\,{t}^{2}s+2\,{s}^{2}t \right) {m_{{w}}}^{2}{m_{{t}}}^{2}\nonumber\\
&&\hspace{0.5cm}
-4\,\left( s+t \right)  \left( 2\,s+u+t \right) {m_{{w}}}^{10}+ \left( 4
\,{t}^{2}u+12\,stu+4\,{s}^{2}u+20\,{s}^{2}t+12\,{t}^{2}s+4\,{s}^{3}
 \right) {m_{{w}}}^{8}\nonumber\\
&&\hspace{0.5cm}-4\,st{m_{{w}}}^{6} \left( 2\,us+3\,st+2\,{s}^{2
}+2\,tu \right) +4\,{s}^{2}{t}^{2} \left( u+s \right) {m_{{w}}}^{4}\bigg\}
 {\rm TriaE} \left( t,{m_H}^{2},{m_{{w}}}^{2},{m_{{t}}}^{2} \right) 
\nonumber\\
&&\hspace{0cm} 
\times\left[4s \left( u+s \right) {m_{{w}}}^{4} \left( 4\,{m_{{t}}}^{2}stu+
 \left(  \left( s+t \right)  \left( -{m_{{w}}}^{2}+{m_{{t}}}^{2}
 \right) +st \right) ^{2} \right) \right]^{-1} 
\label{Aewk_mt}
\end{eqnarray}

where
\begin{eqnarray}
&&{\rm BoxE}^{(d=6)} \left( s,t,{m_H}^{2},{m_{{w}}}^{2},{m_{{t}}}^{2}\right)=
\left( {s}^{2}{t}^{2}-2\, \left( s+t \right) ^{2}{m_{{t}}}^{2}{m_{{w}}}^{2}-2\,st \left( s+t \right) {m_{{w}}}^{2}\nonumber \right. \\
&& \left.
+2\,st\left( t+s+2\,u \right) {m_{{t}}}^{2}+ \left( s+t \right) ^{2}{m_{{w}
}}^{4}+ \left( s+t \right) ^{2}{m_{{t}}}^{4} \right)
\frac {{\rm BoxE} \left( s,t,{m_H}^{2},{m_{{w}}}^{2},m_t^2 \right) }
{2stu \left( -1+2
\,\epsilon \right) }\nonumber\\
&&
+{\frac { \left(  \left( su-tu-st-{t}^{2} \right) {
m_{{w}}}^{2}+ \left( tu+st-su+{t}^{2} \right) {m_{{t}}}^{2}+st \left( 
u+t \right)  \right) {\rm TriaE} \left( s,{m_H}^{2},{m_{{w}}}^{2},
{m_{{t}}}^{2} \right) }{2stu \left( -1+2\,\epsilon \right) }}\nonumber\\
&&
+{\frac {
 \left(  \left( -st-su+tu-{s}^{2} \right) {m_{{w}}}^{2}+ \left( su+st-
tu+{s}^{2} \right) {m_{{t}}}^{2}+st \left( s+u \right)  \right) {\rm
TriaE} \left( t,{m_H}^{2},{m_{{w}}}^{2},{m_{{t}}}^{2} \right) }{2st
u \left( -1+2\,\epsilon \right) }}\nonumber\\
&& 
-\,{\frac { \left( s{\rm TriaF} \left( s
,{m_{{w}}}^{2},{m_{{t}}}^{2} \right) +t{\rm TriaF} \left( t,{m_{{w}}}^
{2},{m_{{t}}}^{2} \right)  \right)  \left(  \left( -t-s \right) {m_{{w
}}}^{2}+ \left( s+t \right) {m_{{t}}}^{2}+st \right) }{2stu \left( -1+2
\,\epsilon \right) }}
\end{eqnarray}

and $m_H^2=s+t+u$. The $m_t\to 0$ limit of this form factor trivially lead to $A_{ewk}$ of eq.~\ref{Aewk}.


\end{appendix}


\bibliographystyle{JHEP}

\begin{thebibliography}{10}

\bibitem{Barate:2003sz}
  R.~Barate {\it et al.}  [LEP Working Group for Higgs boson searches and
                  ALEPH Collaboration and  and],
  Phys.\ Lett.\  B {\bf 565} (2003) 61
  [arXiv:hep-ex/0306033].

\bibitem{Aaltonen:2011gs}
  T.~Aaltonen {\it et al.}  [CDF and D0 Collaboration],
  arXiv:1103.3233 [hep-ex].

\bibitem{Collaboration:2011qi}
  T.~A.~Collaboration,
  arXiv:1106.2748 [hep-ex].

\bibitem{Chatrchyan:2011tz}
  S.~Chatrchyan {\it et al.}  [CMS Collaboration],
  Phys.\ Lett.\  B {\bf 699} (2011) 25
  [arXiv:1102.5429 [hep-ex]].

\bibitem{Marciano:1987gz}
  W.~J.~Marciano,
  Annals N.\ Y.\ Acad.\ Sci.\  {\bf 518} (1987) 180.


\bibitem{Hill:1987wq}
  C.~T.~Hill,
  Annals N.\ Y.\ Acad.\ Sci.\  {\bf 518} (1987) 168.

\bibitem{Glover:1988sr}
  E.~W.~N.~Glover, J.~Ohnemus and S.~S.~D.~Willenbrock,
  Phys.\ Lett.\  B {\bf 206} (1988) 696.

\bibitem{Barger:1986jt}
  V.~D.~Barger, E.~W.~N.~Glover, K.~Hikasa, W.~Y.~Keung, M.~G.~Olsson, C.~J.~Suchyta and X.~R.~Tata,
  Phys.\ Rev.\ Lett.\  {\bf 57} (1986) 1672.


\bibitem{Falkowski:2007hz}
  A.~Falkowski,
  Phys.\ Rev.\  D {\bf 77}, 055018 (2008)
  [arXiv:0711.0828 [hep-ph]].

\bibitem{Giudice:2007fh}
  G.~F.~Giudice, C.~Grojean, A.~Pomarol and R.~Rattazzi,
  JHEP {\bf 0706} (2007) 045
  [arXiv:hep-ph/0703164].

\bibitem{betta}
  E.~Furlan,
  arXiv:1106.4024 [hep-ph].


\bibitem{Cheng:1987rs}
  T.~P.~Cheng and M.~Sher,
  Phys.\ Rev.\  D {\bf 35} (1987) 3484.


\bibitem{Babu:1999me}
  K.~S.~Babu and S.~Nandi,
  Phys.\ Rev.\  D {\bf 62}, 033002 (2000)
  [arXiv:hep-ph/9907213].

\bibitem{Giudice:2008uua}
  G.~F.~Giudice and O.~Lebedev,
  Phys.\ Lett.\  B {\bf 665} (2008) 79
  [arXiv:0804.1753 [hep-ph]].


\bibitem{vanderBij:2007um}
  J.~J.~van der Bij and S.~Dilcher,
  Phys.\ Lett.\  B {\bf 655}, 183 (2007)
  [arXiv:0707.1817 [hep-ph]].

\bibitem{vanderBij:2011wy}
  J.~J.~van der Bij and B.~Pulice,
  arXiv:1104.2062 [hep-ph].
\bibitem{Georgi:2007ek}
  H.~Georgi,
  Phys.\ Rev.\ Lett.\  {\bf 98} (2007) 221601
  [arXiv:hep-ph/0703260].



\bibitem{chaplin}
 S.~Buehler and C.~Duhr,
  arXiv:1106.5739 [hep-ph].


\bibitem{vanHameren:2010cp}
  A.~van Hameren,
  arXiv:1007.4716 [hep-ph].

\bibitem{vanHameren:2009dr}
  A.~van Hameren, C.~G.~Papadopoulos and R.~Pittau,
  JHEP {\bf 0909} (2009) 106
  [arXiv:0903.4665 [hep-ph]].
 
\bibitem{Ellis:2007qk}
  R.~K.~Ellis and G.~Zanderighi,
  JHEP {\bf 0802} (2008) 002
  [arXiv:0712.1851 [hep-ph]].

\bibitem{lhapdf}
{\tt http://hepforge.cedar.ac.uk/lhapdf/} 

\bibitem{Alekhin:2009ni}
  S.~Alekhin, J.~Blumlein, S.~Klein and S.~Moch,
  Phys.\ Rev.\  D {\bf 81} (2010) 014032
  [arXiv:0908.2766 [hep-ph]].

\bibitem{Martin:2009iq}
  A.~D.~Martin, W.~J.~Stirling, R.~S.~Thorne and G.~Watt,
  Eur.\ Phys.\ J.\  C {\bf 63} (2009) 189
  [arXiv:0901.0002 [hep-ph]].

\bibitem{JimenezDelgado:2009tv}
  P.~Jimenez-Delgado and E.~Reya,
  Phys.\ Rev.\  D {\bf 80} (2009) 114011
  [arXiv:0909.1711 [hep-ph]].


\bibitem{Djouadi:1997yw}
  A.~Djouadi, J.~Kalinowski and M.~Spira,
  Comput.\ Phys.\ Commun.\  {\bf 108} (1998) 56
  [arXiv:hep-ph/9704448].




\bibitem{Graudenz:1992pv}
  D.~Graudenz, M.~Spira and P.~M.~Zerwas,
  Phys.\ Rev.\ Lett.\  {\bf 70} (1993) 1372.

\bibitem{Spira:1995rr}
  M.~Spira, A.~Djouadi, D.~Graudenz and P.~M.~Zerwas,
  Nucl.\ Phys.\  B {\bf 453}, 17 (1995)
  [arXiv:hep-ph/9504378].

\bibitem{Dawson:1990zj}
  S.~Dawson,
  Nucl.\ Phys.\  B {\bf 359}, 283 (1991).

\bibitem{Djouadi:1991tka}
  A.~Djouadi, M.~Spira and P.~M.~Zerwas,
  Phys.\ Lett.\  B {\bf 264} (1991) 440.



\bibitem{Harlander:2005rq}
  R.~Harlander and P.~Kant,
  JHEP {\bf 0512}, 015 (2005)
  [arXiv:hep-ph/0509189].

\bibitem{Anastasiou:2006hc}
  C.~Anastasiou, S.~Beerli, S.~Bucherer, A.~Daleo and Z.~Kunszt,
  JHEP {\bf 0701} (2007) 082
  [arXiv:hep-ph/0611236].

\bibitem{Aglietti:2006tp}
  U.~Aglietti, R.~Bonciani, G.~Degrassi and A.~Vicini,
  JHEP {\bf 0701}, 021 (2007)
  [arXiv:hep-ph/0611266].

\bibitem{Ellis:1987xu}
  R.~K.~Ellis, I.~Hinchliffe, M.~Soldate and J.~J.~van der Bij,
  Nucl.\ Phys.\  B {\bf 297} (1988) 221.

\bibitem{Baur:1989cm}
  U.~Baur and E.~W.~N.~Glover,
  Nucl.\ Phys.\  B {\bf 339}, 38 (1990).

\bibitem{Bonciani:2007ex}
  R.~Bonciani, G.~Degrassi and A.~Vicini,
  JHEP {\bf 0711}, 095 (2007)
  [arXiv:0709.4227 [hep-ph]].

\bibitem{Anastasiou:2009kn}
  C.~Anastasiou, S.~Bucherer and Z.~Kunszt,
  JHEP {\bf 0910}, 068 (2009)
  [arXiv:0907.2362 [hep-ph]].



\bibitem{Anastasiou:2010bt}
  C.~Anastasiou, R.~Boughezal and E.~Furlan,
  JHEP {\bf 1006}, 101 (2010)
  [arXiv:1003.4677 [hep-ph]].


\bibitem{Chetyrkin:1997iv}
  K.~G.~Chetyrkin, B.~A.~Kniehl and M.~Steinhauser,
  Phys.\ Rev.\ Lett.\  {\bf 79} (1997) 353
  [arXiv:hep-ph/9705240].

\bibitem{Kramer:1996iq}
  M.~Kramer, E.~Laenen and M.~Spira,
  Nucl.\ Phys.\  B {\bf 511} (1998) 523
  [arXiv:hep-ph/9611272].


\bibitem{Harlander:2001is}
  R.~V.~Harlander and W.~B.~Kilgore,
  Phys.\ Rev.\  D {\bf 64}, 013015 (2001)
  [arXiv:hep-ph/0102241].

\bibitem{Catani:2001ic}
  S.~Catani, D.~de Florian and M.~Grazzini,
  JHEP {\bf 0105}, 025 (2001)
  [arXiv:hep-ph/0102227].


\bibitem{Harlander:2002wh}
  R.~V.~Harlander and W.~B.~Kilgore,
  Phys.\ Rev.\ Lett.\  {\bf 88}, 201801 (2002)
  [arXiv:hep-ph/0201206].

\bibitem{Anastasiou:2002yz}
  C.~Anastasiou and K.~Melnikov,
  Nucl.\ Phys.\  B {\bf 646}, 220 (2002)
  [arXiv:hep-ph/0207004].

\bibitem{Ravindran:2003um}
  V.~Ravindran, J.~Smith and W.~L.~van Neerven,
  Nucl.\ Phys.\  B {\bf 665}, 325 (2003)
  [arXiv:hep-ph/0302135].


\bibitem{Actis:2008ts}
  S.~Actis, G.~Passarino, C.~Sturm and S.~Uccirati,
  Nucl.\ Phys.\  B {\bf 811}, 182 (2009)
  [arXiv:0809.3667 [hep-ph]].
\bibitem{Actis:2008ug}
  S.~Actis, G.~Passarino, C.~Sturm and S.~Uccirati,
  Phys.\ Lett.\  B {\bf 670}, 12 (2008)
  [arXiv:0809.1301 [hep-ph]].

\bibitem{Aglietti:2004nj}
  U.~Aglietti, R.~Bonciani, G.~Degrassi and A.~Vicini,
  Phys.\ Lett.\  B {\bf 595}, 432 (2004)
  [arXiv:hep-ph/0404071].

\bibitem{Keung:2009bs}
  W.~Y.~Keung and F.~J.~Petriello,
  Phys.\ Rev.\  D {\bf 80}, 013007 (2009)
  [arXiv:0905.2775 [hep-ph]].

\bibitem{Anastasiou:2008tj}
  C.~Anastasiou, R.~Boughezal and F.~Petriello,
  JHEP {\bf 0904}, 003 (2009)
  [arXiv:0811.3458 [hep-ph]].

\bibitem{Campbell:2004pu}
  J.~M.~Campbell {\it et al.},
  arXiv:hep-ph/0405302.
\bibitem{Dittmaier:2003ej}
  S.~Dittmaier, M.~1.~Kramer and M.~Spira,
  Phys.\ Rev.\  D {\bf 70} (2004) 074010
  [arXiv:hep-ph/0309204].
\bibitem{Dawson:2003kb}
  S.~Dawson, C.~B.~Jackson, L.~Reina and D.~Wackeroth,
  Phys.\ Rev.\  D {\bf 69} (2004) 074027
  [arXiv:hep-ph/0311067].

\bibitem{Dicus:1998hs}
  D.~Dicus, T.~Stelzer, Z.~Sullivan and S.~Willenbrock,
  Phys.\ Rev.\  D {\bf 59} (1999) 094016
  [arXiv:hep-ph/9811492].
\bibitem{Boos:2003yi}
  E.~Boos and T.~Plehn,
  Phys.\ Rev.\  D {\bf 69}, 094005 (2004)
  [arXiv:hep-ph/0304034].
\bibitem{Maltoni:2003pn}
  F.~Maltoni, Z.~Sullivan and S.~Willenbrock,
  Phys.\ Rev.\  D {\bf 67} (2003) 093005
  [arXiv:hep-ph/0301033].

\bibitem{Harlander:2003ai}
  R.~V.~Harlander, W.~B.~Kilgore,
  Phys.\ Rev.\  {\bf D68}, 013001 (2003).
  [hep-ph/0304035].

\bibitem{bbh@nnlo}
{\tt bbh@nnlo}, numerical implementation of  the calculation in 
Phys. Rev. D 68 (2003) 013001 [hep-ph/0304035], 
http://particle.uni-wuppertal.de/harlander/software/bbh@nnlo/



\bibitem{Beneke:2004km}
  M.~Beneke, A.~P.~Chapovsky, A.~Signer and G.~Zanderighi,
  Nucl.\ Phys.\  B {\bf 686} (2004) 205
  [arXiv:hep-ph/0401002].
 
\bibitem{Denner:2005fg}
  A.~Denner, S.~Dittmaier, M.~Roth and L.~H.~Wieders,
  Nucl.\ Phys.\  B {\bf 724} (2005) 247
  [arXiv:hep-ph/0505042].

\bibitem{Zanderighi:2004qu}
  G.~Zanderighi,
  arXiv:hep-ph/0405124.

\bibitem{Glover:1988rg}
  E.~W.~N.~Glover and J.~J.~van der Bij,
  Nucl.\ Phys.\  B {\bf 321} (1989) 561.


\bibitem{Glover:1988fe}
  E.~W.~N.~Glover and J.~J.~van der Bij,
  Phys.\ Lett.\  B {\bf 219} (1989) 488.


\bibitem{Baur:1990mr}
  U.~Baur and E.~W.~N.~Glover,
  Nucl.\ Phys.\  B {\bf 347} (1990) 12.


\bibitem{Valencia:1992ix}
  G.~Valencia and S.~Willenbrock,
  Phys.\ Rev.\  D {\bf 46} (1992) 2247.

\bibitem{Seymour:1995qg}
  M.~H.~Seymour,
  Phys.\ Lett.\  B {\bf 354} (1995) 409
  [arXiv:hep-ph/9505211].

\bibitem{Passarino:2010qk}
  G.~Passarino, C.~Sturm and S.~Uccirati,
  Nucl.\ Phys.\  B {\bf 834} (2010) 77
  [arXiv:1001.3360 [hep-ph]].





  
\bibitem{Alekhin:2010dd}
  S.~Alekhin, J.~Blumlein, P.~Jimenez-Delgado, S.~Moch, E.~Reya,
  Phys.\ Lett.\  {\bf B697 } (2011)  127-135.
  [arXiv:1011.6259 [hep-ph]].  
  
\bibitem{Anastasiou:2011qw}
  C.~Anastasiou, S.~Buehler, E.~Furlan, F.~Herzog, A.~Lazopoulos,
  [arXiv:1103.3645 [hep-ph]].  
 
\bibitem{Harlander:2009mq}
  R.~V.~Harlander and K.~J.~Ozeren,
  JHEP {\bf 0911} (2009) 088
  [arXiv:0909.3420 [hep-ph]].

\bibitem{Pak:2009dg}
  A.~Pak, M.~Rogal and M.~Steinhauser,
  JHEP {\bf 1002} (2010) 025
  [arXiv:0911.4662 [hep-ph]].
 
\bibitem{ATLAS_TDR}
The ATLAS collaboration, ``Expected Performance of the ATLAS Experiment,", CERN-OPEN-2008-020

\bibitem{deFlorian:2009hc}
  D.~de Florian and M.~Grazzini,
  Phys.\ Lett.\  B {\bf 674} (2009) 291
  [arXiv:0901.2427 [hep-ph]].


\bibitem{Ahrens:2010rs}
  V.~Ahrens, T.~Becher, M.~Neubert and L.~L.~Yang,
  Phys.\ Lett.\  B {\bf 698} (2011) 271
  [arXiv:1008.3162 [hep-ph]].
 

\bibitem{Baglio:2010ae}
  J.~Baglio and A.~Djouadi,
  JHEP {\bf 1103} (2011) 055
  [arXiv:1012.0530 [hep-ph]].

\bibitem{Moch:2005ky}
  S.~Moch and A.~Vogt,
  Phys.\ Lett.\  B {\bf 631} (2005) 48
  [arXiv:hep-ph/0508265].


\bibitem{Alioli:2008tz}
  S.~Alioli, P.~Nason, C.~Oleari and E.~Re,
  JHEP {\bf 0904} (2009) 002
  [arXiv:0812.0578 [hep-ph]].

\bibitem{Hahn:2004fe}
  T.~Hahn,
  Comput.\ Phys.\ Commun.\  {\bf 168} (2005) 78
  [arXiv:hep-ph/0404043].

\bibitem{Whalley:2005nh}
  M.~R.~Whalley, D.~Bourilkov and R.~C.~Group,
  arXiv:hep-ph/0508110.

\bibitem{Brein:2010xj}
  O.~Brein,
  Phys.\ Rev.\  {\bf D81 } (2010)  093006.
  [arXiv:1003.4438 [hep-ph]].



\end{thebibliography}

\end{document}